\documentclass{article}
\usepackage{amsmath,amsthm,amssymb,amsfonts,mathtools,bm,bbm,array,float,mathdots,dsfont,mathrsfs,tensor,setspace,physics}
\usepackage{caption,subcaption,multirow,array,makecell,longtable,lscape,wrapfig,graphicx,comment,placeins}
\usepackage{hyperref,url}
\usepackage[margin=2.5cm]{geometry}
\usepackage[utf8]{inputenc}
\usepackage[numbers,sort&compress]{natbib}
\usepackage{titling}
\usepackage{pgfplots}
\usepackage{tikz-3dplot}
\usepackage{tikz}
\usetikzlibrary{positioning, fit}

\theoremstyle{definition}
\newcommand{\R}{\mathbb{R}}

\newcommand{\AS}[1]{{\color{black} {#1}}}
\newcommand\TSG[1]{{\textcolor{black}{#1}}}

\numberwithin{equation}{section}%subsection originally

\DeclareFontFamily{T1}{calligra}{}
\DeclareFontShape{T1}{calligra}{m}{n}{<->s*[1.44]callig15}{}
\DeclareMathAlphabet\mathcalligra   {T1}{calligra} {m} {n}
\DeclareMathAlphabet\mathzapf       {T1}{pzc} {mb} {it}
\DeclareMathAlphabet\mathchorus     {T1}{qzc} {m} {n}
\DeclareMathAlphabet\mathrsfso      {U}{rsfso}{m}{n}
 
\tdplotsetmaincoords{60}{115}
\pgfplotsset{compat=newest}

\usepackage{dcolumn}% Align table columns on decimal point
\usepackage{bm,url,hyperref}
\usepackage{multirow}

\begin{document}

\begin{titlingpage}

\begin{flushright}
QMUL-PH-25-04\\
\end{flushright}

\vspace{1cm}
    
\centering
{\Large  \textbf{AInstein: \\ Numerical Einstein Metrics via Machine Learning }\par}

\vspace{1cm}
{}\par
\vspace{0.2cm}

{Edward Hirst${}$, Tancredi Schettini Gherardini${}$, Alexander G. Stapleton${}^{*}$}

\vspace{0.6cm}
{\textit{Centre for Theoretical Physics, Queen Mary University of London, E1 4NS, UK} \\
}\par
%\vspace{0.5cm}

\vspace{0.5cm}
{\small
${}^{*}$\texttt{a.g.stapleton@qmul.ac.uk (Corresponding author)}
} \\

\vspace{1.2cm}
%{Report Number: QMUL-PH-25-04}\par
\vspace{1.2cm}
\begin{abstract}
A new semi-supervised machine learning package is introduced which successfully solves the Euclidean vacuum Einstein equations with a cosmological constant, without any symmetry assumptions. 
The model architecture contains subnetworks for each patch in the manifold-defining atlas. Each subnetwork predicts the components of a metric in its associated patch, with the relevant Einstein conditions of the form $R_{\mu \nu} - \lambda g_{\mu \nu} = 0$ being used as independent loss components (here $\mu,\nu = 1, 2, \cdots, n$, where $n$ is the dimension of the Riemannian manifold, and the Einstein constant $\lambda \in \{+1, 0, -1\}$). To ensure the consistency of the global structure of the manifold, another loss component is introduced across the patch subnetworks which enforces the coordinate transformation between the patches, $g' = J^T g J$, for an appropriate analytically known Jacobian $J$.
We test our method for the case of spheres represented by a pair of patches in dimensions 2, 3, 4, and 5. In dimensions 2 and 3, the geometries have been fully classified. However, it is unknown whether a Ricci-flat metric can exist on spheres in dimensions 4 and 5. This work hints against the existence of such a metric.
\end{abstract}
\end{titlingpage}

\section{Introduction}\label{sec:intro}
Finding Einstein metrics on a given manifold has been a central problem in differential geometry for decades. 
An Einstein metric is defined by the condition $Ric(g)=\lambda g$, where $Ric$ is the Ricci curvature tensor, $g$ is the Riemannian metric, and $\lambda$ is a constant. 
These metrics play a prominent role in differential geometry and they are ubiquitous in theoretical physics since they solve Einstein's equations with a cosmological constant. 

Although finding Einstein metrics has been an active area of research for over a century, the field remains vibrant due to numerous unresolved questions. 
Some of them concern the existence (or non-existence) of Einstein metrics on various manifolds, whilst others concern finding appropriate closed-form expressions for those metrics in the cases where their existence has been proven non-constructively. 
Regarding the former type of question, one of the most well-known open problems is whether $S^2 \times S^2$ admits a Ricci-flat (or more generally non-standard Einstein) metric \cite{Besse:1987pua}. 
Similarly, the question of whether $S^n$, with $n>3$, admits non-round Einstein metrics continues to be a major challenge \cite{Berger_2003}. 
Concerning the search for concrete description of metrics which are known to exist, the Calabi-Yau case stands out as the most prominent example \cite{Yau1978OnTR}, but there are many other analogous scenarios, like exotic $7$-spheres \cite{boyer2003einstein, boyer2004einstein}. In addition to its relevance in differential geometry, finding a new Einstein metric numerically represents an impactful result in theoretical physics as well, with immediate applications in ordinary general relativity or higher-dimensional models, like Kaluza--Klein theories or supergravities. These problems accentuate how difficult it is to either construct analytic solutions or prove they cannot exist, especially in the cases where there is little (if any) isometry involved.

This difficulty in verifying existence, and explicitly constructing metrics, has motivated efforts from physicists and mathematicians to explore and develop new methods which are computational in nature. 
Numerical approaches are crucial tools for generating results where analytic techniques are infeasible. 
Already many excellent works have developed numerical schemes to solve Einstein's equations in a variety of scenarios \cite{Figueras:2012xj, Pretorius:2005gq, Dias:2015rxy, Clough_2015, Lehner:2010pn, Dias:2015pda, Chesler:2013lia, Chaurasia:2025}, as far as construction of  black hole/string solutions \cite{Wiseman:2002zc, Kudoh:2003ki, Headrick:2009pv}.
However, these numerical approaches are subject to a curse of dimensionality, where scaling to higher dimensions and more parameters leads to an insurmountable demand for data.

It is here the recent revolution in novel methods of computation statistics can be capitalised on, where copious successes have been seen with application of these techniques across academic fields; techniques of \textit{machine learning}.
In recent years, the first applications of machine learning to numerically approximate metrics have occurred, for complex geometries relevant to string theory, holography, and numerical relativity. 
The most popular compactification spaces for string theory are Calabi-Yau manifolds, where there has been many exceptional works\footnote{Among these works are some very nice packages \cite{Gerdes:2022nzr, Butbaia:2024xgj}, notably \texttt{cymetric} \cite{Larfors:2021pbb} which we take structural inspiration from.} approximating their metrics with machine learning \cite{Ashmore:2019wzb, Douglas:2020hpv, Anderson:2020hux, Jejjala:2020wcc, Larfors:2021pbb, Ashmore:2021ohf, Larfors:2022nep, Berglund:2022gvm, Gerdes:2022nzr, Hendi:2024yin, Ek:2024fgd, Butbaia:2024xgj, Mirjanic:2024gek}; as well as first work towards $G_2$ manifolds \cite{Douglas:2024pmn}.
Other exemplary works numerically solving Einstein's equations for specific manifolds in restricted settings include \cite{deluca2024, Li:2023, chen2024}, where machine learning methods support their approaches.

In this paper, we propose a novel semi-supervised machine learning approach\footnote{The code repository for the package can be found at: \url{https://github.com/xand-stapleton/ainstein}. It is written in \texttt{Python 3} and built on \texttt{TensorFlow} \cite{tensorflow2015whitepaper}.} to approximate general Einstein metrics on a broad class of manifolds. 
We demonstrate its potential by focusing on spheres in various dimensions, with the aim of shedding light on longstanding open problems, providing new perspectives for analysis, and stimulating further research into the numerical and analytical aspects of Einstein geometry. After this successful validation of our method, we aim at applying to other settings with larger relevance in theoretical physics, by looking for black hole solutions and moving to Lorentzian signature.
%\href{https://github.com/xand-stapleton/ainstein}{GitHub}

%%%%%%%%%%%%%%%%%%%%%%%%%%%%%%%%%%%%%%%%%%%%%%%%%
\section{Background}\label{sec:background}
\subsection{Differential Geometry}\label{sec:bkg_dg}
When performing analytic calculations, the use of coordinates in expressions carries some disadvantages. To mention two, it often requires working with cumbersome formulae and can hide the global nature of the objects being described. When tackling a problem via numerical approximation techniques, the situation changes: there is no other choice than to implement coordinate expressions. This prompts the question of which coordinates shall be used to cover the manifolds considered in this work, i.e.~$n$-dimensional spheres. One of the most natural choices consists of the standard stereographic projection atlas. However, since its coordinates span $\mathbb{R}^n$ entirely, sampling and visualising a whole patch becomes non-trivial. For this reason, we use a modified version of the above, where the stereographic projection from $S^n$ to $\mathbb{R}^n$ is followed by a mapping of $\mathbb{R}^n$ to $B^n$, the $n$-dimensional unit open ball. 

Let us define the stereographic atlas as usual. For $S^n$ be defined as the locus $ (\xi_1)^2 + \cdots + (\xi_{n+1})^2 = 1$, and the open covering\footnote{It is standard (and intuitive) to identify the North pole with $(0,...,0,1)$ and the South pole with $(0,...,0,-1)$.} is given by $\{U_1, U_2 \}$, where $ U_1=\{ S^n - \textrm{South Pole} \}$ and $U_2=\{ S^n - \textrm{North Pole} \}$. The two coordinate maps, $\psi_1: U_1 \rightarrow \R^n$ and $\psi_2: U_2 \rightarrow \R^n$, are given respectively by
\begin{equation}
\begin{split}
    (\xi_1, \cdots ,\xi_{n+1}) \mapsto \psi_1  (\xi_1, \cdots ,\xi_{n+1}) & = \frac{1}{1 + \xi_{n+1} } \big(\xi_1, \cdots , \xi_{n}\big) \\ &=: \big ( X_1, \cdots , X_n \big) \, , \\
    (\xi_1, \cdots ,\xi_{n+1}) \mapsto \psi_2  (\xi_1, \cdots ,\xi_{n+1}) &= \frac{1}{1 - \xi_{n+1} } \big(\xi_1,  \cdots ,\xi_{n}\big)\\&=: \big ( \tilde{X}_1 , \cdots , \tilde{X}_n \big) \, .
\end{split}
\end{equation}
Then the map from stereographic coordinates to \textit{ball} coordinates, $\phi_i: \mathbb{R}^n \xrightarrow{} B^n$, reads
\begin{equation}
    \phi_1(X_1, \cdots, X_n) = \left( \frac{X_1}{1+\sqrt{1+|X|^2}}, \cdots , \frac{X_n}{1+\sqrt{1+|X|^2}} \right) =: \left(x_1, \cdots, x_n \right) \, ,
\end{equation}
where $|X|^2 = X_1^2 + \cdots + X_n^2$; and similarly for $\phi_2$, which defines coordinates $(\tilde{x}_1, \cdots, \tilde{x}_n)$ for the second patch. The two ball patches are related by the coordinate transformation
\begin{align}
    \tau(x_1, \cdots , x_n) = \frac{|x| - 1 }{|x| (|x| + 1)} (x_1 , \cdots , x_n) = (\tilde{x}_1, \cdots , \tilde{x}_n ) \, ,
    %y_i = x_i \frac{|x| - 1 }{|x| (1 + |x|)} \, ;
    \label{eq:Change_of_coords}
\end{align}
whilst the entries of the corresponding Jacobian matrix read
\begin{align}
\label{eqn:jacobian}
    J_{ij} = \delta_{ij} \frac{|x|-1}{|x|(1+|x|)} + x_i x_j \frac{1 + 2 |x| - |x|^2 }{ |x|^3 (1 + |x|)^2} \, ,
\end{align}
for $i = 1, 2, \cdots, n$.
It follows from the dependence of the prefactor of \eqref{eq:Change_of_coords} on $|x|$ that co-dimension $1$ spheres centred at the origin in one patch get mapped into co-dimension $1$ spheres in the other. The two radii will be related by the following identity
\begin{align}\label{eq:radii_patchchange}
    r_{\tilde{x}}^2 = \big( \frac{1-r_x}{1+ r_x}\big)^2 \, . 
\end{align}
The \textit{mid-point} radius, which we define to be the radius of the co-dimension $1$ sphere which is mapped to itself under the change of coordinates between the two ball patches, is given by $r_m = \sqrt{2} - 1$.
Consequently, considering the set of points in the ball with radius up to $r_m + \varepsilon$ for both patches is sufficient to have a non-trivial overlap region, and therefore cover the whole manifold. 

On spheres in general dimension, $S^n$, the Einstein equation with positive constant $R_{ij} = \lambda g_{ij}$ for $\lambda = 1$ is solved by the round metric; which in ball coordinates reads
\begin{align}\label{eq:roundmetric}
g_{i j} = \frac{16 (1 - |x|^2)^2 }{(1 + |x|^2)^4} \delta_{i j}  +  \frac{64  }{(1 + |x|^2)^4} x_i x_j \, ,
\end{align}
for both ball patches. The metric is in fact invariant under the change of coordinates between the two patches. For concreteness, in definition of the terms in the Einstein equations, we explicitly state our conventions for the Christoffel symbols and the Ricci tensor in components as\footnote{To be precise, we train the neural network to predict the vielbein, rather than the metric, for convenience. We find this more natural since it lowers the dimension of the output. However, the final stage of the pipeline constructs the metric from the vielbein (according to the Cholesky decomposition - see next section), and the computation of the Ricci tensor is carried out with the standard formulae according to \eqref{eq:Ricci_and_Christoffel}.}
\begin{align}
\label{eq:Ricci_and_Christoffel}
\begin{aligned}
\Gamma_{i j}^k & :=\frac{1}{2} g^{k l}\left(\partial_i g_{j l}+\partial_j g_{i l}-\partial_l g_{i j}\right) \, , \\
R_{j k} & :=\partial_i \Gamma_{j k}^i-\partial_j \Gamma_{k i}^i+\Gamma_{i p}^i \Gamma_{j k}^p-\Gamma_{j p}^i \Gamma_{i k}^p \, .
\end{aligned}
\end{align}
This paper focuses on the Einstein condition above, which can be written globally as $Ric(g) = \lambda g$, for spheres $S^n$ with $n = 2,3,4,5$. Since the Ricci tensor is invariant under conformal scaling, we can restrict to $\lambda \in \{+1, 0, -1\}$ without loss of generality (\cite{Besse:1987pua}). While dimensions $2,3$ are a safe arena to corroborate our method since the metrics are completely classified, dimensions $4,5$ host long-standing open questions regarding the existence of Ricci-flat metrics on spheres for $\lambda = 0$. \\

\subsection{Machine Learning}\label{sec:bkg_ml}
This section outlines the overall structure of the neural network, the regimes in which it may be trained, and the losses which encode the constraints necessary for the model to output a sensible Einstein metric.

The AInstein model is trained to predict the components of the metric $g_{\mu\nu}$ satisfying $R_{\mu\nu} = \lambda g_{\mu\nu}$ given a pair of points in two patches over a given domain. 

Without loss of generality, let $X_\text{Patch 1}$ and $X_\text{Patch 2}$ be a pair of datasets constituting $N$ points of dimension $n$ represented by $n$-tuples from patches $X_\text{Patch 1}$ and $X_\text{Patch 2}$, such that
\begin{align}
X_\text{Patch 1} &:= \{ x_j = (x^0_j, \ldots, x^n_j) \; | \; j \in 0, \ldots, N \} \, , \\
X_\text{Patch 2} &:= \{ \tilde x_j = (\tilde x^0_j, \ldots, \tilde x^n_j) \; | \; j \in 0, \ldots, N \} \, ,
\end{align}
where $j$ indexes the elements of the dataset. Points in $X_\text{Patch 2}$ are related to those in $X_\text{Patch 1}$ by a transition function $T$ such that $\tilde x^i_j = T(x_j^0, \ldots, x_j^n)$. For the specific case of spheres, the map $T$ is identified with $\tau$ in \eqref{eq:Change_of_coords}. Prior to training the network, points in patch 1 are randomly sampled according to the scheme specified in §\ref{app:sampling} in order to generate a set of training data.

As an architecture, we choose a modified multi-layer perceptron network (MLP). In general, an MLP may be defined recursively layer-by-layer, 
\begin{align}
\begin{aligned}
\phi^{(1)}_i &= b^{(1)}_i + w^{(1)}_{ij}x_j \\
h^{(1)}_i &= \sigma(\phi^{(1)}_i) \\
&\vdots \\
\phi^{(l)}_i &= b^{(l)}_i + w^{(l)}_{ij} h^{(l-1)}_j \\
h^{(l)}_i &= \sigma(\phi^{(l)}_i) \, ,
\end{aligned}
\end{align}
where $\phi^{(l)}_i$ is the output of the $l$-th hidden layer pre-activation, and $h^{(l)}_i$ the output of the subsequent activation function used to introduce non-linearity in the network.

Let $\mathcal N_\text{AInstein}$ be a concatenation of a pair of sub-networks $\mathcal N_\text{Patch 1}, \mathcal N_\text{Patch 2}$ each taking input points $x_i$ on their respective patch. More specifically, one may write $\mathcal N_\text{AInstein}$ as
\[
\mathcal N^{\theta_1, \theta_2}_\text{AInstein} := \mathcal N^{\theta_1}_\text{Patch 1} \oplus \mathcal N^{\theta_2}_\text{Patch 2} \, ,
\]
where $\mathcal N^{\theta_1}_\text{Patch 1}$ and $\mathcal{N}^{\theta_2}_\text{Patch 2}$ are the neural networks, parametrised\footnote{The parameters in this case are the set of all weights and biases $w^{(\ell)}$, $b^{(\ell)} \ \forall \ell$.} by variables $\theta_1$ and $\theta_2$, and learn the metric in patches 1 and 2 respectively. For notational simplicity, we choose to henceforth suppress the explicit dependence on $\theta_1$, $\theta_2$, and define $\mathcal N^{\theta_1}_\text{Patch 1} \oplus \mathcal N^{\theta_2}_\text{Patch 2}$ to act such that,
\begin{align}
(\phi_i^{(l)})_\text{AInstein}(x_j, \tilde x_j)&:= (\phi_i^{(l)})_\text{Patch 1}(x_j) \oplus (\phi_i^{(l)})_\text{Patch 2}(\tilde x_j)\\
(h_i^{(l)})_\text{AInstein}(x_j, \tilde x_j) &:= (h_i^{(l)})_\text{Patch 1} (x_j) \oplus (h_i^{(l)})_\text{Patch 2}(\tilde x_j),
\end{align}
for $x_j \in X_\text{Patch 1}$, $\tilde x_j \in X_\text{Patch 2}$.

Assuming the network has depth $D$, we denote the output of the final layer of each subnetwork by
\begin{equation}
    \phi^D_\text{Patch 1}(x) := (g_{ij})_x^\text{Patch 1} \quad \phi^D_\text{Patch 2}(\tilde x):= (g_{ij})^\text{Patch 2}_{\tilde x},
\end{equation}
where it is understood that the patch label is associated with both the (sub)network and the data on which it acts.

\begin{figure}[ht]
\centering
\resizebox{0.5\textwidth}{!}{\begin{tikzpicture}[
    roundnode/.style={circle, draw=black, thick, minimum size=1cm},
    squarednode/.style={rectangle, draw=black, thick, minimum size=1cm, anchor=center},
    ->, thick]

    % First subnetwork with circles
    \node[roundnode] (A1) {$x_{\text{Patch 1}}^i$};
    \node[below=0.4cm of A1] (dots1) {$\vdots$};
    \node[roundnode, below=0.4cm of dots1] (A2) {$x_{\text{Patch 1}}^d$};
    
    % Draw a box around both circles manually
    \node[draw, rectangle, minimum height=6cm, minimum width=1.8cm, anchor=center] (Abox) at ($(A1)!0.5!(A2)$) {}; 
    
    % Hidden layer for Patch 1
    \node[squarednode, right=3.5cm of Abox, yshift=1.5cm] (H1) {$\{H^{(l)}_{\text{Patch 1}}\}$};

    % Create an intermediate point for two-segment connection
    \coordinate (MidH1) at ($(Abox.east) + (2,1.5)$);
    
    % Connect Abox to H1 with two lines
    \draw (Abox.east) -- (MidH1) -- (H1.west);

    % Box for T_patch_2, positioned below H1
    \node[squarednode, right=1.9cm of Abox, yshift=-1.5cm] (T2) {$T_{\text{Patch 2}}$};

    % Arrow from center of right edge of box to center of left edge of T2
    \draw (Abox.east) -- (T2.west);

    % Hidden layer for Patch 2 (aligned horizontally with H1)
    \node[squarednode, right=2.5cm of T2, anchor=center] (H2) {$\{H^{(l)}_{\text{Patch 2}}\}$};
    
    % Center H2 relative to T2 and ensure horizontal alignment
    \draw (T2.east) -- (H2.west);
    
    % Compute vertical center of H1 and H2
    \coordinate (Hmid) at ($(H1)!0.5!(H2)$);
    
    % Concatenation block centered between H1 and H2
    \node[squarednode, right=3.5cm of Hmid] (Concat) {Concat};

    % Intermediate point for two-segment connection from H1
    \coordinate (MidH1Concat) at ($(H1.east) + (1.5,0)$);

    % Connect H1 to Concat with two lines
    \draw (H1.east) -- (MidH1Concat) -- (Concat.west);

    % Connect H2 to Concat with a straight line
    \draw (H2.east) -- (Concat.west);

    % Output layer centered below Concat
    \node[roundnode, right=1.5cm of Concat] (O) {O};

    % Connect Concat to Output
    \draw (Concat.east) -- (O.west);

\end{tikzpicture}}
\caption{Overview sketch of the AInstein architecture. Here, $T$ is a patch transition function layer which converts the points in patch 1 to their equivalents in patch 2, $\{H^{(l)}_\text{Patch $p$}\}$ a set of hidden layers with non-linear activations for each patch, and `Concat' a concatenation layer, then followed by a Cholesky transform on the output of the hidden states in the pipeline, producing the metrics on both patches.}
\label{fig:machine_learning_arch}
\end{figure}
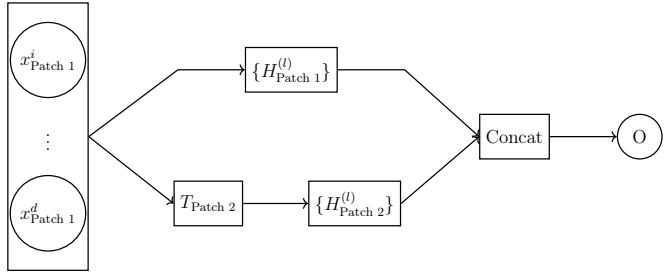

The full model provides an output $\mathcal N_\text{AInstein}(x, \tilde x) = (g_{ij})_x^\text{Patch 1} \oplus (g_{ij})_{\tilde{x}}^\text{Patch 2}$., where this predicted metric is evaluated at points $(x, \tilde x)$ from each patch respectively. The sub-networks are trained simultaneously subject to the loss function defined in equation \eqref{eq:whole_loss}. The architecture is depicted in Figure \ref{fig:machine_learning_arch}.

The specified model, $\mathcal N_\text{AInstein}$, may be trained in two regimes: the \textit{supervised} regime and the \textit{semi-supervised} regime. The main contribution of this work arises from training the model in the semi-supervised regime subject to the losses presented in §\ref{ssub:losses}. To enhance training convergence, rather than initializing the network's weights from a random configuration, one can leverage the identical architecture shared between the supervised and unsupervised models. Specifically, the initialization can be derived from the parameters obtained by training the supervised model on a known function. %Such a technique is philosophically akin to fine-tuning a foundation model.

\subsubsection{Supervised Reference Models}
In the supervised regime of training the $\mathcal{N}_\text{AInstein}$ architecture, the outputs of the function are known in the training data.
Therefore for every input point $x$ the output metric is known for that point $g_{ij}(x)$, such that the training seeks to minimise a mean squared error loss between the known metric components at each point and the components predicted by the model.

The known Einstein metric on the sphere we consider is the round metric, for $\lambda = +1$, as defined in \eqref{eq:roundmetric}.
By training the same architecture in a supervised manner, using explicitly computed round metric components as output, the architecture is trained to model this round metric.
This is important as the test loss scores on this trained metric set an important baseline for comparison, with full knowledge of the output metric values for the manifold points, how well can an Einstein metric be modelled with the allocated computational resources.
These loss scores are hence reported alongside the semi-supervised test losses, dictating the loss order which indicates the architecture has learnt a metric function which truly exists.

Further to a supervised training of the round metric, the supervised architecture can be used to design intelligent starting points for the model.
With random initialisation of the parameters $(\theta_1,\theta_2)$ the initial function represented by $\mathcal{N}_\text{AInstein}$ is far from smooth which leads to a blow up of Einstein loss values, however if we could pick parameter values which represented a smoother function the loss order would initialise within a computable range and encourage sensible learning.
To do this we choose to train a supervised model to predict the identity function in each patch ($g_{ij}=\delta_{ij}$), an ansatz which is completely flat and hence also smooth, despite substantially violating the overlap condition.
This is trained again with a mean squared error loss, now with network outputs which match $\delta_{ij}$ for \textit{every} input point.
After training, the parameters $(\theta_1,\theta_2)$ are saved, and used to initialise the $\mathcal{N}_\text{AInstein}$ function in the semi-supervised training (as well as the supervised training of the round metric to ensure fair comparison).

\subsubsection{Semi-Supervised Loss Components}
\label{ssub:losses}

As with all deep learning tasks, we must supply the network with a loss function to use during training. This loss acts on both subnetworks simultaneously, and contains a set of designed loss components, from which we consider their weighted sum, which is minimised where the output and $\mathcal{N}_\text{AInstein}$ function represents a sensible Einstein metric.

The loss may be written as
% \begin{align}
% \mathcal L_\text{AInstein}[\theta_1, \theta_2](g_x^\text{Patch 1}, g_{\tilde x}^\text{Patch 2}) &:= f_1 \mathcal L^\text{Einstein}_\text{Patch 1}[\theta_1, \theta_2] (g_x^\text{Patch 1}) + f_2\mathcal L^\text{Finiteness}_\text{Patch 1}[\theta_1, \theta_2](g_x^\text{Patch 1}) \\ &+ f_3\mathcal L^\text{Overlap}_\text{Patch 1}[\theta_1, \theta_2](g_x)^\text{Patch 1} + \nonumber (\text{Patch 1} \leftrightarrow \text{Patch 2}),
% \end{align}
%

\begin{align}
\label{eq:whole_loss}
\mathcal{L}_\text{AInstein}[\theta_1, \theta_2](g_x^\text{Patch 1}, g_{\tilde{x}}^\text{Patch 2})
:= \ &f_1 \bigg( \mathcal{L}^\text{Einstein}_\text{Patch 1}[\theta_1] (g_x^\text{Patch 1}) + \mathcal{L}^\text{Einstein}_\text{Patch 2}[\theta_2] (g_{\tilde{x}}^\text{Patch 2}) \bigg) \nonumber \\
 + &f_2 \bigg(\mathcal{L}^\text{Overlap}[\theta_1, \theta_2](g_x^\text{Patch 1},g_{\tilde{x}}^\text{Patch 2})\bigg)\nonumber \\
+ &f_3 \bigg(\mathcal{L}^\text{Finiteness}_\text{Patch 1}[\theta_1](g_x^\text{Patch 1}) + \mathcal{L}^\text{Finiteness}_\text{Patch 2}[\theta_2](g_{\tilde{x}}^\text{Patch 2})\bigg)\;,
\end{align}
where $f_i$ are the respective loss term multipliers, specifying the relative importance of the loss components; practically we used $(f_1,f_2,f_3) = (1,10,1)$.
Each loss component implicitly contains a filter, which weights the contribution of points depending on which part of the patches are most important to that loss, this improves the global metric learning and additionally improves numerical stability; more information is provided in §\ref{app:filters}. \AS{Notice that the loss components can be separated into three distinct components: those which depend on the output of the patch 1 sub-network; those which depend on the output of the patch 2 subnetwork; and those which depend on both. As such, in order to increase the training speed (particularly when the dimensionality of the manifold is high), one may employ a federated learning regime \cite{kairouz2021advancesopenproblemsfederated} and segregate the learning into three parts, and moderated by some master node. Of course, since the total loss also depends on the overlap region in both patches, such a federated learning approach remains bottlenecked by the slowest loss term evaluation. The implementation of such an optimisation is beyond the scope of this work, but remains an open avenue for future exploration.} We now describe these loss components in detail.

\paragraph{Einstein loss}
To satisfy the Einstein condition of the solution, we impose the following loss term:
\begin{equation}
    \mathcal L^\text{Einstein}_\text{Patch $p$}[\theta_p](g_x^\text{Patch $p$}) := ||\lambda (g_{ij})_x^\text{Patch $p$} -  (R_{ij})_{x}^\text{Patch $p$}|| \, ,
    \label{eq:Einstein_loss}
\end{equation}
where $p \in \{1, 2 \}$, $\lambda$ is the Einstein constant, $R_{ij}$ is the \textit{Ricci tensor}, and $||\cdot||$ represents the Euclidean 2-norm. By inspection, it is evident the Einstein loss term penalises metrics which deviate far from $\lambda R_{ij}$ evaluated at point $x$. 
This loss term is weighted according to the point's radial coordinate, as described in §\ref{app:filters}, to prioritise points in the patch region used in defining the global metric model.

\paragraph{Overlap loss}
This loss component enforces the gluing condition of the patches, ensuring the metric evaluated on points in one patch is consistent with the companion metric evaluated on equivalent points in the other.

Concretely, let $x_j \in X_\text{Patch 1}$ possess an associated point\footnote{Here we consider that $x_j$ and $\tilde x_j$ are finite.} $\tilde x_j \in X_\text{Patch 2}$ related by $\tilde x_j = T(x_j)$, where $T$ is an appropriate transition function. Equivalently, the metric is related by the Jacobian matrix $J$, meaning one may write an overlap loss as
\begin{equation}
\begin{split}
   &\mathcal L^{\text{Overlap}}[\theta_1, \theta_2](g^\text{Patch 1}_{x},g^\text{Patch 2}_{\tilde{x}}) :=  ||(g_{ij})_x^\text{Patch 1} - J_{ki} (g_{kl})^\text{Patch 2}_{\tilde x} J_{lj}|| \\
    & \quad + ||J_{ki}(g_{kl})_x^\text{Patch 1}J_{lj} -  (g_{ij})^\text{Patch 2}_{\tilde x}|| \;,
\end{split}
\end{equation}
where $J$ is the analytically known Jacobian matrix corresponding to the change of coordinates between the two patches. For the case of spheres, it is given by \eqref{eqn:jacobian}, and is equal to its inverse. % and transpose?
This loss term is also weighted according to the point's radial coordinate, as described in §\ref{app:filters}, but in a different way to prioritise points in the overlap region.

\paragraph{Finiteness loss}
In order to ensure finiteness, and discourage the machine learning algorithm from approaching the ``zero-metric'' ($g_{ij} \sim 0$), we introduce a loss which takes the form:\footnote{By inspection, we found that the neural network tended to minimise the loss \eqref{eq:Einstein_loss} by simultaneously making the components of the metric and of the Ricci tensor smaller and smaller. This is clearly a numerical artifact: the violation of the Einstein condition should be small compared to the components of the metric, not just in absolute terms.}

\begin{equation}
\begin{split}
\mathcal L^\text{Finiteness}_\text{Patch p}[\theta_p](g^\text{Patch p}_{x}) := 1 &+ \left(h \, e^{-\left(\frac{F - c_f}{w_f}\right)^{t_f}}- h \right)^2 
+ \left(\frac{F - (c_f + w_f )}{ s } \right) \cdot \frac{1+\tanh \left(\frac{F - (c_f + w_f )}{2}\right)}{2} \\ 
&+ \left(\frac{-F + (c_f - w_f)}{s}\right) \cdot \frac{1+\tanh \left(\frac{-F + (c_f - w_f )}{2}\right)}{2} \, ,
\end{split}
\end{equation}
for parameters $(F,h,c_f,w_f,t_f,s)$. Here $F = \sum_{i,j} ||(g_{ij})^\text{Patch p}_x||$ is the sum of the absolute value of all the components of the metric; $h$ controls the ``height'' of the well with centre $c_f$, and width $w_f$. Moreover, $t_f$ controls how vertical the walls are, and $s$ determines the gradient of the ``slopes'' which emanate from the well. A plot of this filter, as used in the finiteness loss is shown in Figure \ref{fig:filters_f}. The motivation for this filter is to avoid the components of the predicted metric getting arbitrarily close to zero. Since it involves the sum of the components of the predicted metric, we supplement it with a dimension-dependent normalisation.

\paragraph{Filters}
Many of the loss components, if implemented na\"{i}vely for the case of spheres, result in training behaviours which are unpredictable and numerically unstable. This is because of the pathological behaviour of the metric and the Jacobian as one approaches the boundary of the unit ball (see \eqref{eqn:jacobian} and \eqref{eq:roundmetric}). As such, we introduce a set of loss `filters' which smoothly\footnote{It is important each filter is smooth and differentiable to enable derivatives to be taken for back-propagation.} suppress each loss component's contribution depending on the location of the point being evaluated in the patch.

To explain our choices in more detail, let us consider the usual stereographic projection (or simple modifications of it); both patches cover the whole sphere with the exception of their associated pole. Since the overlap consists of the whole sphere excluding the poles, in the coordinates of each patch, the overlap region is the whole of $\mathbb{R}^n$ (or $B^n$, if working with ball coordinates) excluding the origin. If the patches are made smaller, the overlap region shrinks. As discussed in §\ref{sec:bkg_dg} (see \eqref{eq:radii_patchchange} and the following comments), we can choose our charts to consist of the $n$-ball with radius $r_m +  \varepsilon$, and the corresponding overlap region is the annulus between $\frac{1 - r_m -  \varepsilon}{1 + r_m +  \varepsilon}$ and $r_m +  \varepsilon$. This is what is implemented in our code, with the choice of $\varepsilon$ being one of the hyperparameters. With these charts, one needs to introduce a filter that devalues contributions from points whose radius is larger than $r_m +  \varepsilon$ when evaluating the Einstein condition; because they are not contained within the patch. When calculating the overlap loss, another filter should devalue points outside of the annulus overlap region described above.
These filters are described further in §\ref{app:filters}. \TSG{We emphasise that these filters are \textit{fixed}, i.e.~do not change during training, just like the Jacobian. This is because they represent a way of encoding the manifold's differentiable structure numerically; the aim of AInstein is to find Einstein metrics on a \textit{given} differentiable manifold. If one were to update the Jacobian and filters during training, then the differentiable manifold would be constantly changing, which hard to justify and interpret mathematically, since it would be equivalent to defining new metrics without specifying the underlying manifold.}

% Finally, we artificially add a filter which penalises metrics with entries which are too small or too large, to exclude numerical artifacts. These are discussed further in the §\ref{app:filters}.

\subsubsection{Global Test Loss}\label{ssub:global_loss}
As described in §\ref{ssub:losses}, the final global model of the trained metrics are restricted to patches of radii $r_m + \varepsilon$, using an overlap region of radii $\in [\frac{1 - r_m -  \varepsilon}{1 + r_m +  \varepsilon}, r_m +  \varepsilon]$.
Where the training loss used includes weighted contributions from \textit{all} points (which improves the learning), our final testing evaluation is restricted to just these patch regions required for global definition.

The training filters are hence converted into hard cutoffs, and the finiteness loss which has non-geometric motivation is ignored.
Hence, the global test loss, as reported in the results of §\ref{sec:results}, is defined 
\begin{align}
\label{eq:global_loss}
\mathcal{L}_\text{Global}[\theta_1, \theta_2](g_x^\text{Patch 1}, g_{\tilde{x}}^\text{Patch 2}) :=
&f_1 \bigg( \mathcal{L}^\text{Einstein}_\text{Patch 1}[\theta_1] (g_x^\text{Patch 1} \; \big| \; ||x|| < r_m +\varepsilon)\\ \nonumber
&\qquad + \mathcal{L}^\text{Einstein}_\text{Patch 2}[\theta_2] (g_{\tilde{x}}^\text{Patch 2} \; \big| \; ||\tilde{x}|| < r_m+\varepsilon) \bigg) \\
+ &f_2 \bigg(\mathcal{L}^\text{Overlap}[\theta_1, \theta_2] (g_x^\text{Patch 1},g_{\tilde{x}}^\text{Patch 2} \; \big| \; ||x|| \in \bigg[\frac{1 - r_m -  \varepsilon}{1 + r_m +  \varepsilon}, r_m +  \varepsilon \bigg])\bigg) \;, \nonumber
\end{align}
where $||\cdot||$ indicates the 2-norm of the input point, which equals its radial coordinate.

%%%%%%%%%%%%%%%%%%%%%%%%%%%%%%%%%%%%%%%%%%%%%%%%%
\section{Results}\label{sec:results}
To train the network and obtain Einstein metrics, data must first be generated.
To match the architecture style described in §\ref{sec:bkg_ml}, where the input is a point's coordinates in one patch and the output is the metric vielbeins for all patches, the data need only be generated for the first patch.

The patches are represented in ball coordinates such that for the sphere $S^n$ each patch is a unit $B^n$, which we parameterise by $n$ Euclidean coordinates evaluating in the range $x_i \in (-1,1)$.
The patch is sampled using a modified beta distribution, designed to prioritise the patch overlap region and minimise numerical instabilities; more details are given in §\ref{app:sampling}, including exemplary plots of the distributions in 2d.

For training, the number of points sampled were $(10^4, 10^4, 10^5, 10^5)$ for dimensions $(2,3,4,5)$ respectively; consistently resampled across all runs and Einstein constants, where testing used $10^4$ independently sampled points.
Traditionally, exponential increases in the sampling size is desired as data dimension increases, which makes the displayed results for higher dimensions all the more impressive. 
Consequently, we would also expect performances to improve further with more training data.

Once the patch data has been sampled, the NN architecture is initialised. 
For the hyperparameters introduced in §\ref{sec:bkg_ml} and listed in §\ref{app:hyperparams}, the model parameters are set such that the metrics are identity matrices. 
To do this a supervised model with the same architecture is first trained on independent inputs sampled equivalently, paired with output vielbeins which produce the identity matrix for every point in both patches.
Four networks of this form are trained for each of the four considered dimensions, and their parameters are used to initiate each model of that respective dimension in the subsequent learning.
These start points\footnote{Preliminary investigations used random initialisations for the model parameters, but since the metrics they represented were so far from being smooth the Einstein loss condition blew up, obstructing sensible learning and often exceeding the floating point memory limit.} are by nature smooth, and violate the Einstein equation in each patch to an order comparable with the metric components, but are exceptionally far from satisfying the overlap gluing conditions between the patches, hence representing \textit{non-geometric} starting points.
It is worth emphasising here that these identity function start point are completely independent of the problem, or any knowledge of solutions, they can be quickly and cheaply defined for any dimension and proved surprisingly effective. 

With the data sampled, and architectures initialised, 10 independent runs were performed for each investigation (over varying dimension and Einstein constant), and final performances were evaluated with the Global test loss described in §\ref{ssub:global_loss}.
However, an additional means of assessing the test loss measures was also devised.
Since the round metric is known to exist as a solution for $\lambda = +1$ in all dimensions, and the explicit metric form can be computed for any input patch point using \eqref{eq:roundmetric}, a supervised model can be trained to explicitly model this metric.
This is done by training the same architecture, also initialised with the same pre-trained identity functions, with MSE loss on input-output pairs of the point coordinates in patch 1 and the round metric vielbein coordinates for that point in both patches.
These were trained with the same hyperparameters for each dimension, had Global test loss scores equivalently independently evaluated, and provide an important comparison baseline for the main investigation test loss scores.
These baselines represent the feasible limit of solutions to the Einstein equations from these techniques with the compute resources provided. %remove this statement?

\subsection{Local Einstein Geometries}
As a warm-up, to test the effectiveness of the Einstein loss, and the code functionality, we begin with a single patch.
By working in a single patch without boundary conditions, the architecture is being trained to find Einstein metrics on a space which is topologically equivalent to $\mathbb{R}^n$.
The solutions to the Einstein equations in the cases of $\lambda \in \{-1,0,+1\}$ are known, and represent spherical, flat, and hyperbolic spaces, often expressed with trigonometric functions, and which here would be restricted to the ball patch (hence `local').
%The ball patch used can be continuously mapped to $\mathbb{R}^n$.

The data is sampled in the same way, except the NN metric architecture is set up with only a single patch subnetwork, outputting the metric vielbein for the input patch alone.
Since there is only one patch, the overlap loss is redundant, and hence ignored.
This leaves the Einstein loss and finiteness loss as the only terms in the training loss, where each is now only for the single patch.
The multiplier weightings of these two losses are set as equal to mirror the behaviour in \eqref{eq:whole_loss} for the full training loss, and the global test loss has only a single contributing term: the Einstein loss for the patch. Experimental neural network loss values are provided for seed averaged runs in Table \ref{tab:local_losses}.

\begin{figure*}[!t]
    \centering
    \begin{subfigure}{0.32\textwidth}
        \centering
        \includegraphics[width=0.98\textwidth]{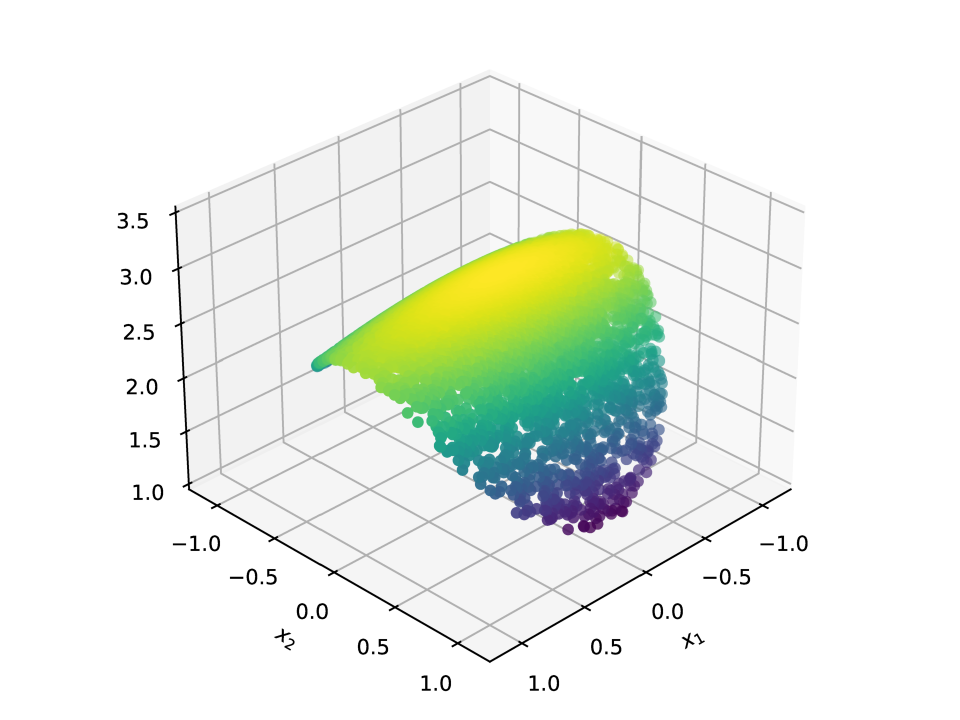}
        \caption{$g_{00}$ ($\lambda = +1$)}
    \end{subfigure} 
    \begin{subfigure}{0.32\textwidth}
        \centering
        \includegraphics[width=0.98\textwidth]{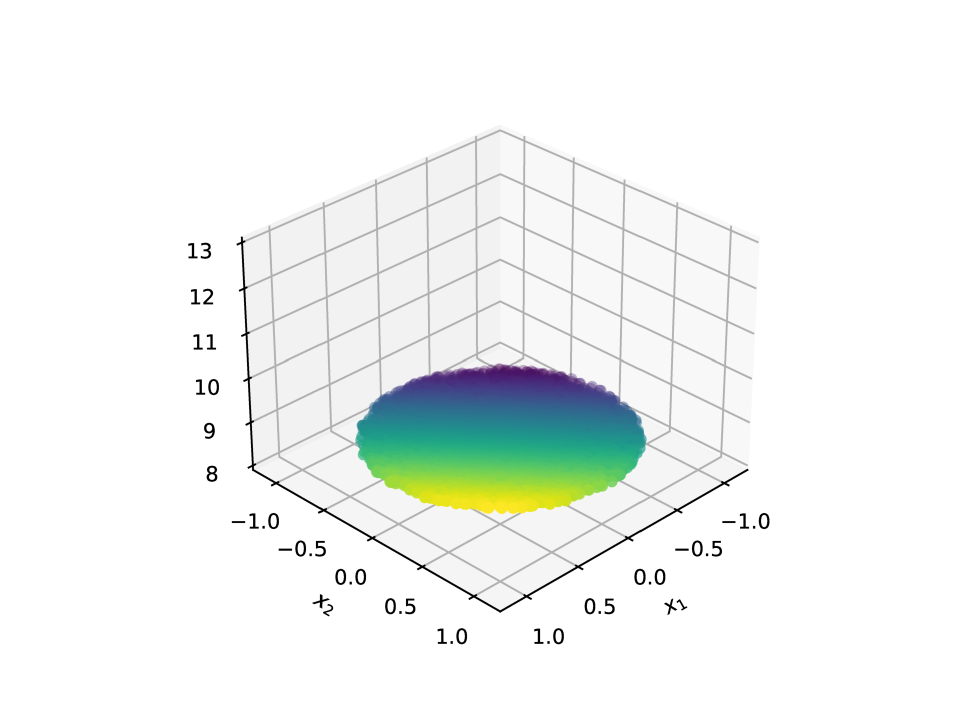}
        \caption{$g_{00}$ ($\lambda = 0$)}
    \end{subfigure}
    \begin{subfigure}{0.32\textwidth}
        \centering
        \includegraphics[width=0.98\textwidth]{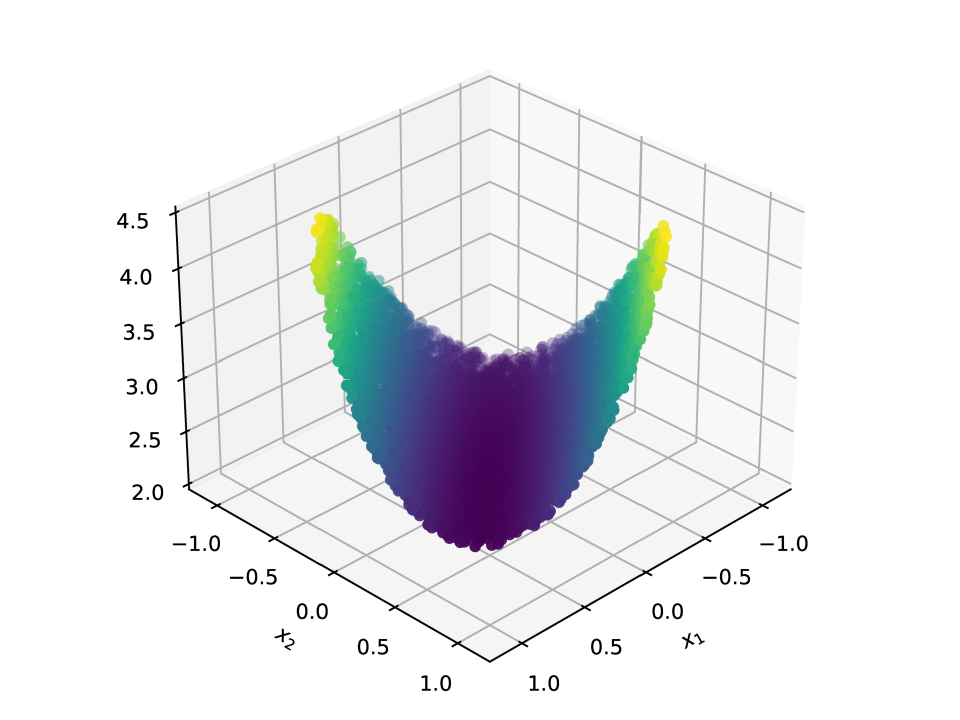}
        \caption{$g_{00}$ ($\lambda = -1$)}
    \end{subfigure}\\
    \begin{subfigure}{0.32\textwidth}
        \centering
        \includegraphics[width=0.98\textwidth]{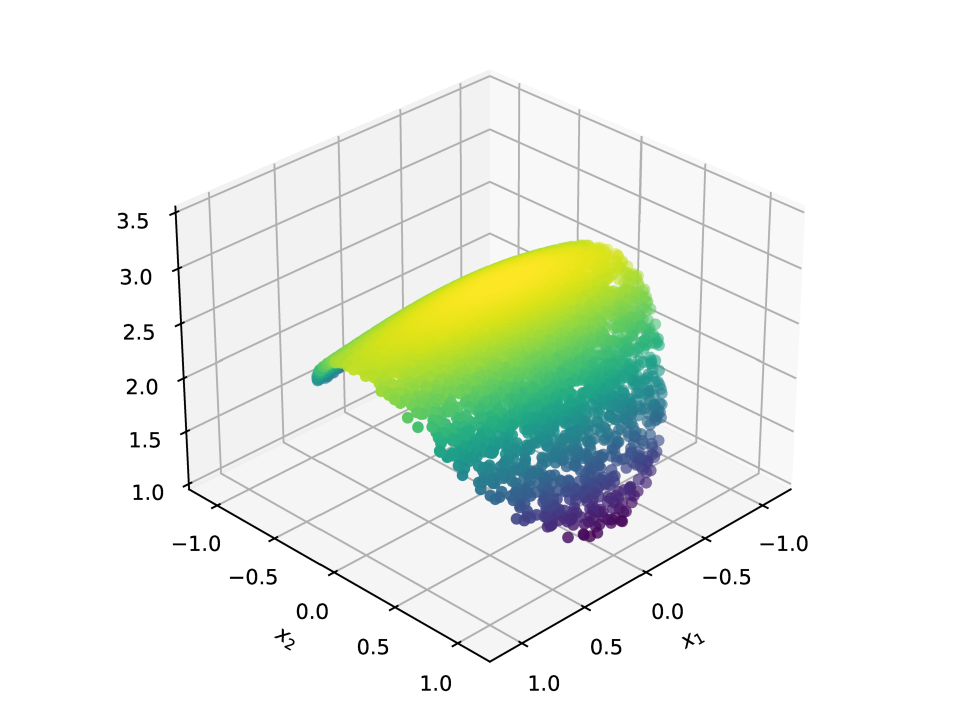}
        \caption{$R_{00}$ ($\lambda = +1$)}
    \end{subfigure} 
    \begin{subfigure}{0.32\textwidth}
        \centering
        \includegraphics[width=0.98\textwidth]{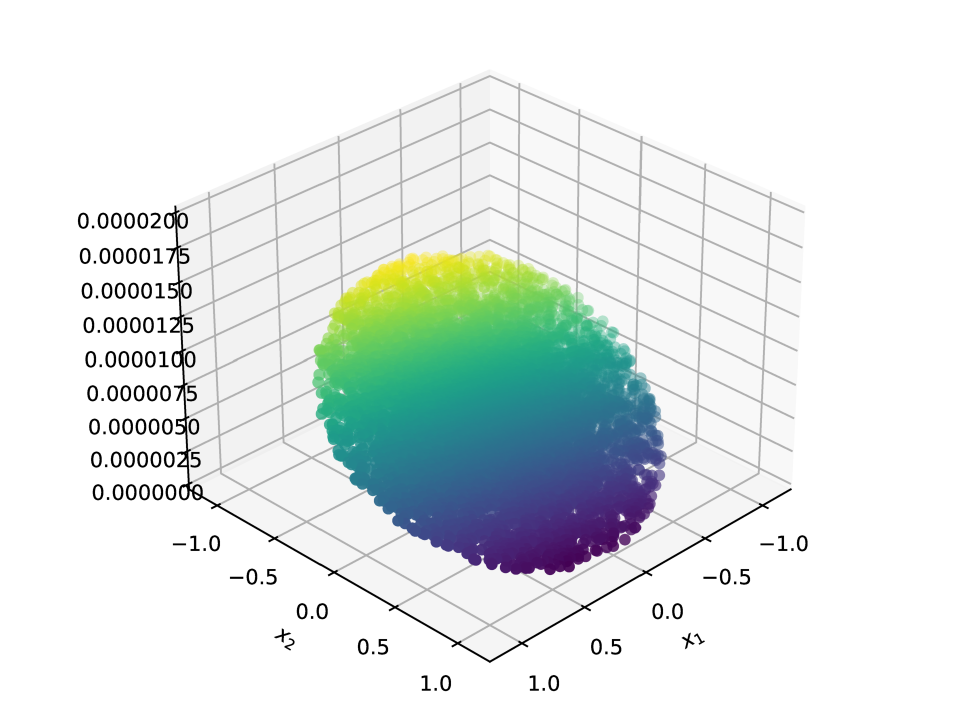}
        \caption{$R_{00}$ ($\lambda = 0$)}
    \end{subfigure}
    \begin{subfigure}{0.32\textwidth}
        \centering
        \includegraphics[width=0.98\textwidth]{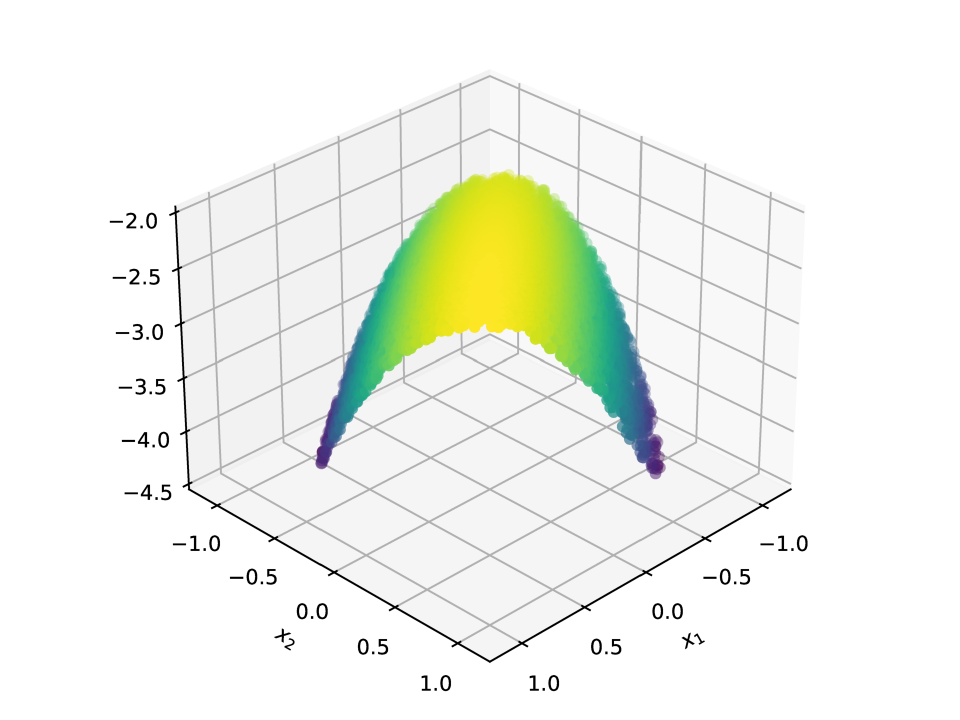}
        \caption{$R_{00}$ ($\lambda = -1$)}
    \end{subfigure}
    \caption{Visualisations of the $(0,0)$ components of the learnt metrics and their respective Ricci tensors, in 2d on a single patch. These metrics solve the Einstein equation with Einstein constants of $\lambda \in \{+1, 0, -1\}$ respectively. We emphasise the $R_{00}$ $(\lambda = 0)$ scale is $\sim 10^{-5}$, indicating Ricci-flat.}
    \label{fig:vis_2d1p}
\end{figure*}

Training with the same hyperparameters, as stated in §\ref{app:hyperparams}, 10 runs for each $\lambda$ value were performed for a 2d ball patch, starting from the same identity initialisation\footnote{Since the architecture has changed by removing one subnetwork for the second patch, technically a new 1-patch version of the supervised identity function was trained to be used for initialisation.}.
The trained metrics were evaluated on independent sample sets, and test Einstein losses computed, reported in Table \ref{tab:local_losses}.
Visualisation of the $(0,0)$ components for a single run are shown in Figure \ref{fig:vis_2d1p}, were the other components had similar behaviour.

\begin{table}[!t]
\centering
{\small
\begin{tabular}{|c|ccc|}
\hline
\multirow{2}{*}{\begin{tabular}[c]{@{}c@{}}Loss\\ Component\end{tabular}} & \multicolumn{3}{c|}{Einstein Constant $\lambda$}                            \\ \cline{2-4} 
& \multicolumn{1}{c|}{$+1$} & \multicolumn{1}{c|}{$0$} & $-1$ \\ \hline
Einstein & \multicolumn{1}{c|}{0.038 $\pm$ 0.016}    & \multicolumn{1}{c|}{0.000 $\pm$ 0.000}    &   0.025 $\pm$ 0.017  \\ \hline
\end{tabular}
}
\caption{Global test loss results averaged over 10 runs, for NN approximations of Einstein metrics with the respective curvatures on single patches in 2d; note overlap loss not applicable, so the global loss's only contribution is from the Einstein loss. All losses are reported to 3 decimal places with standard deviations across their 10 runs.}
\label{tab:local_losses}
%{+,0,-} e_losses: [3.78433230e-02, 6.07318622e-07, 2.48807400e-02] \pm [1.61231648e-02, 1.37953385e-06, 1.69388458e-02]
\end{table}

The losses in Table \ref{tab:local_losses} are all very low, significantly $<1$. 
The $\lambda=0$ case is especially low since the identity initialisation already satisfies this Einstein condition, but the others which are initialised not satisfying the condition modify their metrics to satisfy the condition well, reaching similar performance scores up to error.
The visualisations on Figure \ref{fig:vis_2d1p} are especially insightful, the shapes show trigonometric-like behaviour, matching the expected style.
The computed Ricci tensors on the test data for $\lambda=+1$ look identical to the metric, for $\lambda=0$ are near-identically 0 throughout the patch, and for $\lambda=-1$ is the negation of the metric.

These performances validate this machine learning approach nicely, and set up scope for development to more non-trivial manifolds, with boundary conditions or further patches -- the latter we focus on now.

\subsection{Global Einstein Geometries on Spheres}
Extending the setup to a more non-trivial manifold, one wishes to consider multiple patches satisfying a gluing condition on their overlap.
In this work, we do this by considering spheres, $S^n$, covered by an atlas with 2 patches, as described in §\ref{sec:bkg_dg}.
The gluing condition, associated to the transition function of the atlas, is defined over the patches with a weighting that prioritises an overlap region for radii $\sim r_m$, and is packaged within an overlap loss term.
This is coupled with the Einstein loss and the finiteness loss defined for both patches in the full training loss, as described in §\ref{sec:bkg_ml}.

In performing these investigations, again 10 runs were trained for each investigation spanning the Einstein constants $\lambda \in \{+1,0,-1\}$ and dimensions $\{2,3,4,5\}$. 
The architectures were initialised using the parameters from a supervised pre-trained identity function for each patch, and run with hyperparameters as specified in §\ref{app:hyperparams}.
The trained metrics were evaluated using the Global test loss, which had only Einstein and overlap contributions as described in \eqref{eq:global_loss}.
The Einstein contribution was computed on each patch for test points within a restricted radii of $r_m+0.1$, and the overlap contribution was computed for the test points with radii in the range $[\frac{1-(r_m+0.1)}{1+(r_m+0.1)}, r_m+0.1]$, which selects the same points for both patches.
We emphasise that a restriction of $r_m + \varepsilon$ for $0 < \varepsilon << 1$ is sufficient to give a global description of the manifolds, but to ensure sufficient test data for each loss this was expanded to an upper width given by $\varepsilon = 0.1$.
The proportion of test points in each patch and the overlap region was remarkably consistent across each runs metric testing for all $\lambda$ and dimension.
The average proportions of test points in the (restricted patch 1, restricted patch 2, overlap region) were $(0.594, 0.594, 0.188)$, which can be multiplied by $10^4$ to get the number of points contributing to each loss term. \TSG{Note that these figures change if one modifies either the parameters of the beta distribution or the value of $\varepsilon$. They are tuned in order to obtain a significant fraction of samples belonging to the overlap region; the reason behind it being very evident: one would like to ensure that the network sees enough points for which the overlap condition applies. With very few samples in the overlap region, the global structure of the manifold is not prioritised.} %[0.59375583, 0.59441, 0.18816583]

The average Global test losses are shown in Table \ref{tab:global_test_losses}, with a breakdown into the sublosses in §\ref{app:extra_results}.
In addition to reporting Global test losses for the considered dimensions and $\lambda$ values run with the semi-supervised architecture, results are also reported for supervised models trained to model the analytic round metric defined in \eqref{eq:roundmetric} which solves the Einstein equations with $\lambda=+1$.
The supervised Global test loss scores set a threshold for learning a true metric, as we know the round metric to exist in all dimensions.

\begin{table*}[!t]
\centering
{\small
\begin{tabular}{|c|ccc!{\vrule width 1.5pt}c|}
\hline
\multirow{2}{*}{Dim} & \multicolumn{3}{c!{\vrule width 1.5pt}}{Einstein Constant $\lambda$} & \multirow{2}{*}{\begin{tabular}[c]{@{}c@{}}Supervised\\ $\lambda = +1$\end{tabular}} \\ \cline{2-4} 
& \multicolumn{1}{c|}{$+1$} & \multicolumn{1}{c|}{$0$} & $-1$ &  \\ \hline
2                          & \multicolumn{1}{c|}{0.083 $\pm$ 0.023} & \multicolumn{1}{c|}{2.881 $\pm$ 0.113} & \multicolumn{1}{c!{\vrule width 1.5pt}}{4.364 $\pm$ 0.093} & 0.096 $\pm$ 0.013 \\ \hline
3                          & \multicolumn{1}{c|}{0.151 $\pm$ 0.027} & \multicolumn{1}{c|}{5.560 $\pm$ 0.160} & 8.641 $\pm$ 0.183 & 0.195 $\pm$ 0.020 \\ \hline
4                          & \multicolumn{1}{c|}{0.150 $\pm$ 0.018} & \multicolumn{1}{c|}{8.494 $\pm$ 0.121} & 14.928 $\pm$ 1.317 & 0.248 $\pm$ 0.024 \\ \hline
5                          & \multicolumn{1}{c|}{0.244 $\pm$ 0.039} & \multicolumn{1}{c|}{10.810 $\pm$ 0.185} & 18.798 $\pm$ 2.024 & 0.518 $\pm$ 0.063 \\ \hline
\end{tabular}
}
\caption{Global test loss results averaged over 10 runs, for NN approximations of Einstein metrics with the respective curvatures on spheres in dimensions 2-5 (2-patches). For comparison, the right-hand column shows the respective global test losses for the \textit{supervised} NN model approximation of the analytic round metric (which satisfies the Einstein equation for $\lambda = +1$). All losses are reported with standard deviations across their 10 runs.} %trained for 500 epochs
\label{tab:global_test_losses}
\end{table*}

\begin{comment}
\begin{table}[!t]
\centering
\begin{tabular}{|c|c|}
\hline
Dimension & Loss Value \\ \hline
2         & $0.096 \pm 0.013$        \\ \hline
3         & $0.195 \pm 0.020$        \\ \hline
4         & $0.248 \pm 0.024$        \\ \hline
5         & $0.518 \pm 0.063$        \\ \hline
\end{tabular}
\caption{Global test loss results averaged over 10 runs,. Losses computed for \textit{supervised} NN approximations of the analytic round metric on spheres of dimensions 2-5 (2-patches). Losses are reported with standard deviations across the 10 runs.}
\label{tab:supervised_global_test_losses}
\end{table}
\end{comment}

Interpreting the losses, one can see in each dimension for the case of $\lambda=+1$ the semi-supervised architecture has learnt to approximate an Einstein metric exceptionally well.
Where in the supervised case the output is explicitly known, in the semi-supervised case the only conditions informing the learning are the values of the Einstein and other losses, and the model starts from an identity initialisation which is far from satisfying the Einstein and overlap conditions (training loss often starts $>10^4$).
It is therefore exceptionally impressive that the model can approximate these $\lambda=+1$ Einstein metrics so well, even exceeding the performance scores of the supervised model\footnote{Despite the supervised model's training being informed by the exact metric values at each training data-point, the Ricci tensor is so highly sensitive that the semi-supervised architecture can better learn the metric, even without the explicit knowledge of its values.}.

Where existence of Einstein metrics with $\lambda=+1$ is known and proven for spheres $S^n$ in any dimension, those with $\lambda = 0,-1$ are forbidden in dimensions $2,3$ \cite{Besse:1987pua}.
The runs where the model attempts to find a metric with $\lambda = 0,-1$ in those dimensions satisfyingly fail: all losses are large ($>1$, and over an order of magnitude above that of the supervised model), and these set the opposing loss score baselines for comparison where a metric does not exist.
Of greater significance are the $\lambda=0$ and $\lambda=-1$ cases for higher dimensions; especially the former, from a physics perspective. Specifically, the existence of Ricci-flat metrics on $S^{4,5}$ is an open problem which excitingly this machine learning approach can provide a new numerical perspective on.
%The $\lambda=0$ Global test loss results in 2d \& 3d are expectedly high, nearer to the order of the equivalent $\lambda=-1$ runs, corroborating the lack of existence of Ricci-flat Einstein metrics in these dimensions.
Therefore, of new insight are the results for 4d \& 5d, which are not conclusive\footnote{One may comment that 10 runs is not particularly many for finding a likely obscure metric, we add here that $\sim 50$ more runs were performed for the Ricci-flat search in further hope of finding suitable metrics, all with similar performance scores; and we plan to continue submitting runs in search of evidence for their existence.}, with losses of order $10$, and thus much larger than the supervised model losses.
These results hence provide new numerical evidence hinting \textit{against} this longstanding open problem of Ricci-flat metric existence on the spheres $S^4$ and $S^5$, and no examples of Einstein metric with negative Einstein constant are found either. \AS{One should, however, exercise care not to draw emphatic conclusions about existence: in higher dimensions the loss landscape is, of course, more complex. One potential modification for future work is to apply more sophisticated optimisers to the problem. More optimistically, however, we emphasise that the AInstein package was indeed able to find the known positive curvature solution in dimensions 4 and 5 with comparable global losses to the lower dimensional cases.}

\subsubsection{Visualisations}
\begin{figure*}
    \centering
    \begin{subfigure}{0.24\textwidth}
        \centering
        \includegraphics[width=0.98\textwidth]{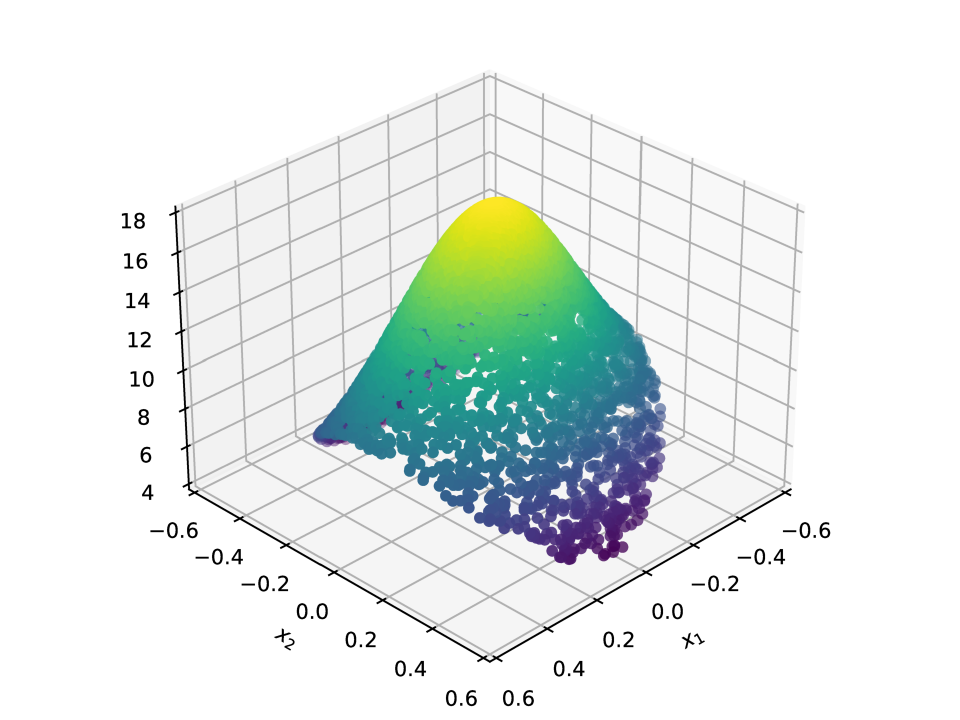}
        \caption{$g_{00}$ Patch 1}
        \label{fig:vis_2dpos_g001}
    \end{subfigure} 
    \begin{subfigure}{0.24\textwidth}
        \centering
        \includegraphics[width=0.98\textwidth]{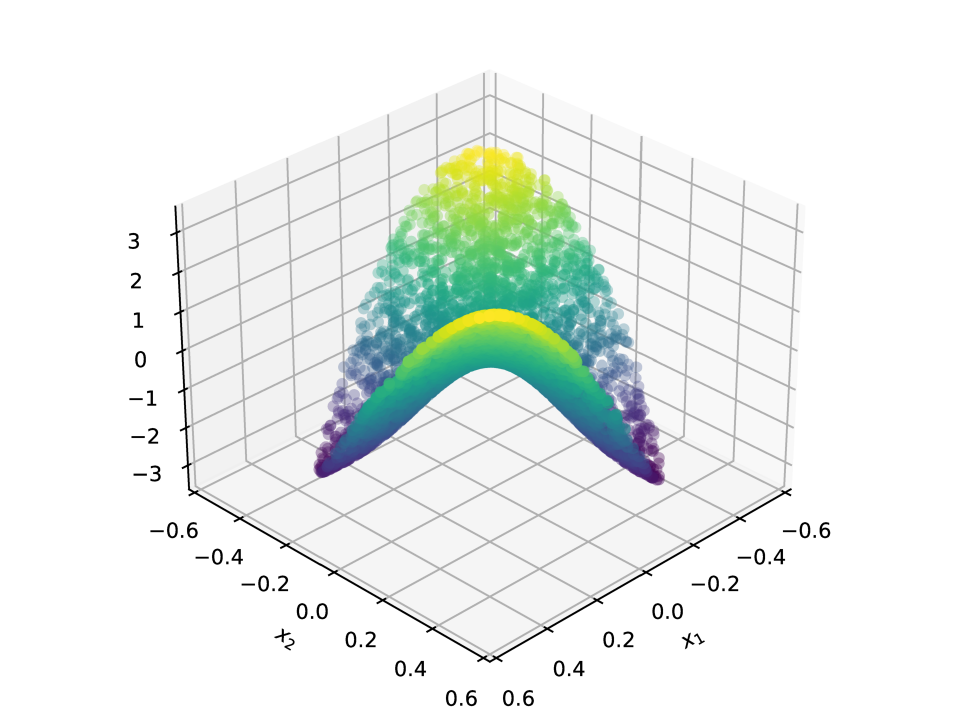}
        \caption{$g_{01}$ Patch 1}
    \end{subfigure} 
    \begin{subfigure}{0.24\textwidth}
        \centering
        \includegraphics[width=0.98\textwidth]{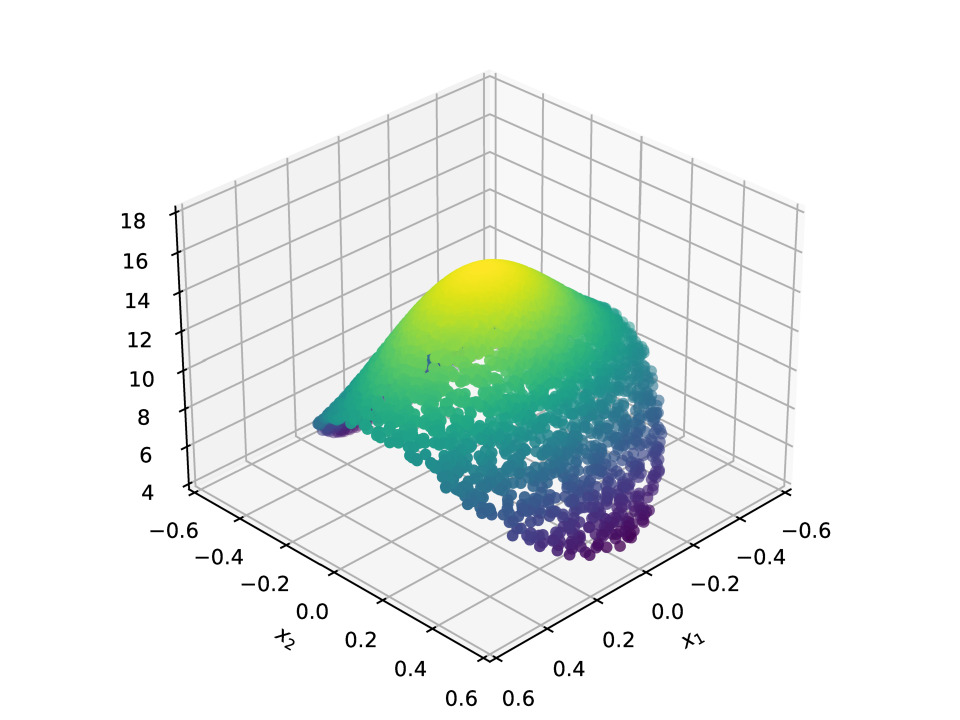}
        \caption{$g_{00}$ Patch 2}
    \end{subfigure} 
    \begin{subfigure}{0.24\textwidth}
        \centering
        \includegraphics[width=0.98\textwidth]{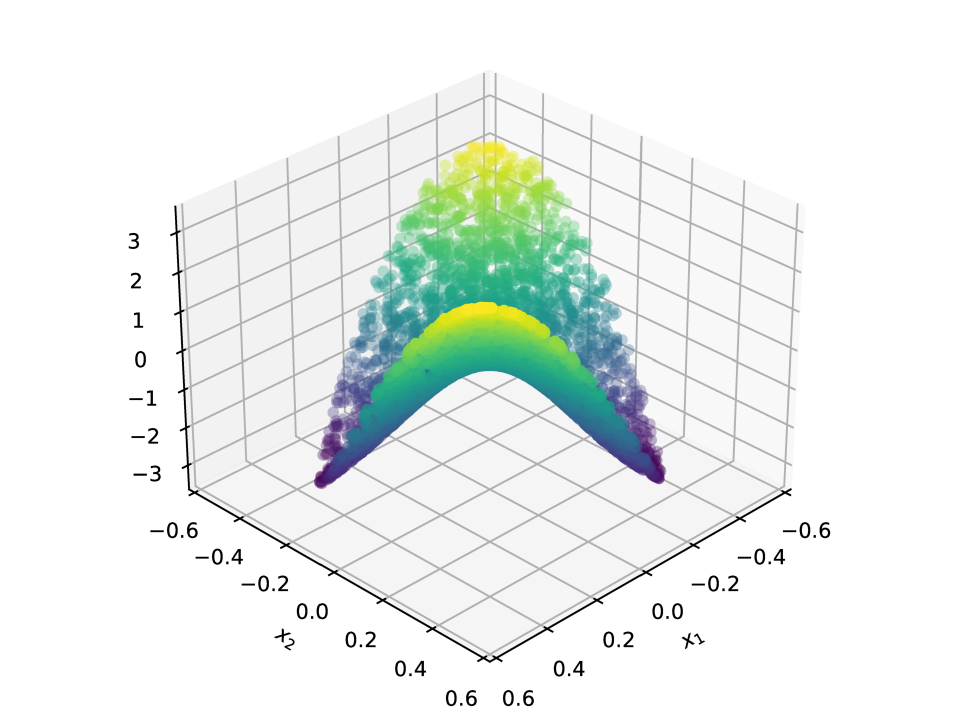}
        \caption{$g_{01}$ Patch 2}
    \end{subfigure}\\
    \begin{subfigure}{0.24\textwidth}
        \centering
        \includegraphics[width=0.98\textwidth]{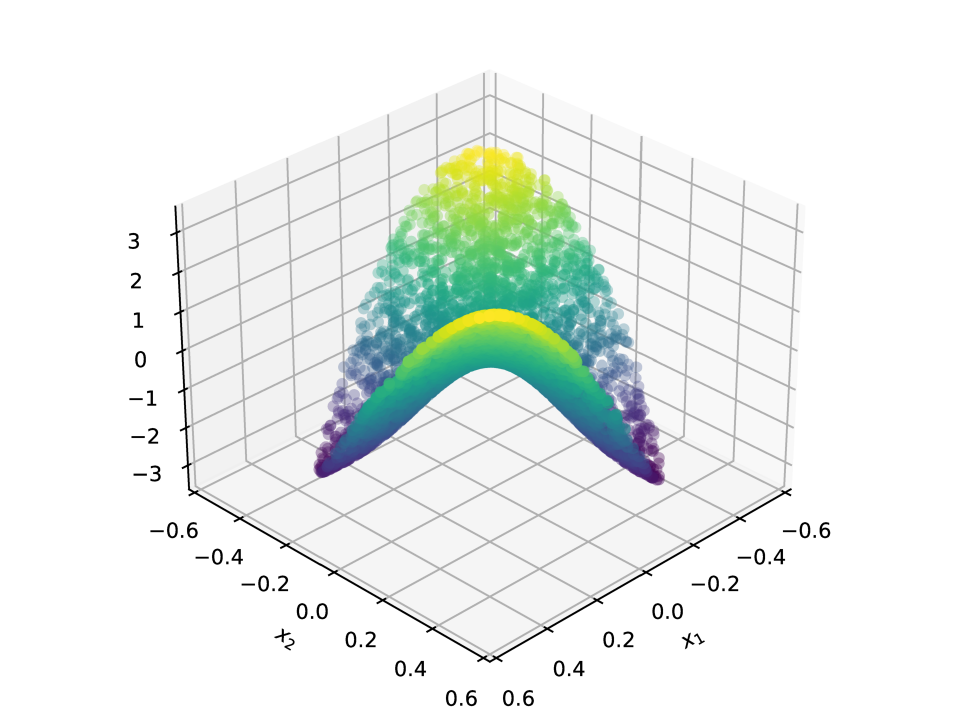}
        \caption{$g_{10}$ Patch 1}
    \end{subfigure} 
    \begin{subfigure}{0.24\textwidth}
        \centering
        \includegraphics[width=0.98\textwidth]{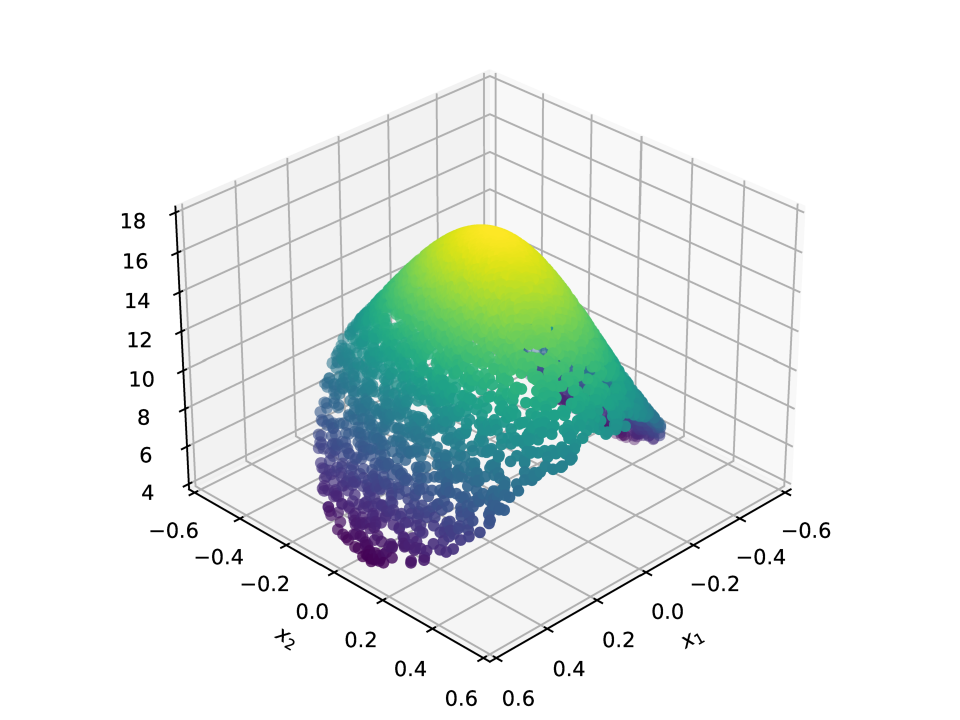}
        \caption{$g_{11}$ Patch 1}
    \end{subfigure} 
    \begin{subfigure}{0.24\textwidth}
        \centering
        \includegraphics[width=0.98\textwidth]{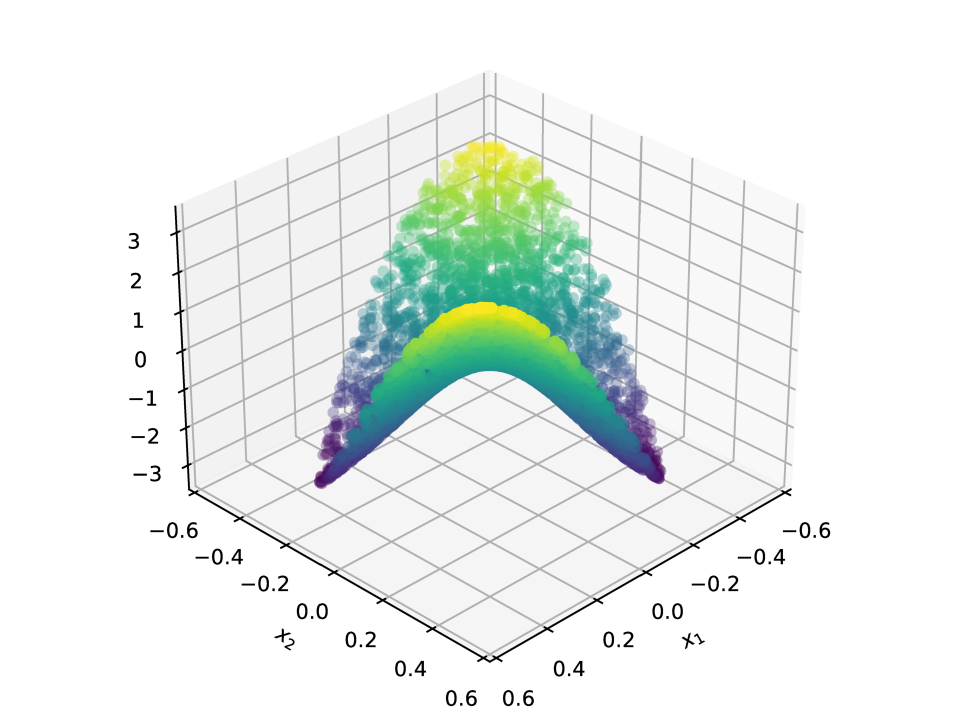}
        \caption{$g_{10}$ Patch 2}
    \end{subfigure} 
    \begin{subfigure}{0.24\textwidth}
        \centering
        \includegraphics[width=0.98\textwidth]{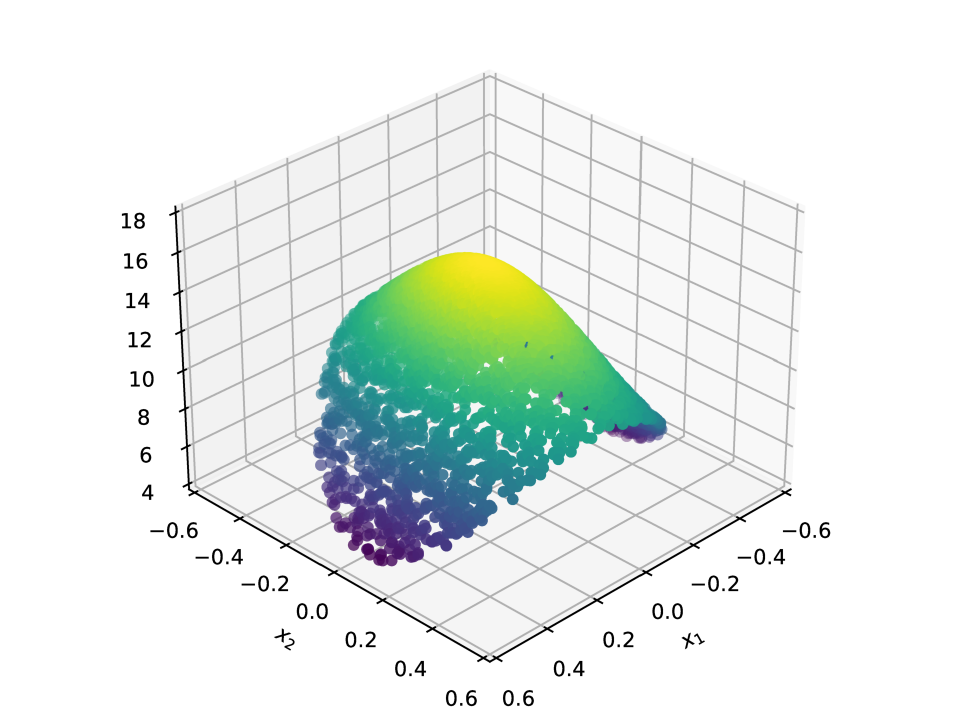}
        \caption{$g_{11}$ Patch 2}
    \end{subfigure}
    \caption{Visualisations of the learnt metrics, $g_{ij}$, in 2d, on the 2 patches, trained with positive Einstein constant (such that $R_{ij} = g_{ij}$).}
    \label{fig:vis_2dpos_g}
\end{figure*}

\begin{figure*}
    \centering
    \begin{subfigure}{0.24\textwidth}
        \centering
        \includegraphics[width=0.98\textwidth]{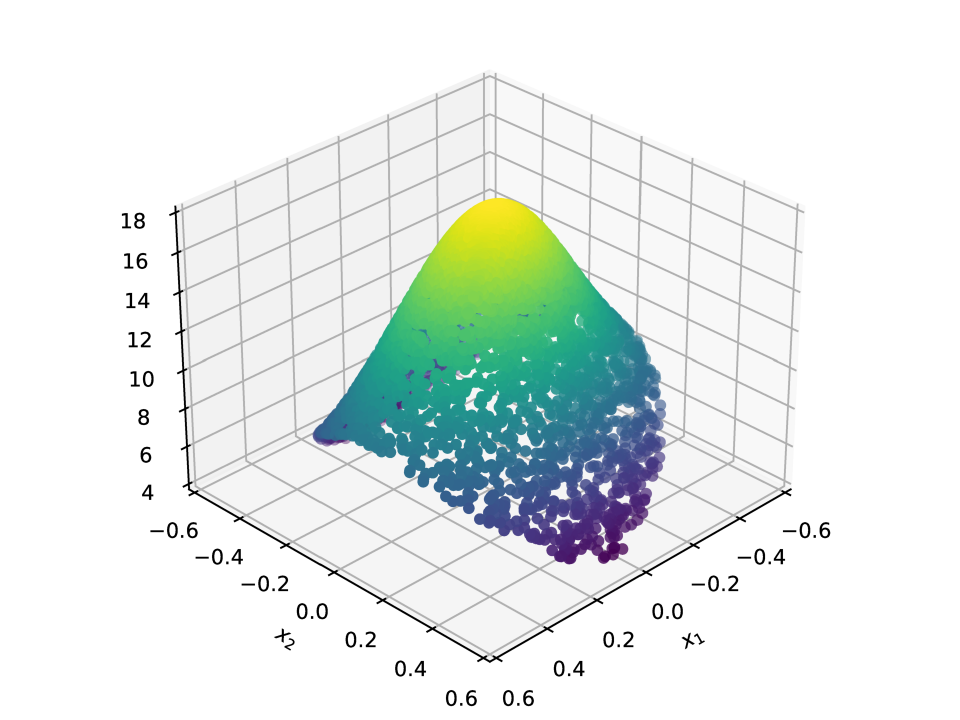}
        \caption{$R_{00}$ Patch 1}
    \end{subfigure} 
    \begin{subfigure}{0.24\textwidth}
        \centering
        \includegraphics[width=0.98\textwidth]{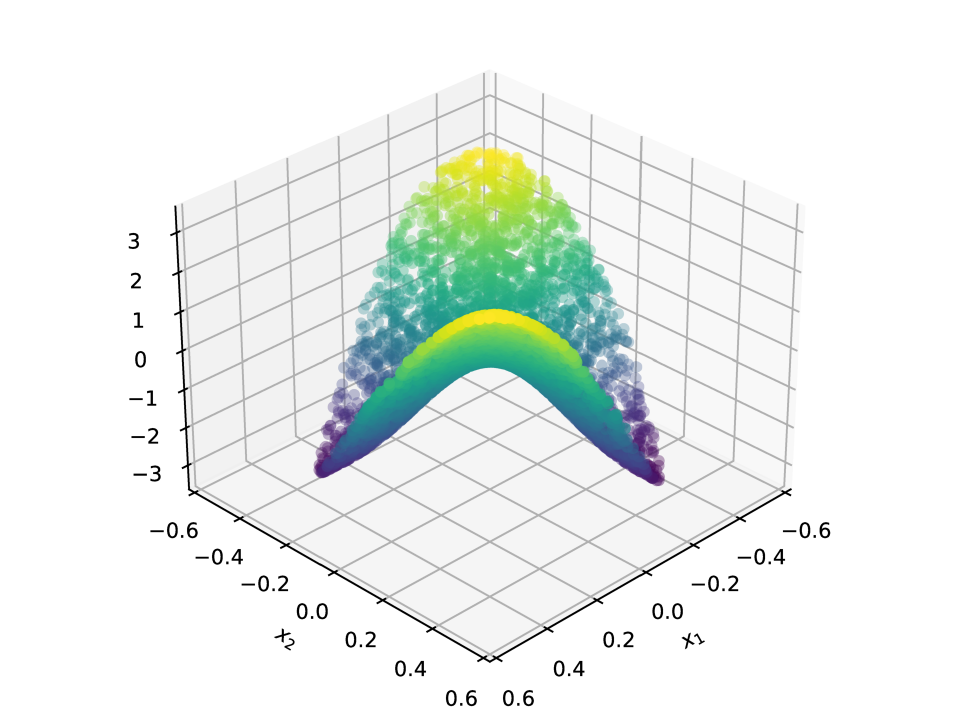}
        \caption{$R_{01}$ Patch 1}
    \end{subfigure} 
    \begin{subfigure}{0.24\textwidth}
        \centering
        \includegraphics[width=0.98\textwidth]{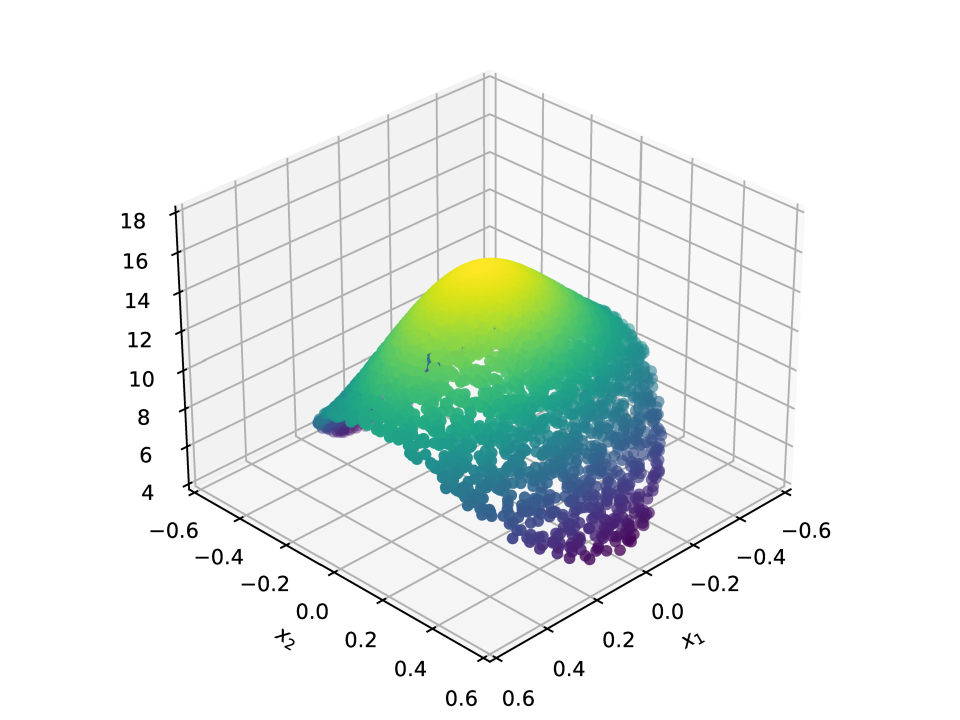}
        \caption{$R_{00}$ Patch 2}
    \end{subfigure} 
    \begin{subfigure}{0.24\textwidth}
        \centering
        \includegraphics[width=0.98\textwidth]{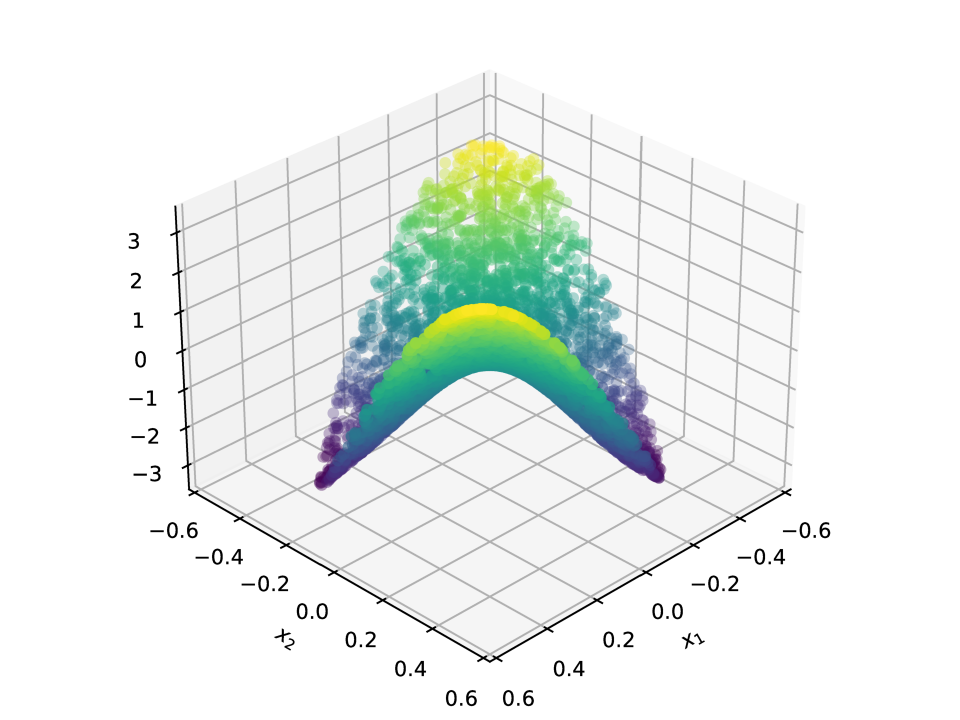}
        \caption{$R_{01}$ Patch 2}
    \end{subfigure}\\
    \begin{subfigure}{0.24\textwidth}
        \centering
        \includegraphics[width=0.98\textwidth]{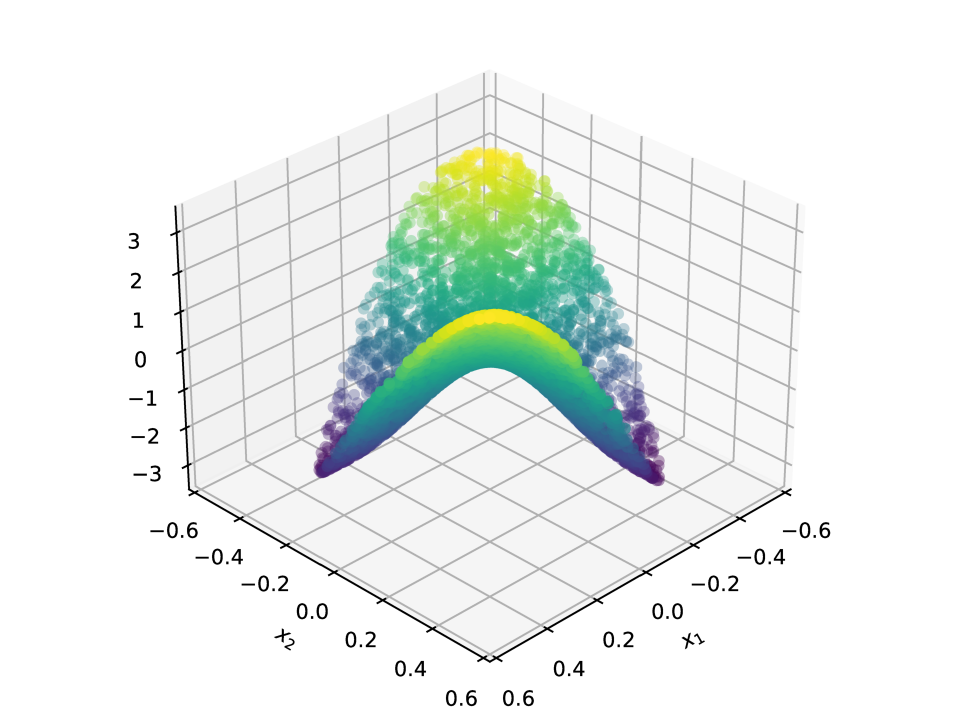}
        \caption{$R_{10}$ Patch 1}
    \end{subfigure} 
    \begin{subfigure}{0.24\textwidth}
        \centering
        \includegraphics[width=0.98\textwidth]{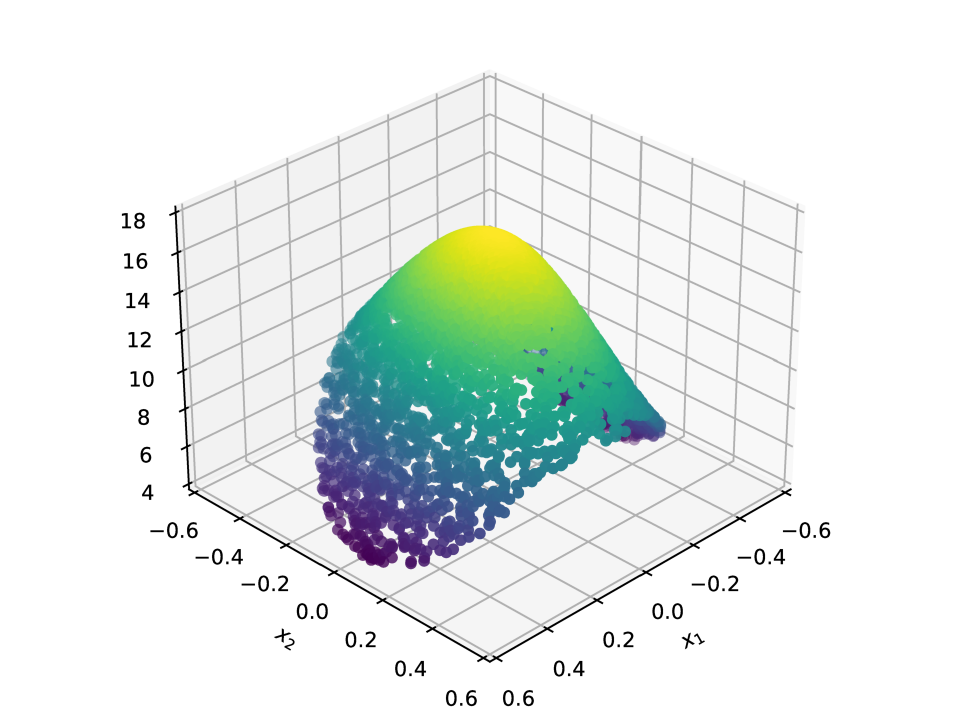}
        \caption{$R_{11}$ Patch 1}
    \end{subfigure} 
    \begin{subfigure}{0.24\textwidth}
        \centering
        \includegraphics[width=0.98\textwidth]{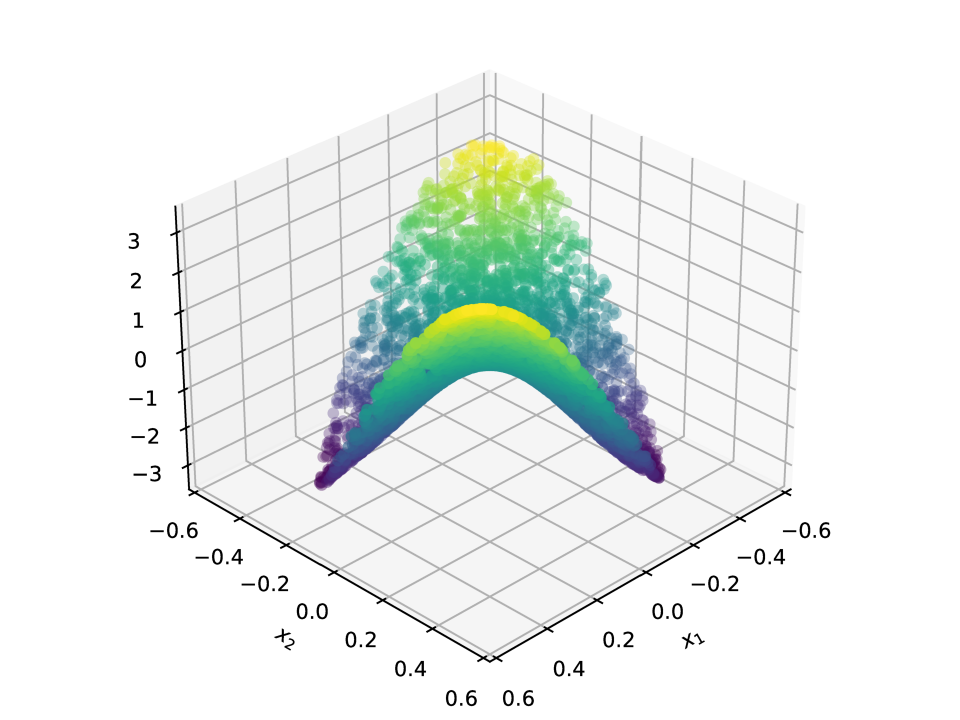}
        \caption{$R_{10}$ Patch 2}
    \end{subfigure} 
    \begin{subfigure}{0.24\textwidth}
        \centering
        \includegraphics[width=0.98\textwidth]{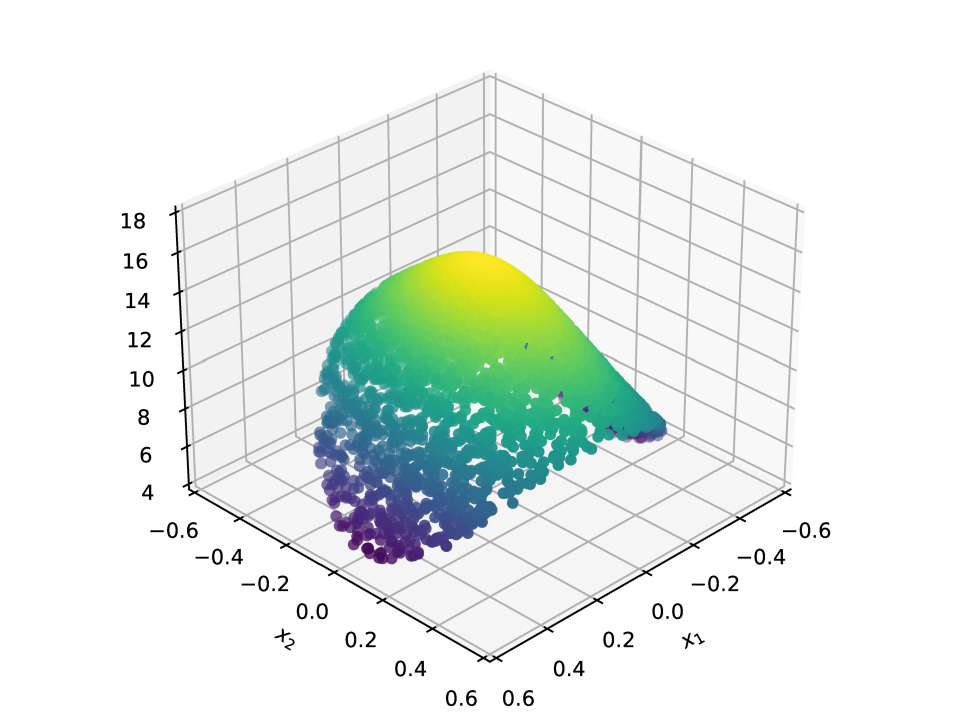}
        \caption{$R_{11}$ Patch 2}
    \end{subfigure}
    \caption{Visualisations of the Ricci tensors, $R_{ij}$, of the learnt metrics in 2d, on the 2 patches, trained with positive Einstein constant (such that $R_{ij} = g_{ij}$).}
    \label{fig:vis_2dpos_R}
\end{figure*}

To make tangible the metric learning provided by this package and the respective semi-supervised models, we present here visualisations of an example run of the 2d $\lambda=+1$ investigation.
Figure \ref{fig:vis_2dpos_g} shows the 4 metric components ($g_{ij}$) in both patches, whilst Figure \ref{fig:vis_2dpos_R} shows the 4 respective Ricci components ($R_{ij}$) in both patches also.
The plot data uses the same test data, with the same patch restriction to radii $r_m+0.1 \sim 0.51$ to reflect the required patch and overlap elements for building the global manifold.
We emphasise that the behaviour was consistent across the 10 runs, and note that equivalent visualisations for the 2d $\lambda \in \{0,-1\}$ investigations are shown in §\ref{app:extra_vis}.

Since the $\lambda=+1$ investigation involves solving the Einstein equation $R_{ij}=g_{ij}$, one expects a solution to have identical metric and Ricci components over the patch, these Figures \ref{fig:vis_2dpos_g} \& \ref{fig:vis_2dpos_R} demonstrate this especially well, with matching components between metric and Ricci, equally good in both patches.
These visualisations corroborate the strong learning of the $\lambda=+1$ Einstein metrics, confirming that the low losses observed for $\lambda=+1$ in Table \ref{tab:global_test_losses} do truly represent good Einstein metrics\footnote{We add that visualisations were also generated in higher-dimensions, using 2d sections of the patches, and matching was equivalently good.}.

\begin{figure}[hbtp]
    \centering
    \begin{subfigure}{0.23\textwidth}
        \centering
        \includegraphics[width=0.98\textwidth]{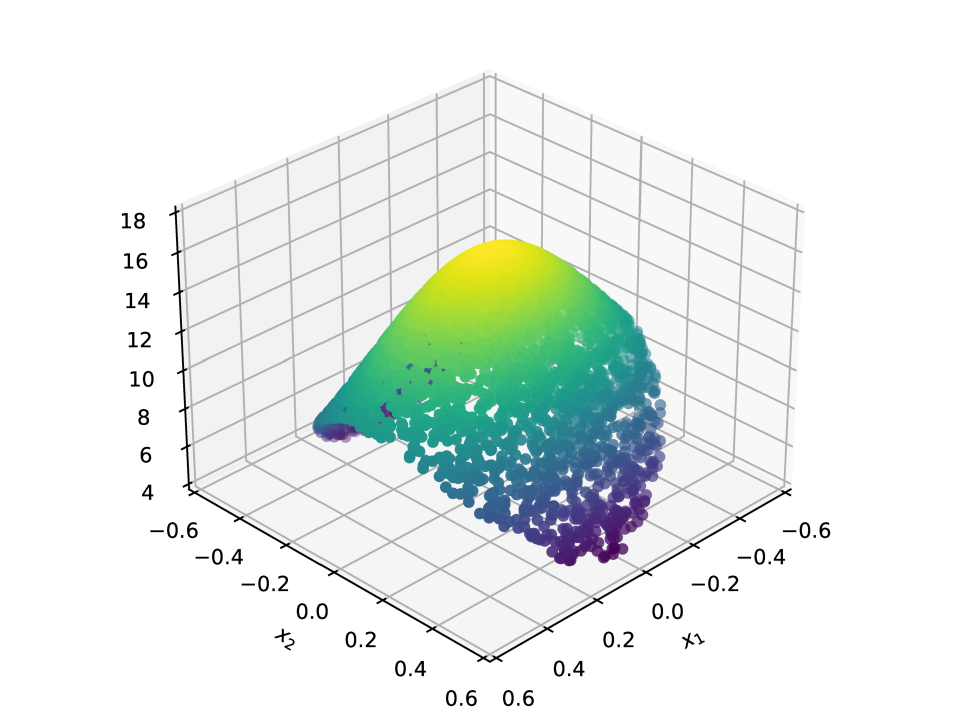}
        \caption{$g_{00}$ Analytic}
    \end{subfigure} 
    \begin{subfigure}{0.23\textwidth}
        \centering
        \includegraphics[width=0.98\textwidth]{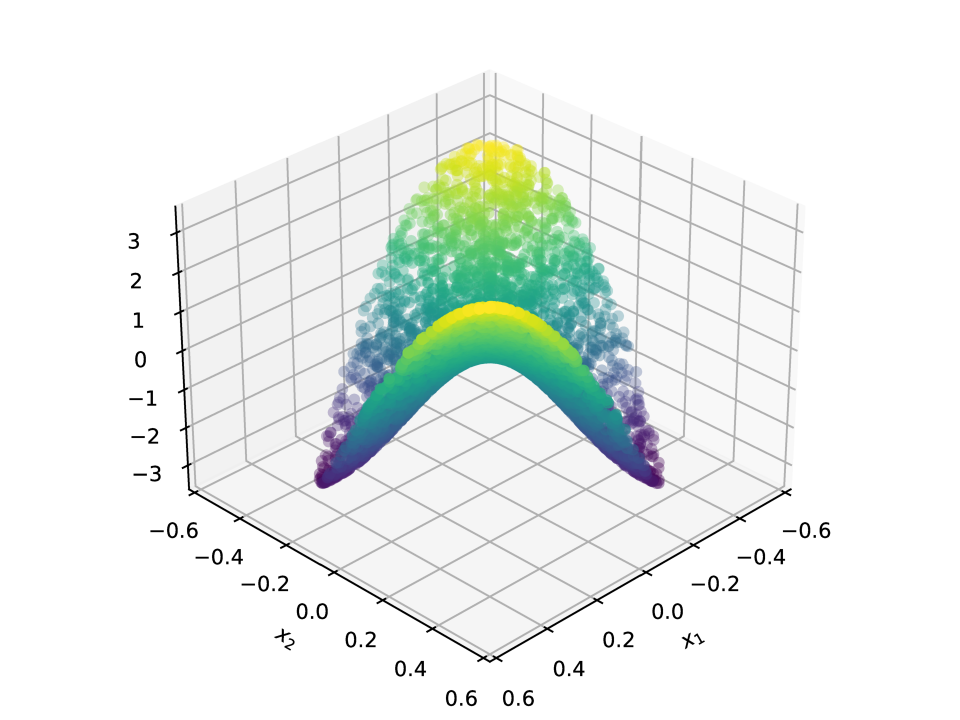}
        \caption{$g_{01}$ Analytic}
    \end{subfigure}\\
    \begin{subfigure}{0.23\textwidth}
        \centering
        \includegraphics[width=0.98\textwidth]{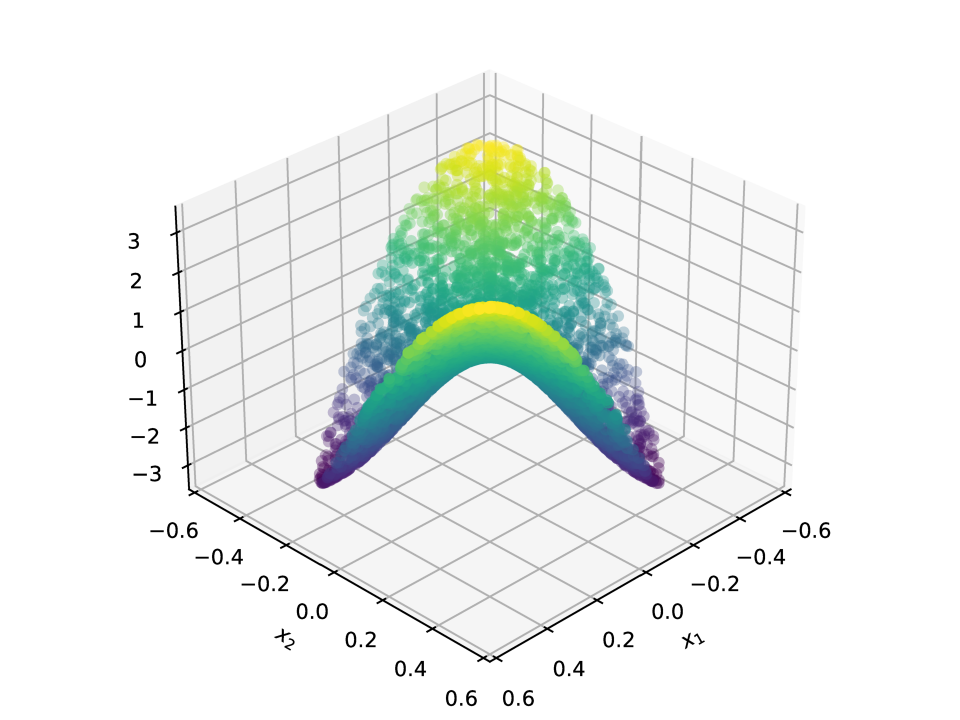}
        \caption{$g_{10}$ Analytic}
    \end{subfigure} 
    \begin{subfigure}{0.23\textwidth}
        \centering
        \includegraphics[width=0.98\textwidth]{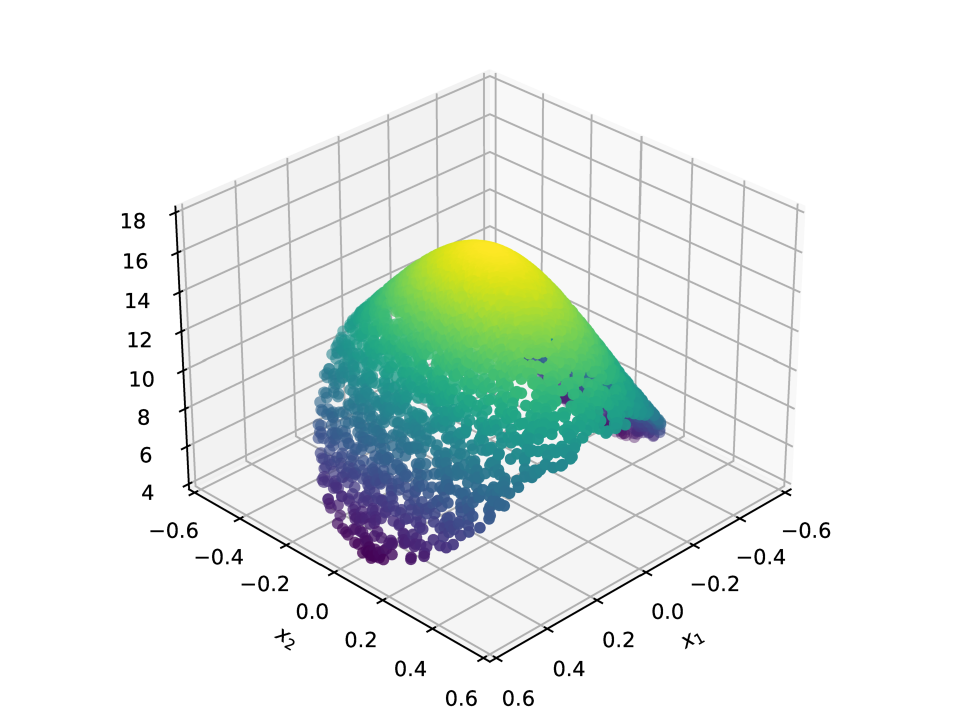}
        \caption{$g_{11}$ Analytic}
    \end{subfigure} 
    \caption{Visualisations of the analytic round metric, $g_{ij}$, in 2d on a ball patch. This metric solves the Einstein metric equation with positive Einstein constant ($R_{ij} = g_{ij}$), such that each metric component $g_{ij}$ equals its equivalent Ricci component $R_{ij}$.}
    \label{fig:vis_analytic}
\end{figure}

As a final comparison, in Figure \ref{fig:vis_analytic}, the metric components of the analytic round metric of \eqref{eq:roundmetric} are computed and plotted in the same visualisation style.
This metric is the same in both patches, and these plotted metric values were computed in the same way as the outputs used for the training of the supervised models whose test scores are shown in Table \ref{tab:global_test_losses}.
Of note is that these visualisations are strikingly similar to those in Figure \ref{fig:vis_2dpos_g}, indicating that the 2d $\lambda=+1$ model learnt by the semi-supervised model is this known analytic round metric, yet learned \textit{better} without the knowledge of the metric values, relying only on solving the Einstein equation directly.

% OLD

%%%%%%%%%%%%%%%%%%%%%%%%%%%%%%%%%%%%%%%%%%%%%%%%%
\newpage

\section{Conclusions}\label{sec:conc}
In this paper, we introduced a numerical scheme, based on semi-supervised machine learning, which approximates Einstein metrics on arbitrary manifolds. Results in this work restricted investigations to spheres of various dimensions, as a source of open questions regarding the existence of Einstein metrics. 

We presented an architecture which mimics the patching structure of a manifold, consisting of two parallel subnetworks. The input data are the coordinates of points in one patch; they are fed directly to the first sub-network, and they are transformed into coordinates of the second patch before being fed to the second subnetwork. Then, each subnetwork predicts the components of the metric, which we label as $g^1$ and $g^2$. The first loss component computes the Einstein condition for each patch independently, as $|\lambda g^{1,2} - Ric(g^{1,2})|$.
The second loss component ensures the correct transformation property of the metric under a change of coordinates; this is, schematically, $J^Tg^1J = g^2$, where $J$ is the Jacobian of the change of coordinates between the two patches. Such loss is evaluated to prioritise points belonging to the overlap region of the two patches. 
Finally, an artificial component is also added to the loss function to prevent the convergence to metrics with very low entries.

As mentioned, we applied our method to the case of spheres in dimension $2,3,4,5$, which admit a natural description in terms of two patches. While essentially all geometric properties have been fully understood in the former two dimensions, many questions are open in the latter two dimensions, especially regarding the existence of Ricci-flat metrics. 

For all of our runs, we initialise the neural network with a non-geometric configuration, where the metric is flat in both patches - therefore violating the patching condition. This is done in order not to introduce any bias in the process. Our findings show that the semi-supervised model trained with $\lambda = +1$ is able to converge to the round metric on $S^{2,3,4,5}$, consistently and with absolute errors in the order of $10^{-1}$ for both the Einstein condition and the overlap condition. To confirm that the output metric coincides with the maximally symmetric round one, we perform a qualitative as well as a quantitative verification. The former one consists of inspecting (sections of) the output for the various components, and comparing it with the analytic prediction; this is reported in many of the plots. The latter one is provided by comparing the performance of the semi-supervised model with a fully supervised model trained to approximate the exact analytic form of the round metric, for the same amount of data and training epochs. We find that the semi-supervised results always outperform the supervised ones, corroborating the convergence properties of our method. When applied to the cases $\lambda = 0, -1$ in dimensions $2$ and $3$, the error increases consistently by at least one order of magnitude. This is in accordance with known results which disprove the existence of Einstein metrics with zero or negative constant on $S^{2,3}$. The results concerning $S^{4,5}$ are analogous, with a marginal increase in the error across all values of $\lambda$. Since the method does not rely on any analytic assumption regarding symmetry or Killing vectors, these results provide numerical evidence towards the non-existence of Einstein metrics on $S^{4,5}$, which is a long-standing open problem in differential geometry. As we mentioned, this investigation allowed us to validate and test our method, before applying it to a wide variety of problems that are naturally suitable for \textit{AInstein}.

The advantages of our method compared to traditional algorithms are numerous. First of all, the stochastic nature of neural networks allows for a more dynamical exploration of the landscape of metrics. Moreover, we observe an exceptionally good scaling of the number of samples with the manifold's dimension. Instead of the traditional $D^n$ associated with finite-difference methods, we find that almost no scaling is required for our purposes. As another key advantage, the neural network architecture can be adapted to predict not just one metric, but a family of them, which would allow exploration of moduli spaces of metrics. The simplest scenario for testing this feature is the case of $T^2$. Finally, the general construction of our code allows application to manifolds which are described by more than two patches; these are hard to deal with if one uses current algorithms.\AS{We comment however, that in some cases, traditional numerical methods may converge faster and produce metrics with higher accuracy \cite{Figueras:2012xj, Pretorius:2005gq, Dias:2015rxy, Clough_2015, Lehner:2010pn, Dias:2015pda, Chesler:2013lia, Chaurasia:2025, Wiseman:2002zc,Kudoh:2003ki,Headrick:2009pv}. It is important to emphasise this work is a `proof-of-concept' which demonstrates the power of the AInstein approach; it may, in principle, be employed where traditional numerical methods no longer converge. Accordingly, we opted to omit a direct empirical comparison with numerical methods to avoid obscuring the placement of this work; such a comparison is left as future work for a more optimised incarnation of the AInstein algorithm which remains under development.}

In addition to tackling questions regarding the existence of certain metrics, our method could also be used to find numerical approximations for metrics lacking an analytic description\footnote{To be fully rigorous, we can foresee coupling this scheme to computer-assisted proofs' techniques (see \cite{808365d6fe2f449e8be7d40295302da1} for instance).}. In this light, it is our intention to apply it to the case of exotic $7$-spheres, for which existence results have been proven in \cite{boyer2003einstein, boyer2004einstein}, and recent progresses in the understanding of their geometry have been presented in \cite{Gherardini:2023uyx, berman2024curvatureexotic7sphere}. Going to higher dimensions would, of course, result in more expensive computations. For this reason, we also plan to develop the AInstein package in two directions\footnote{\AS{Other ML interpretability and optimisation tools exist, for example using Fisher information to examine which parameters and inputs of the network are most dominantly utilised in the output metric \cite{Berman_2023, amari2016information}.}} : making it better suited to GPUs and fixing the diffeomorphism freedom at the level of the loss function, following the prescription outlined in \cite{Figueras:2012xj}. \TSG{Regarding the latter, such an addition to the loss function would discourage the network to explore the redundant directions in the space of metrics, which are associated with mere coordinate redefinitions.} In order to make our method more robust, we are also working to reformulate the finiteness loss (as well as a few other features of the code) in a more geometric fashion. Finally, we are currently working on applying our method to problems of great relevance in theoretical physics, including investigating Ricci-flat solutions with the Euclidean Schwarzschild's topology, and modifying the code to include Lorentzian-signature metrics. \TSG{Concerning the latter, a few subtleties naturally arise. Firstly, the Cholesky decomposition needs to be adjusted in order to achieve the right signature. Secondly, in the case of convergence to an Einstein metric, we would need to carefully study the associated causal structure, by identifying time-like, space-like and null regions. This is left for future work, where we aim to employ Penrose diagrams to facilitate the interpretation of any numerical result.}
%In addition to the applications mentioned above, this numerical scheme could also be used to generate starting configurations to then be fed to traditional algorithms (see shooting methods like [***]). 

%%%%%%%%%%%%%%%%%%%%%%%%%%%%%%%%%%%%%%%%%%%%%%%%%
\section*{Acknowledgements}
We wish to thank Michael Douglas, Fabian Ruehle and Tomás Silva for their helpful comments during the ``Mathematics and Machine Learning Program'' at Harvard University. We also wish to thank David Berman for bringing us together, and Toby Wiseman for his insights on how this method compares to traditional algorithms. AGS and EH acknowledge support from Pierre Andurand over the course of this research. TSG is supported by the Science and Technology Facilities Council (STFC) Consolidated Grants ST/T000686/1 ``Amplitudes, Strings \& Duality'' and ST/X00063X/1 ``Amplitudes, Strings \& Duality''. This research utilised Queen Mary's Apocrita HPC facility \cite{apocrita}, supported by QMUL Research-IT.

\section*{Conflicts of interest}
The authors feel there are no conflicts of interest to report in this work.

\section*{Author contributions}
All authors contributed equally.

\section*{Data availability}

The code repository for this project's package can be found at: \url{https://github.com/xand-stapleton/ainstein}. It is written in \texttt{Python 3} and built on \texttt{TensorFlow} \cite{tensorflow2015whitepaper}.

%%%%%%%%%%%%%%%%%%%%%%%%%%%%%%%%%%%%%%%%%%%%%%%%%

%%
\newpage
\bibliographystyle{JHEP} 
\bibliography{chonker}
%%%%%%%%%%%%%%%%%%%%%%%%%%%%%%%%%%%%%%%%%%%%%%%%%
\newpage
\appendix
% Reset counters
% \setcounter{table}{0}
% \setcounter{figure}{0}
% \setcounter{equation}{0}

% \renewcommand{\theequation}{A\arabic{equation}}
% \renewcommand{\thefigure}{A\arabic{figure}}
% \renewcommand{\thetable}{A\arabic{table}}

\section{Further Results}\label{app:extra_results}
This appendix expands on the results of §\ref{sec:results}, displaying further breakdown of the test losses as the performance measures of the learning, as well as further example visualisations for 2d runs with other values of $\lambda$.

\subsection{Losses}\label{app:extra_losses}
The results in Table \ref{tab:global_test_losses} display the Global test losses, averaged over the 10 runs with standard deviations, for each of the investigations performed.
The Global test loss has 3 components, the Einstein loss in each of the 2 patches, and the overlap loss, calculated with respective multiplier weightings as described in §\ref{sec:bkg_ml}.

In Table \ref{tab:full_loss_results}, the average values of the constituent losses used in computing each Global test loss across the investigations are shown, again with standard deviations over the 10 runs.

\begin{table*}[h!]
\centering
\begin{tabular}{|cc|ccc!{\vrule width 1.5pt}c|}
\hline
\multicolumn{2}{|c|}{\multirow{2}{*}{Dim}}       & \multicolumn{3}{c!{\vrule width 1.5pt}}{Einstein Constant $\lambda$}  & \multirow{2}{*}{\begin{tabular}[c]{@{}c@{}}Supervised\\ $\lambda = +1$\end{tabular}} \\ \cline{3-5} 
\multicolumn{2}{|c|}{}                                 & \multicolumn{1}{c|}{$+1$}         & \multicolumn{1}{c|}{$0$}         & $-1$    &     \\ \hline
\Xhline{2\arrayrulewidth}
\multicolumn{1}{|c|}{\multirow{4}{*}{2}} & Global      & \multicolumn{1}{c|}{0.083 $\pm$ 0.023} & \multicolumn{1}{c|}{2.881 $\pm$ 0.113} & \multicolumn{1}{c!{\vrule width 1.5pt}}{4.364 $\pm$ 0.093} & 0.096 $\pm$ 0.013 \\ \cline{2-6} 
\multicolumn{1}{|c|}{}                   & Einstein patch 1 & \multicolumn{1}{c|}{0.077 $\pm$ 0.032} & \multicolumn{1}{c|}{11.992 $\pm$ 0.522} & \multicolumn{1}{c!{\vrule width 1.5pt}}{19.728 $\pm$ 0.772} & 0.219 $\pm$ 0.034 \\ \cline{2-6} 
\multicolumn{1}{|c|}{}                   & Einstein patch 2 & \multicolumn{1}{c|}{0.073 $\pm$ 0.021} & \multicolumn{1}{c|}{12.391 $\pm$ 0.674} & \multicolumn{1}{c!{\vrule width 1.5pt}}{19.596 $\pm$ 0.341} & 0.198 $\pm$ 0.034 \\ \cline{2-6} 
\multicolumn{1}{|c|}{}                   & Overlap     & \multicolumn{1}{c|}{0.076 $\pm$ 0.021} & \multicolumn{1}{c|}{0.731 $\pm$ 0.030} & \multicolumn{1}{c!{\vrule width 1.5pt}}{0.868 $\pm$ 0.019} & 0.064 $\pm$ 0.013 \\ \hline
\Xhline{2\arrayrulewidth}
\multicolumn{1}{|c|}{\multirow{4}{*}{3}} & Global      & \multicolumn{1}{c|}{0.151 $\pm$ 0.027} & \multicolumn{1}{c|}{5.560 $\pm$ 0.160} & \multicolumn{1}{c!{\vrule width 1.5pt}}{8.641 $\pm$ 0.183} & 0.195 $\pm$ 0.020 \\ \cline{2-6} 
\multicolumn{1}{|c|}{}                   & Einstein patch 1 & \multicolumn{1}{c|}{0.217 $\pm$ 0.052} & \multicolumn{1}{c|}{25.631 $\pm$ 1.019} & \multicolumn{1}{c!{\vrule width 1.5pt}}{41.246 $\pm$ 1.392} & 0.434 $\pm$ 0.058 \\ \cline{2-6} 
\multicolumn{1}{|c|}{}                   & Einstein patch 2 & \multicolumn{1}{c|}{0.188 $\pm$ 0.053} & \multicolumn{1}{c|}{25.444 $\pm$ 0.838} & \multicolumn{1}{c!{\vrule width 1.5pt}}{42.160 $\pm$ 1.042} & 0.439 $\pm$ 0.059 \\ \cline{2-6} 
\multicolumn{1}{|c|}{}                   & Overlap     & \multicolumn{1}{c|}{0.126 $\pm$ 0.021} & \multicolumn{1}{c|}{1.008 $\pm$ 0.018} & \multicolumn{1}{c!{\vrule width 1.5pt}}{1.164 $\pm$ 0.021} & 0.127 $\pm$ 0.018 \\ \hline
\Xhline{2\arrayrulewidth}
\multicolumn{1}{|c|}{\multirow{4}{*}{4}} & Global      & \multicolumn{1}{c|}{0.150 $\pm$ 0.018} & \multicolumn{1}{c|}{8.494 $\pm$ 0.121} & \multicolumn{1}{c!{\vrule width 1.5pt}}{14.928 $\pm$ 1.317} & 0.248 $\pm$ 0.024 \\ \cline{2-6} 
\multicolumn{1}{|c|}{}                   & Einstein patch 1 & \multicolumn{1}{c|}{0.343 $\pm$ 0.070} & \multicolumn{1}{c|}{40.827 $\pm$ 0.939} & \multicolumn{1}{c!{\vrule width 1.5pt}}{74.663 $\pm$ 4.943} & 0.640 $\pm$ 0.092 \\ \cline{2-6} 
\multicolumn{1}{|c|}{}                   & Einstein patch 2 & \multicolumn{1}{c|}{0.303 $\pm$ 0.051} & \multicolumn{1}{c|}{41.170 $\pm$ 1.059} & \multicolumn{1}{c!{\vrule width 1.5pt}}{74.845 $\pm$ 3.626} & 0.603 $\pm$ 0.043 \\ \cline{2-6} 
\multicolumn{1}{|c|}{}                   & Overlap     & \multicolumn{1}{c|}{0.100 $\pm$ 0.012} & \multicolumn{1}{c|}{1.144 $\pm$ 0.081} & \multicolumn{1}{c!{\vrule width 1.5pt}}{1.470 $\pm$ 0.700} & 0.148 $\pm$ 0.018 \\ \hline
\Xhline{2\arrayrulewidth}
\multicolumn{1}{|c|}{\multirow{4}{*}{5}} & Global      & \multicolumn{1}{c|}{0.244 $\pm$ 0.039} & \multicolumn{1}{c|}{10.810 $\pm$ 0.185} & \multicolumn{1}{c!{\vrule width 1.5pt}}{18.798 $\pm$ 2.024} & 0.518 $\pm$ 0.063 \\ \cline{2-6} 
\multicolumn{1}{|c|}{}                   & Einstein patch 1 & \multicolumn{1}{c|}{0.615 $\pm$ 0.132} & \multicolumn{1}{c|}{53.410 $\pm$ 1.641} & \multicolumn{1}{c!{\vrule width 1.5pt}}{97.398 $\pm$ 10.361} & 2.032 $\pm$ 0.291 \\ \cline{2-6} 
\multicolumn{1}{|c|}{}                   & Einstein patch 2 & \multicolumn{1}{c|}{0.595 $\pm$ 0.181} & \multicolumn{1}{c|}{54.186 $\pm$ 1.487} & \multicolumn{1}{c!{\vrule width 1.5pt}}{97.198 $\pm$ 12.189} & 1.552 $\pm$ 0.356 \\ \cline{2-6} 
\multicolumn{1}{|c|}{}                   & Overlap     & \multicolumn{1}{c|}{0.148 $\pm$ 0.022} & \multicolumn{1}{c|}{1.131 $\pm$ 0.066} & \multicolumn{1}{c!{\vrule width 1.5pt}}{1.218 $\pm$ 0.077} & 0.211 $\pm$ 0.016 \\ \hline
\end{tabular}
%full results:
%g: (array([0.09574733, 0.1945125 , 0.24771331, 0.51756187]),array([0.0133581 , 0.02022955, 0.02431109, 0.06255164]))
%e1: (array([0.21917214, 0.43420727, 0.6403612 , 2.03241056]), array([0.03434699, 0.05837891, 0.09248077, 0.29129287]))
%e2: (array([0.19794844, 0.43907716, 0.60320144, 1.5520142 ]), array([0.03407959, 0.05908973, 0.04258567, 0.3561177 ]))
%o: (array([0.06361   , 0.12663531, 0.14812838, 0.21087558]), array([0.01320912, 0.0182617 , 0.0179256 , 0.01564018]))
\caption{Global test loss results, with decompositions into the constituent sublosses: Einstein loss patch 1, Einstein loss patch 2, Overlap loss; averaged over 10 runs. Losses computed for NN approximations of Einstein metrics with the respective curvatures on spheres in dimensions 2-5 (2-patches). For comparison, the right-hand column shows the respective global test losses for the \textit{supervised} NN model approximation of the analytic round metric (which satisfies the Einstein equation for $\lambda = +1$). ll losses are reported with standard deviations across the 10 runs in each case.}
\label{tab:full_loss_results}
\end{table*}

One can see that the overlap loss is naturally lower, which is a good sign of consistency, since the patching condition is essential for ensuring the global metric definition is consistent; this is what motivated the higher multiplier weighting of this overlap loss component.
The Einstein losses within each investigation are approximately equal between the 2 patches, supporting the symmetric treatment of the patches.
Furthermore, the $\lambda=+1$ investigations all have low values across the loss components, particularly with both Einstein losses $<1$.
Conversely, the Einstein losses in the $\lambda \in \{0,-1\}$ investigations are all much higher, demonstrating further the geometric obstruction to learning Einstein metric's with these Einstein constants in these dimensions.

\subsection{Visualisations}\label{app:extra_vis}
\begin{figure*}[hbtp!]
    \centering
    \begin{subfigure}{0.24\textwidth}
        \centering
        \includegraphics[width=0.98\textwidth]{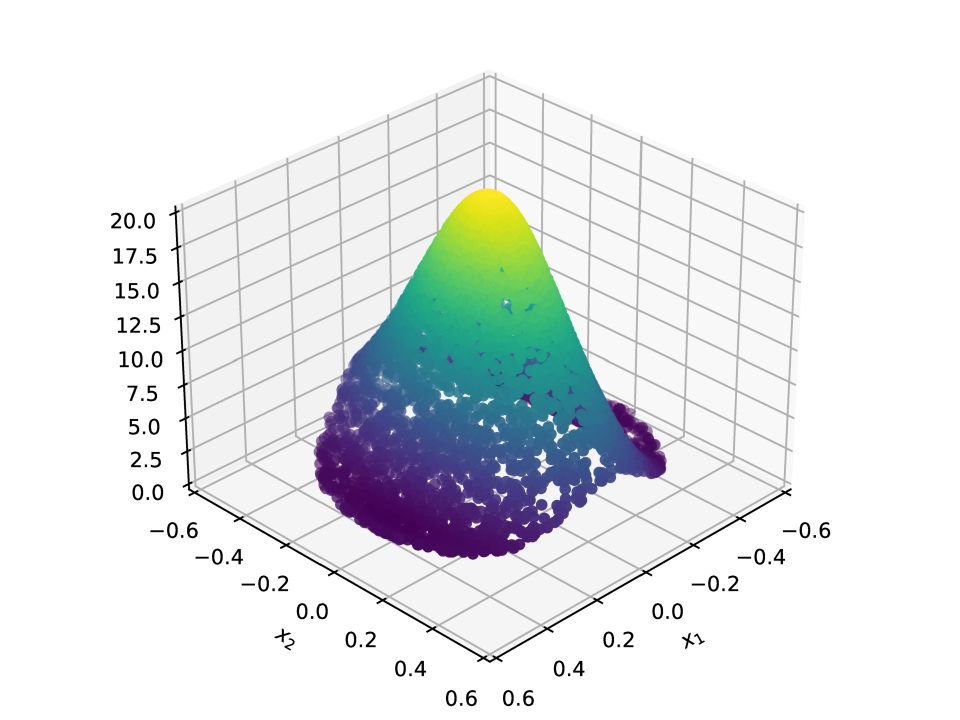}
        \caption{$g_{00}$ Patch 1}
        \label{fig:vis_2d0_g001}
    \end{subfigure} 
    \begin{subfigure}{0.24\textwidth}
        \centering
        \includegraphics[width=0.98\textwidth]{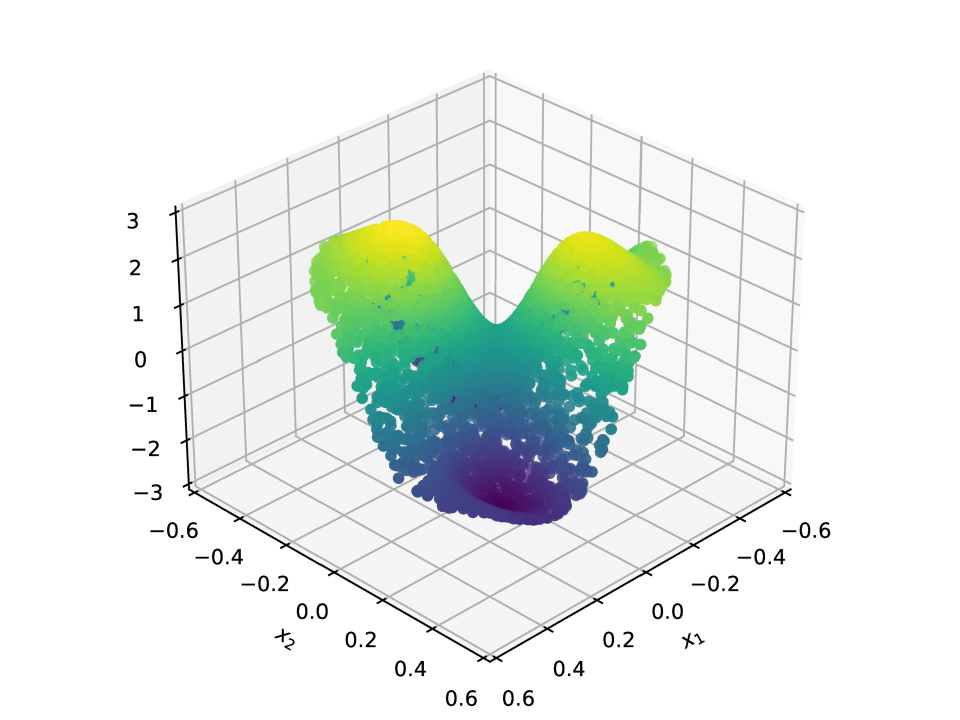}
        \caption{$g_{01}$ Patch 1}
    \end{subfigure} 
    \begin{subfigure}{0.24\textwidth}
        \centering
        \includegraphics[width=0.98\textwidth]{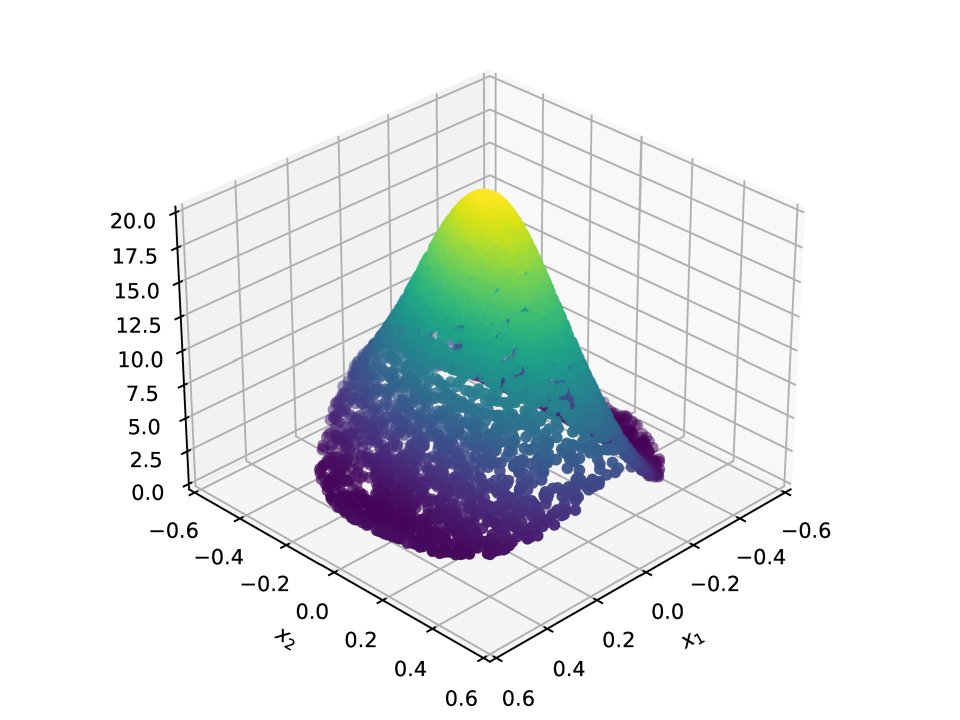}
        \caption{$g_{00}$ Patch 2}
    \end{subfigure} 
    \begin{subfigure}{0.24\textwidth}
        \centering
        \includegraphics[width=0.98\textwidth]{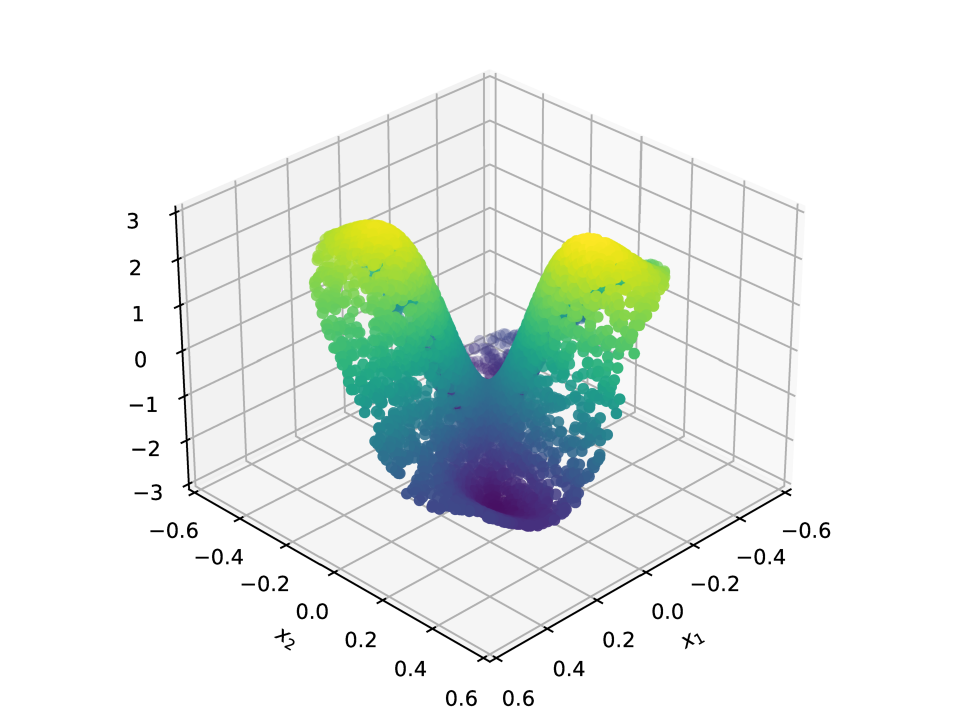}
        \caption{$g_{01}$ Patch 2}
    \end{subfigure}\\
    \begin{subfigure}{0.24\textwidth}
        \centering
        \includegraphics[width=0.98\textwidth]{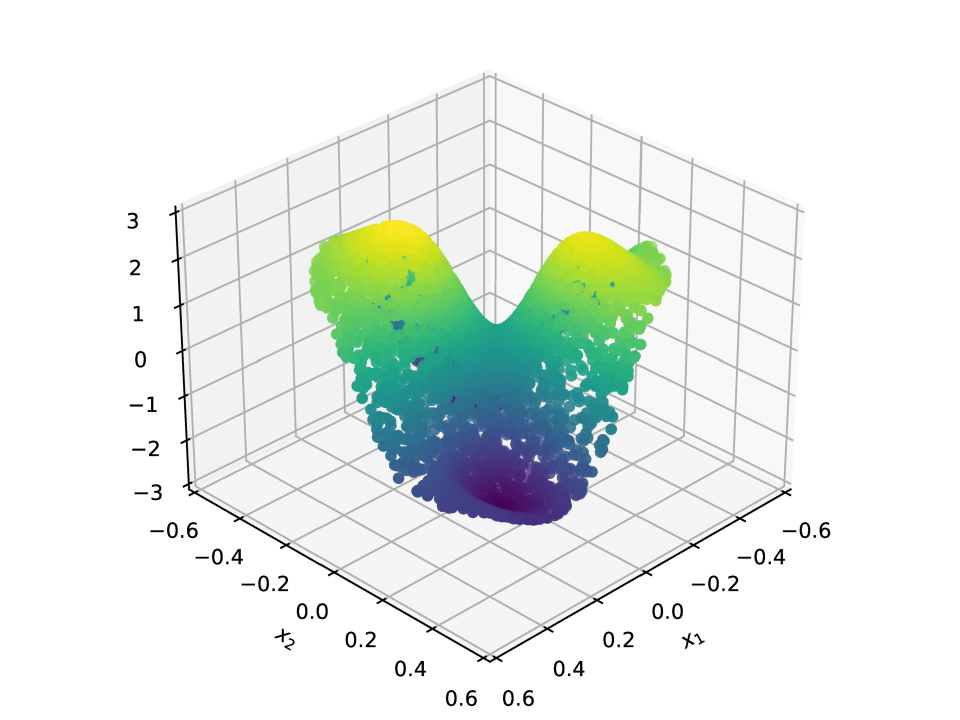}
        \caption{$g_{10}$ Patch 1}
    \end{subfigure} 
    \begin{subfigure}{0.24\textwidth}
        \centering
        \includegraphics[width=0.98\textwidth]{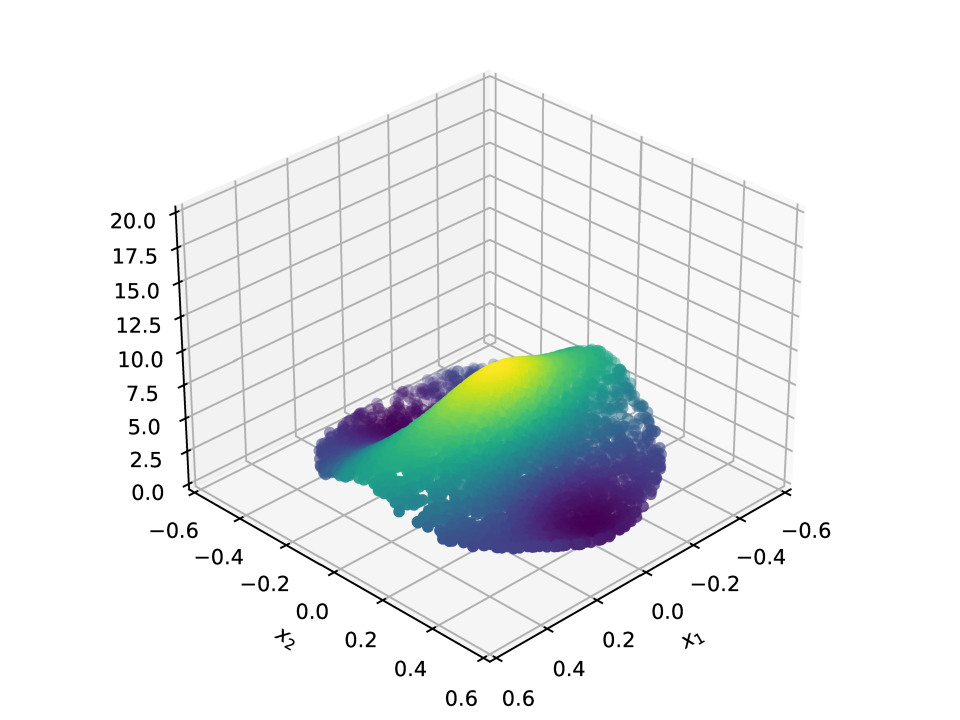}
        \caption{$g_{11}$ Patch 1}
    \end{subfigure} 
    \begin{subfigure}{0.24\textwidth}
        \centering
        \includegraphics[width=0.98\textwidth]{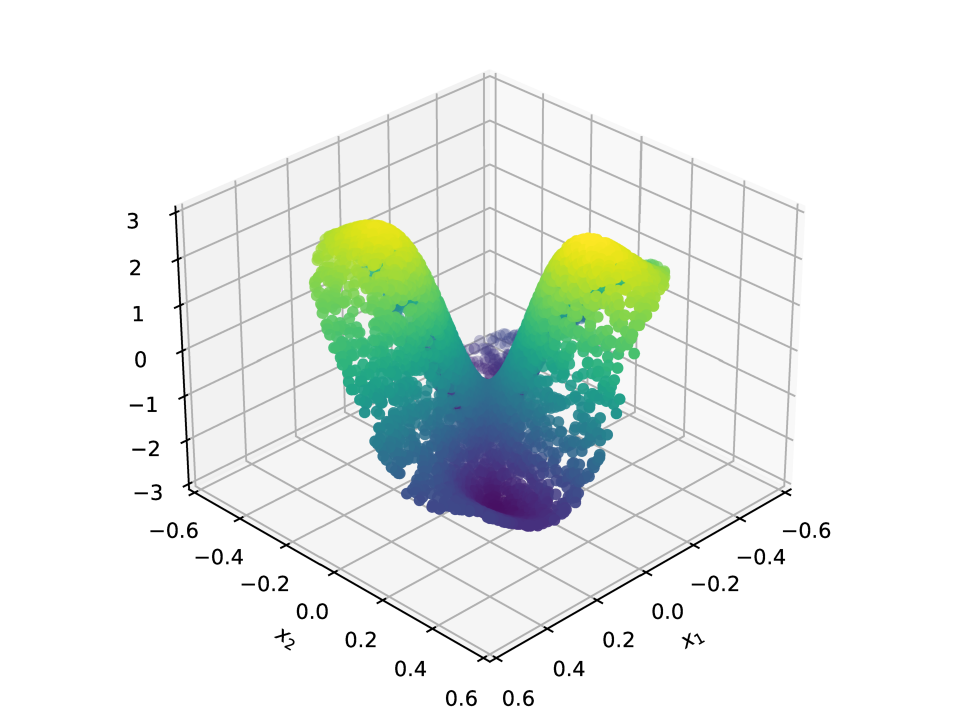}
        \caption{$g_{10}$ Patch 2}
    \end{subfigure} 
    \begin{subfigure}{0.24\textwidth}
        \centering
        \includegraphics[width=0.98\textwidth]{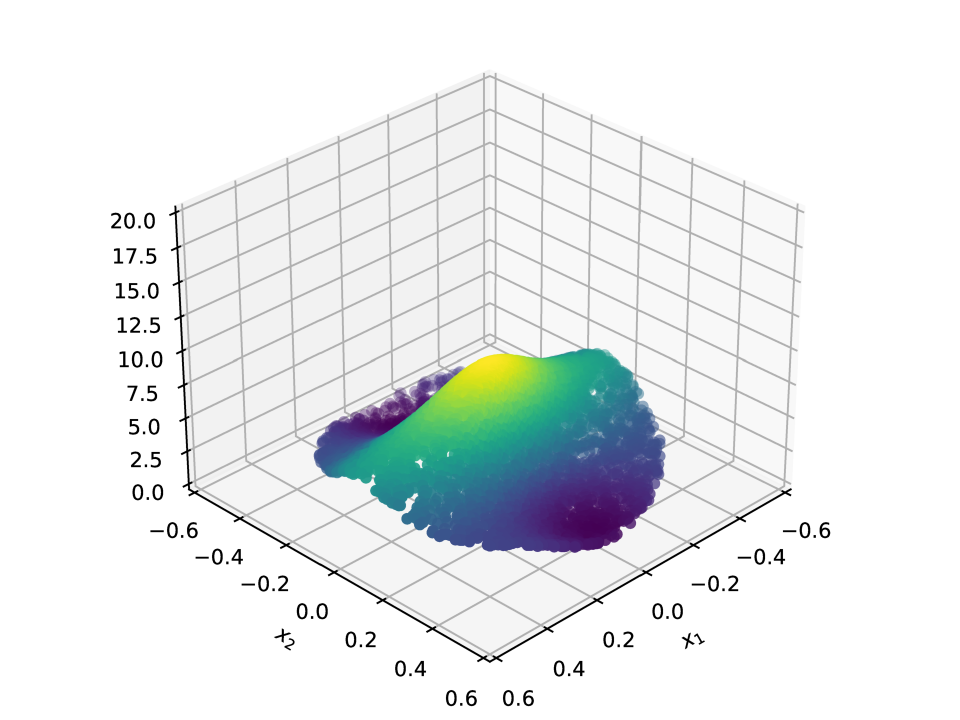}
        \caption{$g_{11}$ Patch 2}
    \end{subfigure}
    \caption{Visualisations of the learnt metrics, $g_{ij}$, in 2d, on the 2 patches, trained with zero Einstein constant (such that $R_{ij} = 0$), and the metric's goal is to be Ricci-flat.}
    \label{fig:vis_2d0_g}
\end{figure*}

\begin{figure*}[hbtp!]
    \centering
    \begin{subfigure}{0.24\textwidth}
        \centering
        \includegraphics[width=0.98\textwidth]{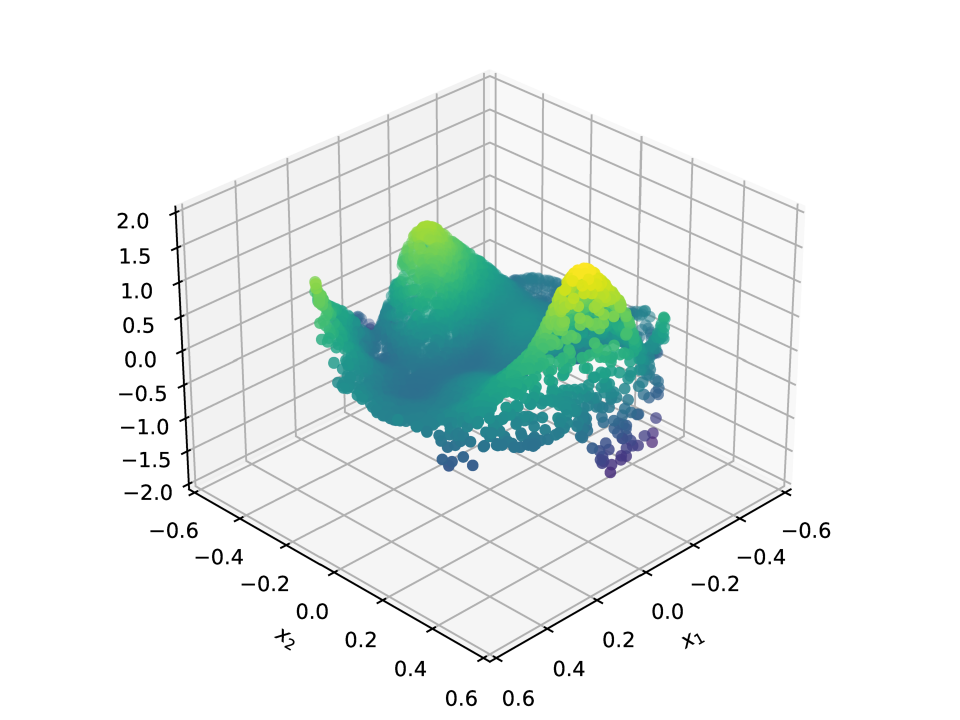}
        \caption{$R_{00}$ Patch 1}
    \end{subfigure} 
    \begin{subfigure}{0.24\textwidth}
        \centering
        \includegraphics[width=0.98\textwidth]{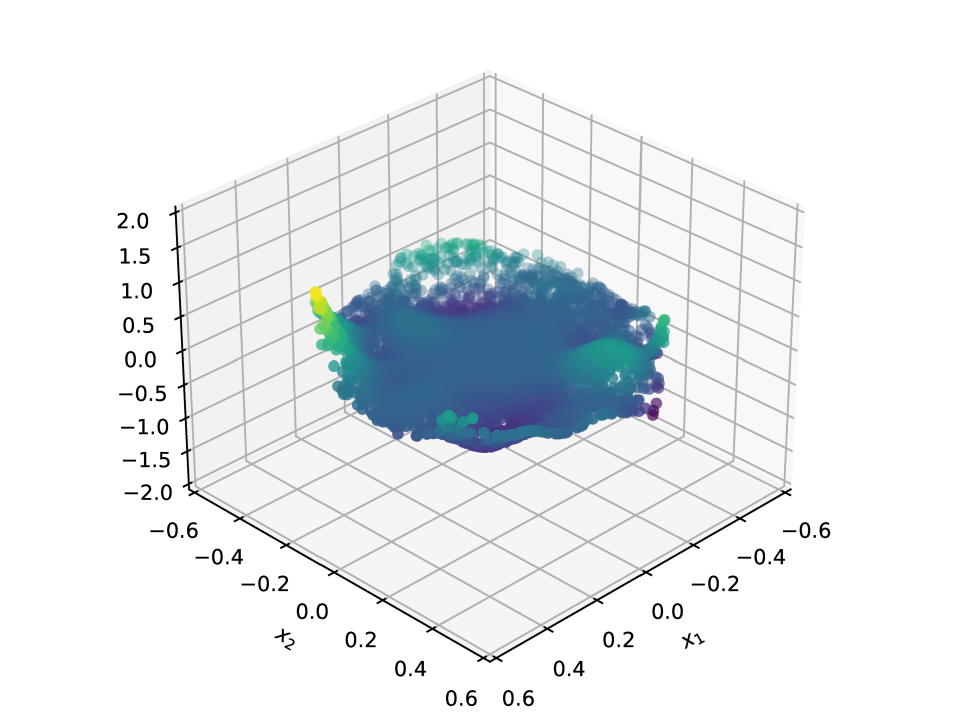}
        \caption{$R_{01}$ Patch 1}
    \end{subfigure} 
    \begin{subfigure}{0.24\textwidth}
        \centering
        \includegraphics[width=0.98\textwidth]{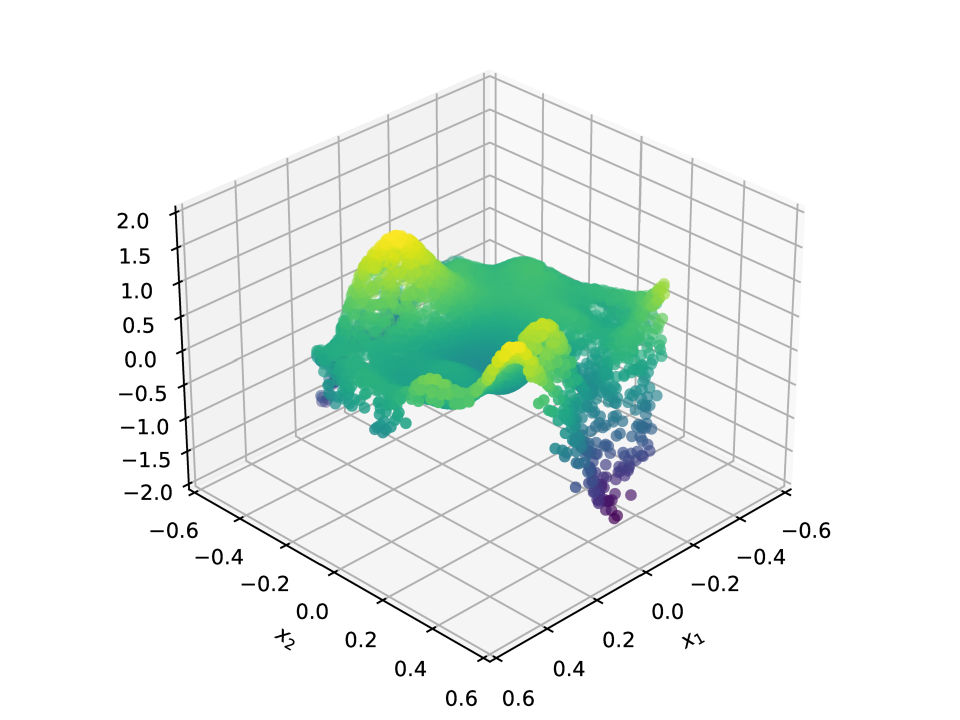}
        \caption{$R_{00}$ Patch 2}
    \end{subfigure} 
    \begin{subfigure}{0.24\textwidth}
        \centering
        \includegraphics[width=0.98\textwidth]{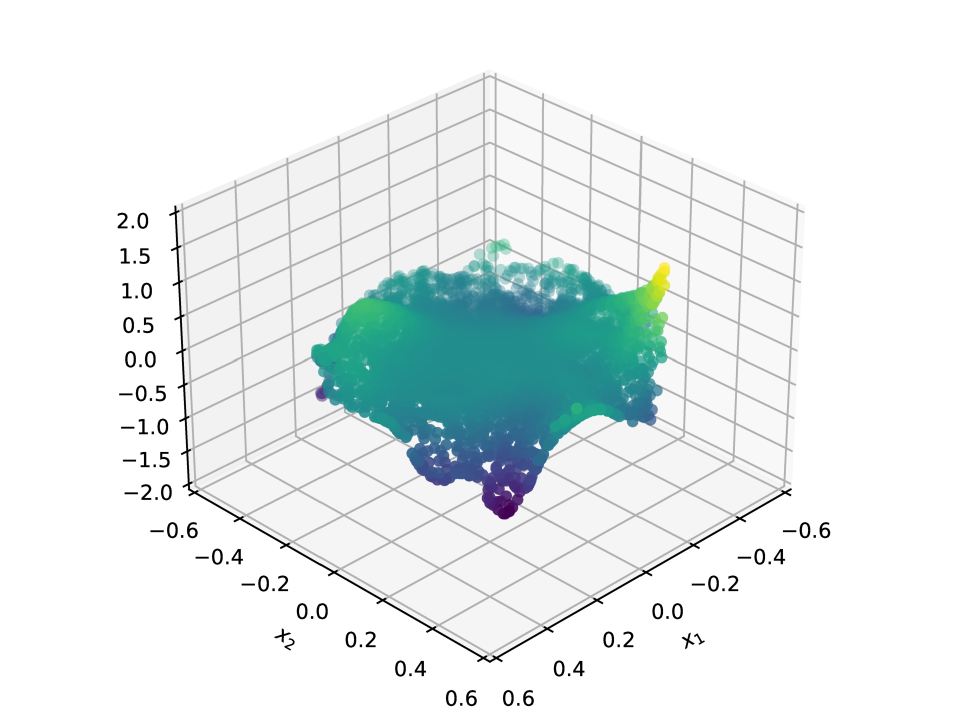}
        \caption{$R_{01}$ Patch 2}
    \end{subfigure}\\
    \begin{subfigure}{0.24\textwidth}
        \centering
        \includegraphics[width=0.98\textwidth]{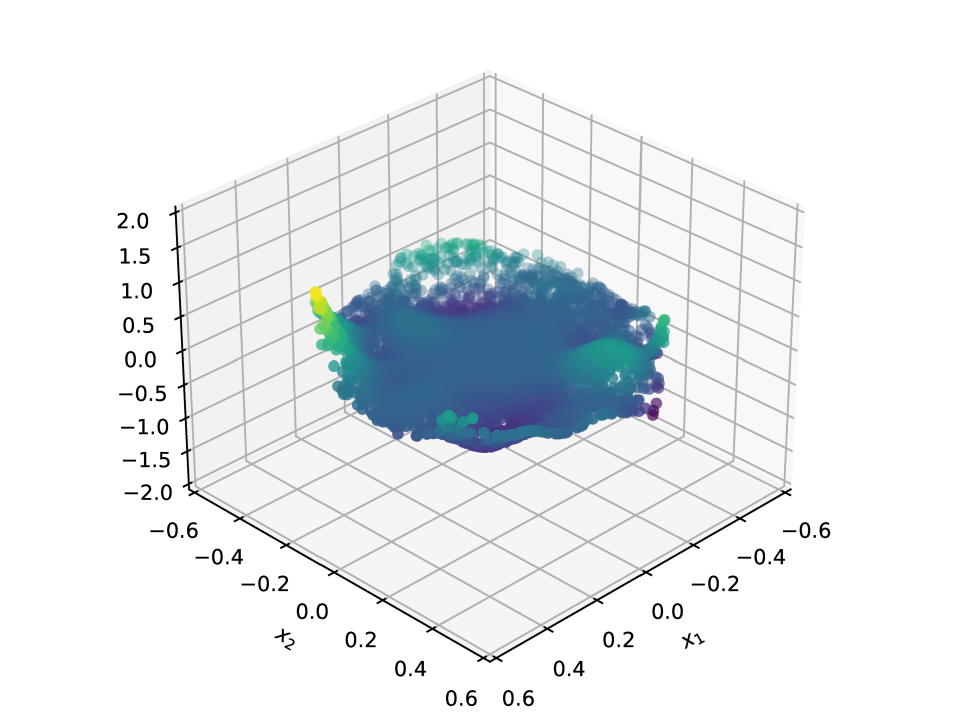}
        \caption{$R_{10}$ Patch 1}
    \end{subfigure} 
    \begin{subfigure}{0.24\textwidth}
        \centering
        \includegraphics[width=0.98\textwidth]{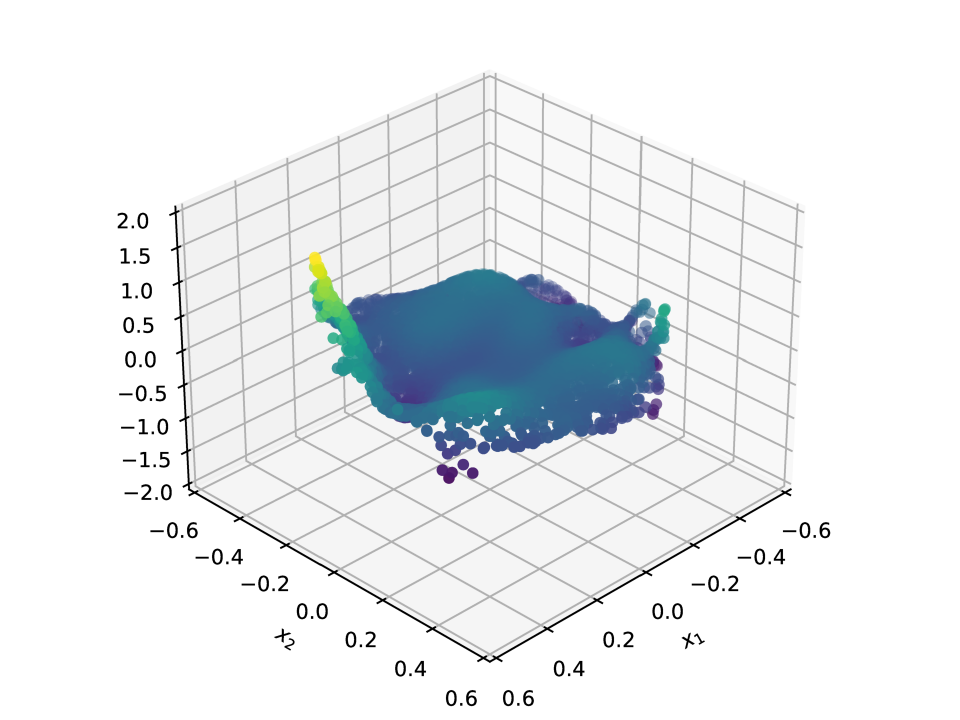}
        \caption{$R_{11}$ Patch 1}
    \end{subfigure} 
    \begin{subfigure}{0.24\textwidth}
        \centering
        \includegraphics[width=0.98\textwidth]{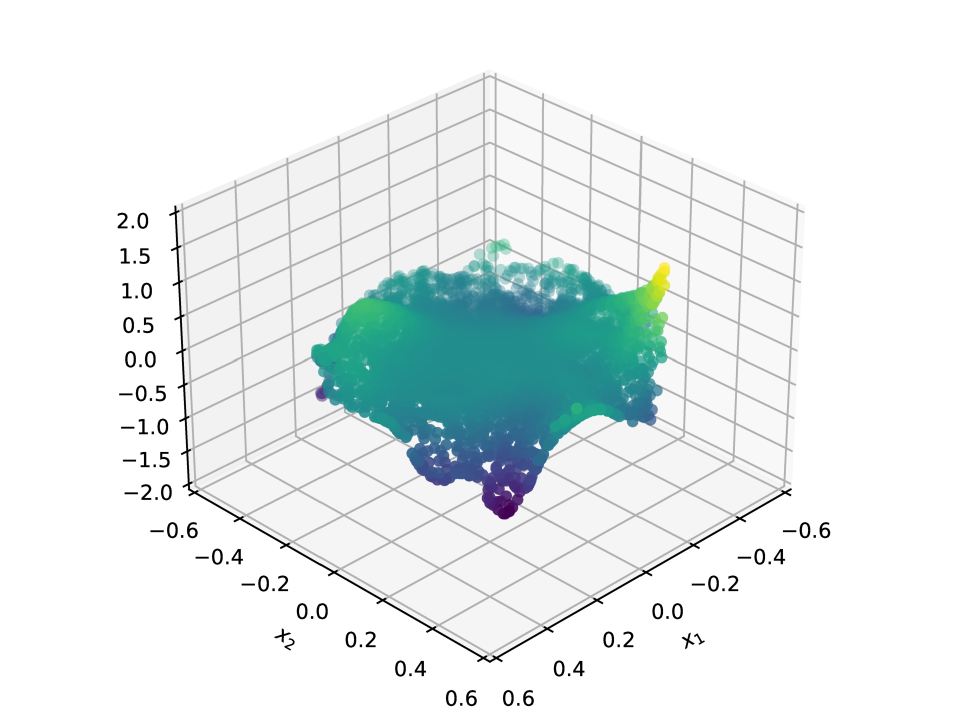}
        \caption{$R_{10}$ Patch 2}
    \end{subfigure} 
    \begin{subfigure}{0.24\textwidth}
        \centering
        \includegraphics[width=0.98\textwidth]{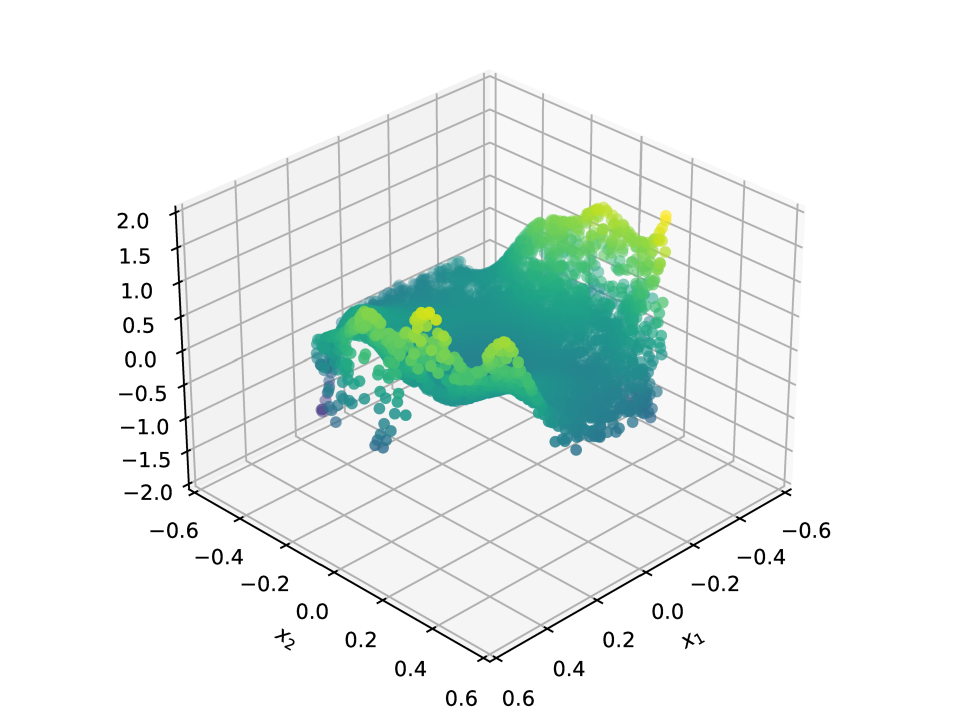}
        \caption{$R_{11}$ Patch 2}
    \end{subfigure}
    \caption{Visualisations of the Ricci tensors, $R_{ij}$, of the learnt metrics in 2d, on the 2 patches, trained for zero Einstein constant (such that $R_{ij} = 0$), and the metric's goal is to be Ricci-flat.}
    \label{fig:vis_2d0_R}
\end{figure*}

\begin{figure*}[hbtp!]
    \centering
    \begin{subfigure}{0.24\textwidth}
        \centering
        \includegraphics[width=0.98\textwidth]{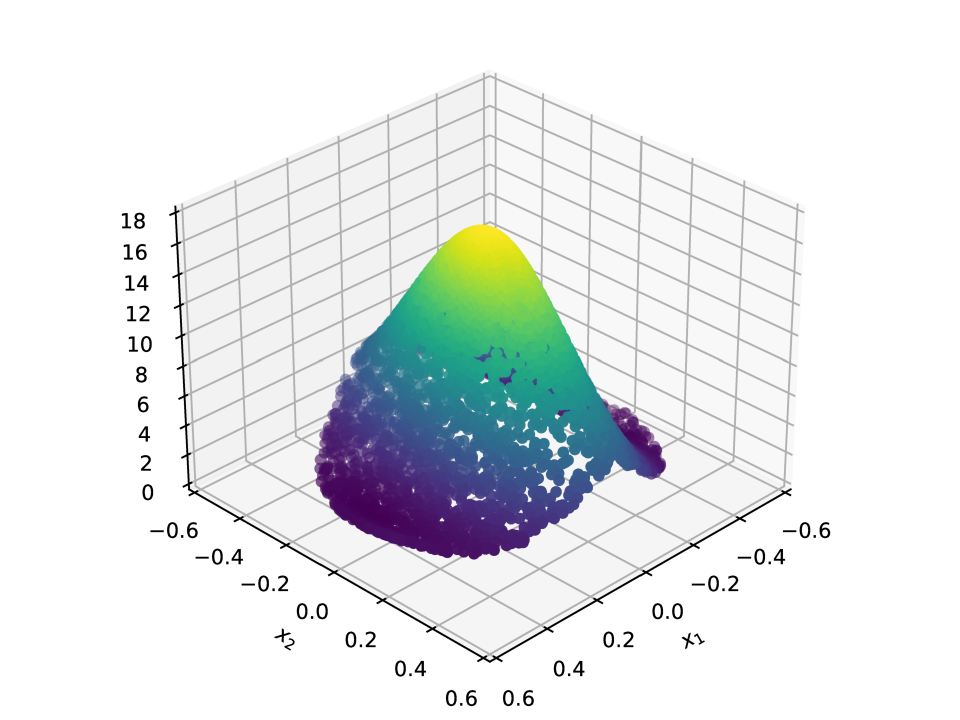}
        \caption{$g_{00}$ Patch 1}
        \label{fig:vis_2dneg_g001}
    \end{subfigure} 
    \begin{subfigure}{0.24\textwidth}
        \centering
        \includegraphics[width=0.98\textwidth]{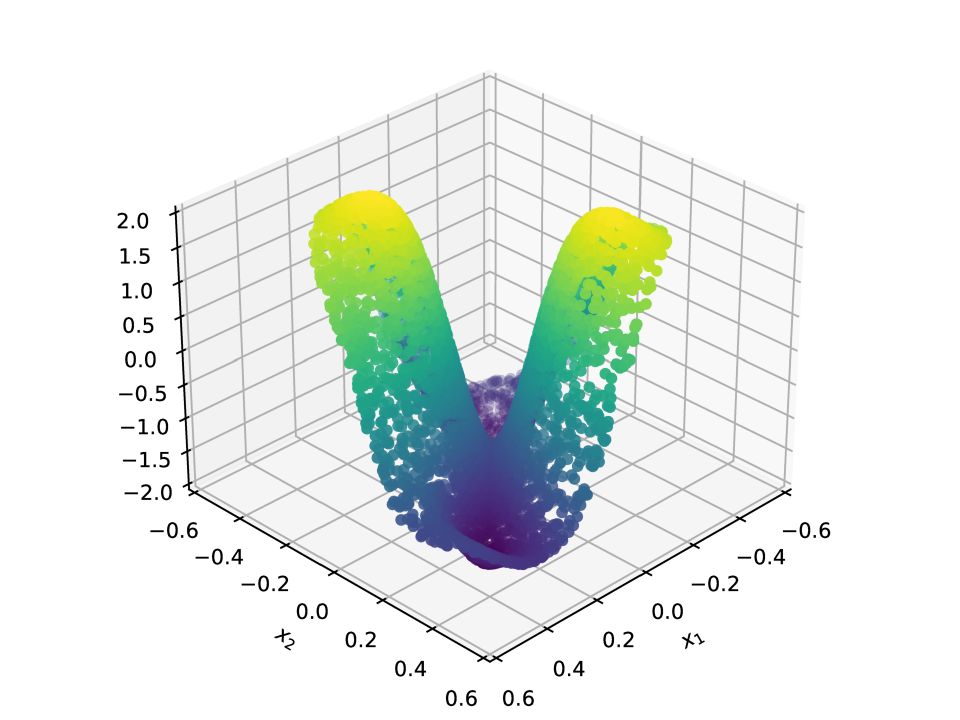}
        \caption{$g_{01}$ Patch 1}
    \end{subfigure} 
    \begin{subfigure}{0.24\textwidth}
        \centering
        \includegraphics[width=0.98\textwidth]{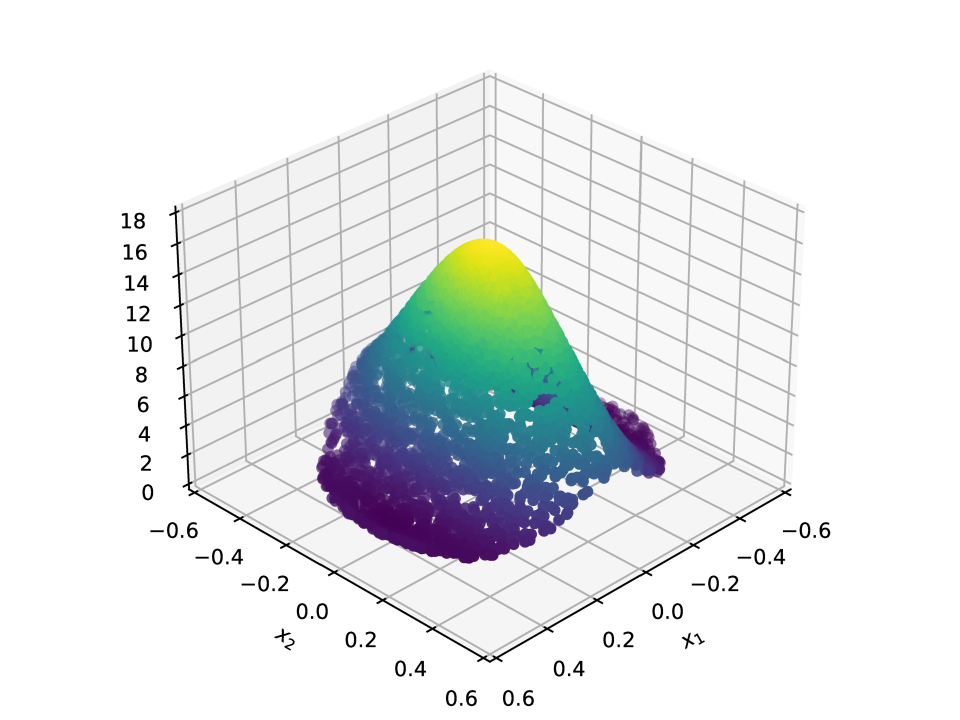}
        \caption{$g_{00}$ Patch 2}
    \end{subfigure} 
    \begin{subfigure}{0.24\textwidth}
        \centering
        \includegraphics[width=0.98\textwidth]{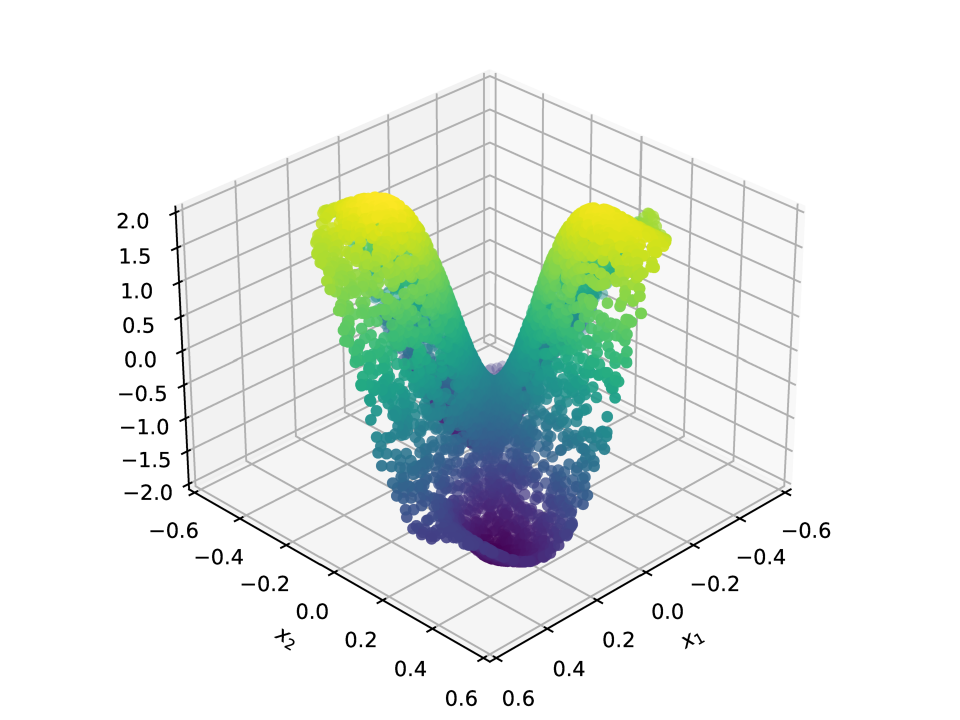}
        \caption{$g_{01}$ Patch 2}
    \end{subfigure}\\
    \begin{subfigure}{0.24\textwidth}
        \centering
        \includegraphics[width=0.98\textwidth]{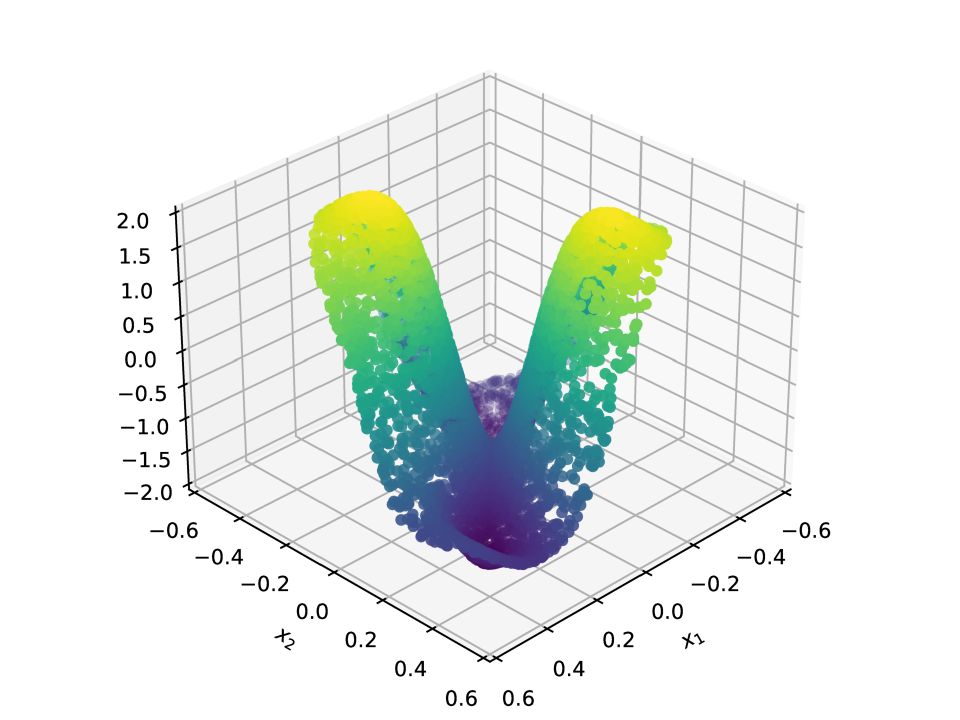}
        \caption{$g_{10}$ Patch 1}
    \end{subfigure} 
    \begin{subfigure}{0.24\textwidth}
        \centering
        \includegraphics[width=0.98\textwidth]{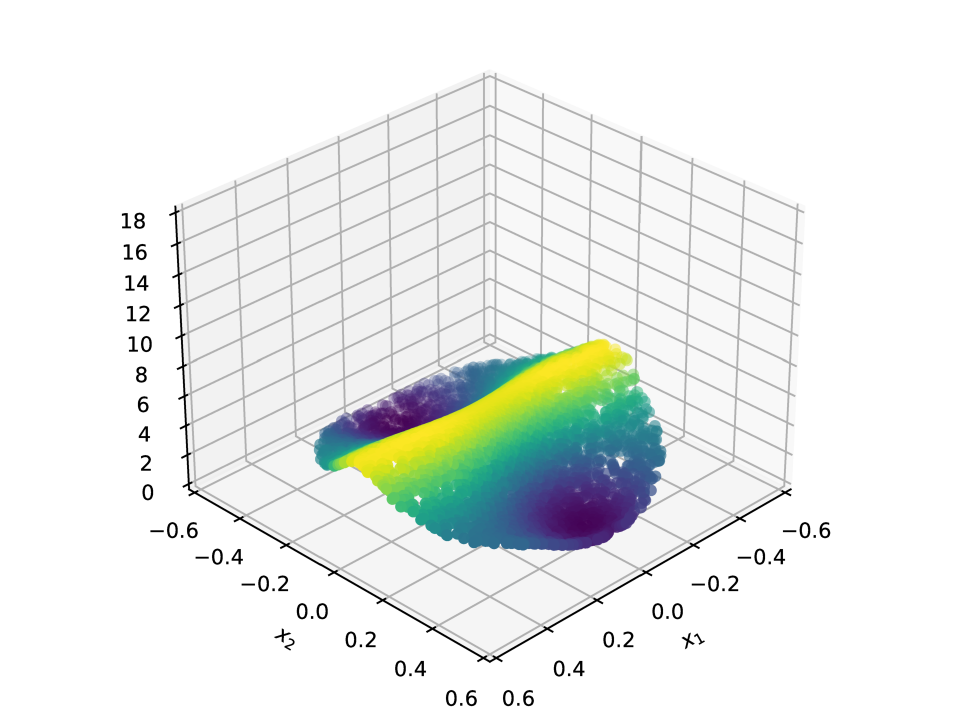}
        \caption{$g_{11}$ Patch 1}
    \end{subfigure} 
    \begin{subfigure}{0.24\textwidth}
        \centering
        \includegraphics[width=0.98\textwidth]{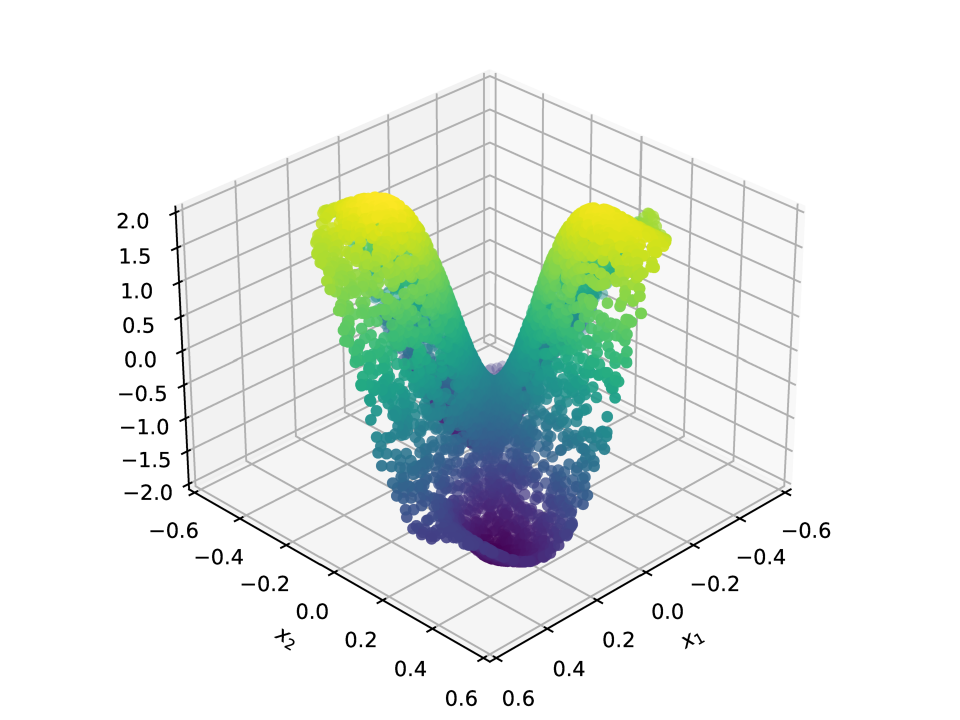}
        \caption{$g_{10}$ Patch 2}
    \end{subfigure} 
    \begin{subfigure}{0.24\textwidth}
        \centering
        \includegraphics[width=0.98\textwidth]{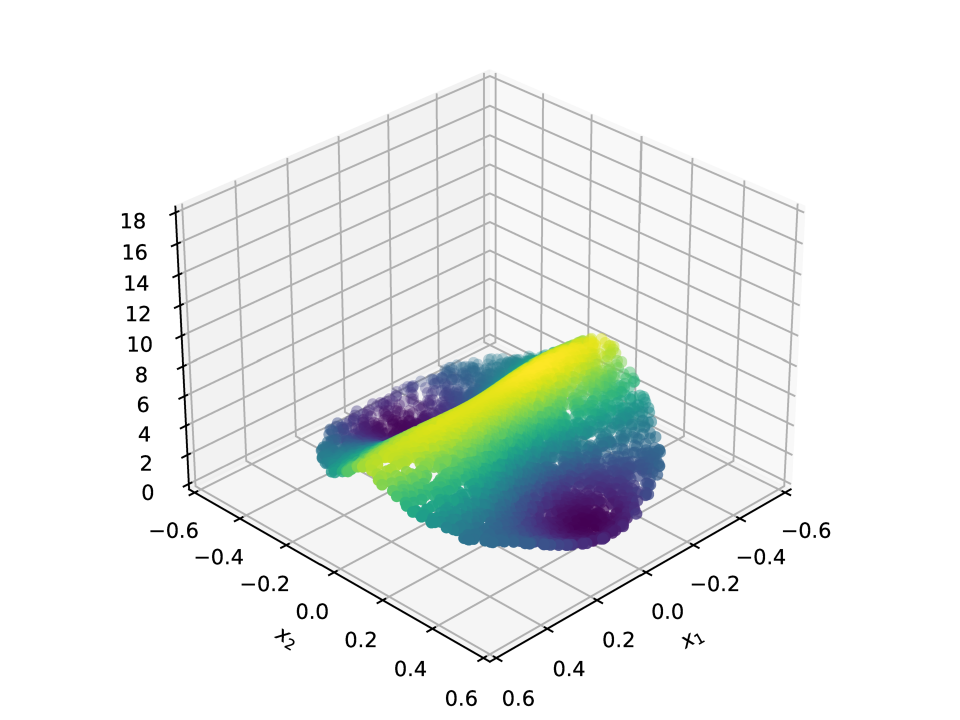}
        \caption{$g_{11}$ Patch 2}
    \end{subfigure}
    \caption{Visualisations of the learnt metrics, $g_{ij}$, in 2d, on the 2 patches, trained with negative Einstein constant (such that $R_{ij} = -g_{ij}$).}
    \label{fig:vis_2dneg_g}
\end{figure*}

\begin{figure*}[hbtp!]
    \centering
    \begin{subfigure}{0.24\textwidth}
        \centering
        \includegraphics[width=0.98\textwidth]{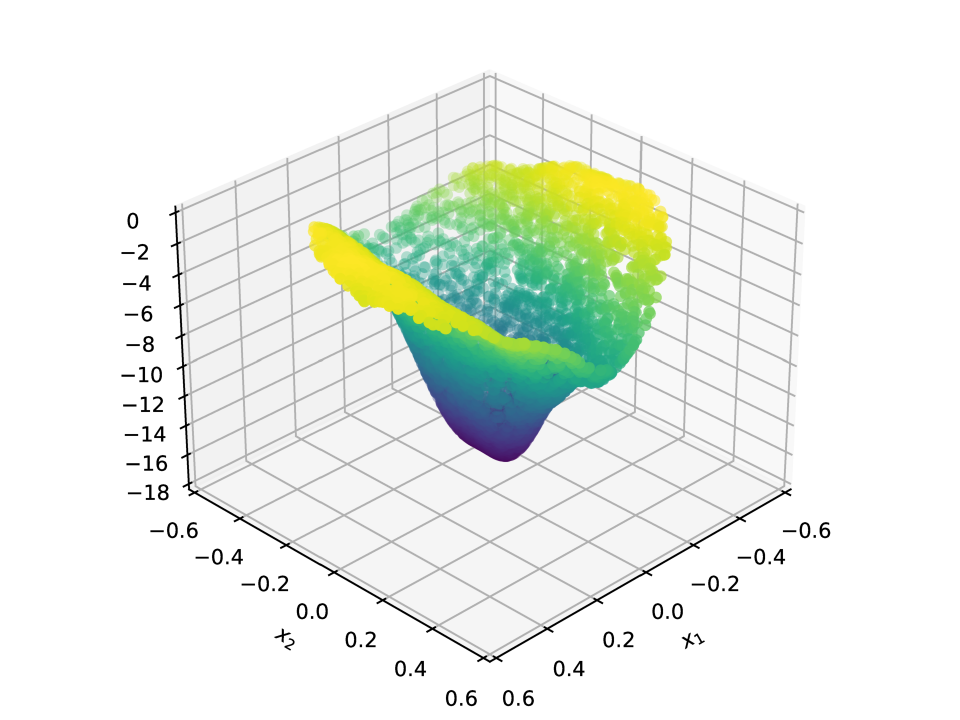}
        \caption{$R_{00}$ Patch 1}
    \end{subfigure} 
    \begin{subfigure}{0.24\textwidth}
        \centering
        \includegraphics[width=0.98\textwidth]{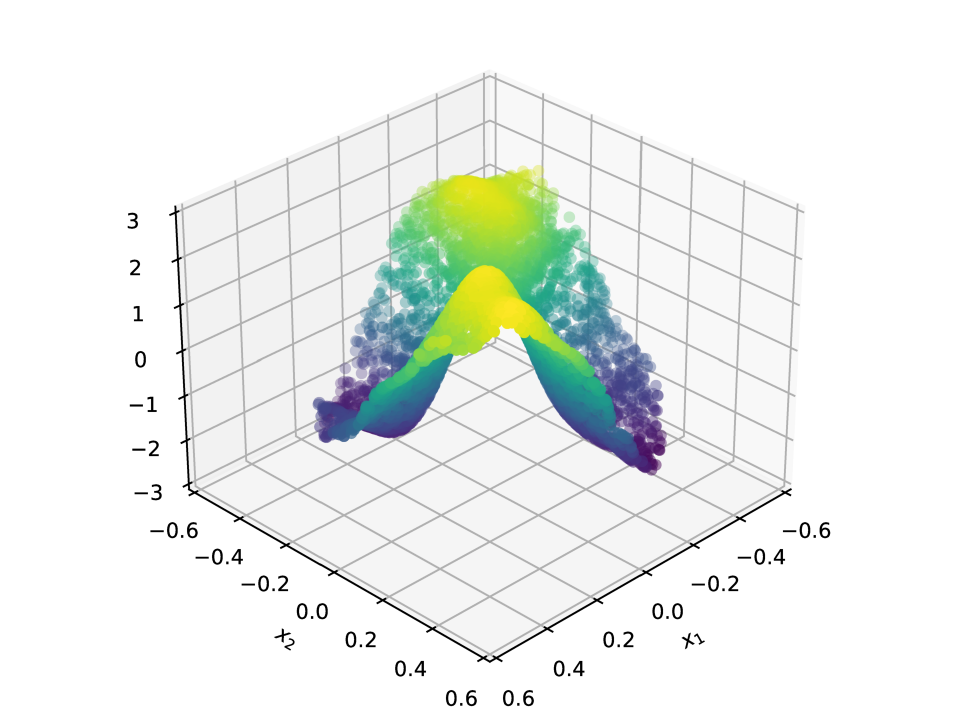}
        \caption{$R_{01}$ Patch 1}
    \end{subfigure} 
    \begin{subfigure}{0.24\textwidth}
        \centering
        \includegraphics[width=0.98\textwidth]{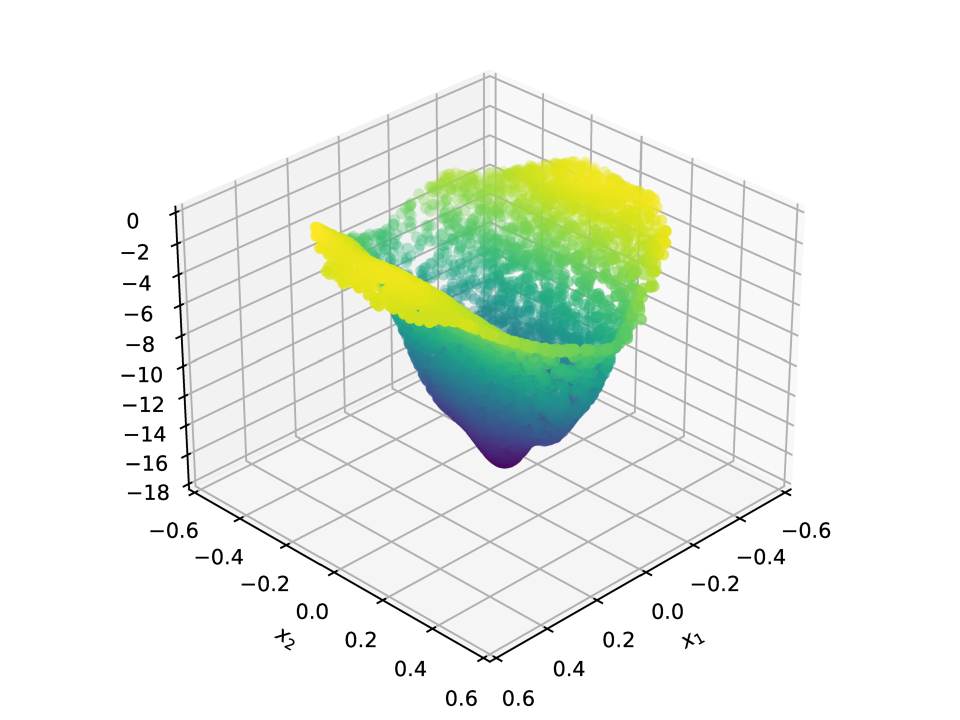}
        \caption{$R_{00}$ Patch 2}
    \end{subfigure} 
    \begin{subfigure}{0.24\textwidth}
        \centering
        \includegraphics[width=0.98\textwidth]{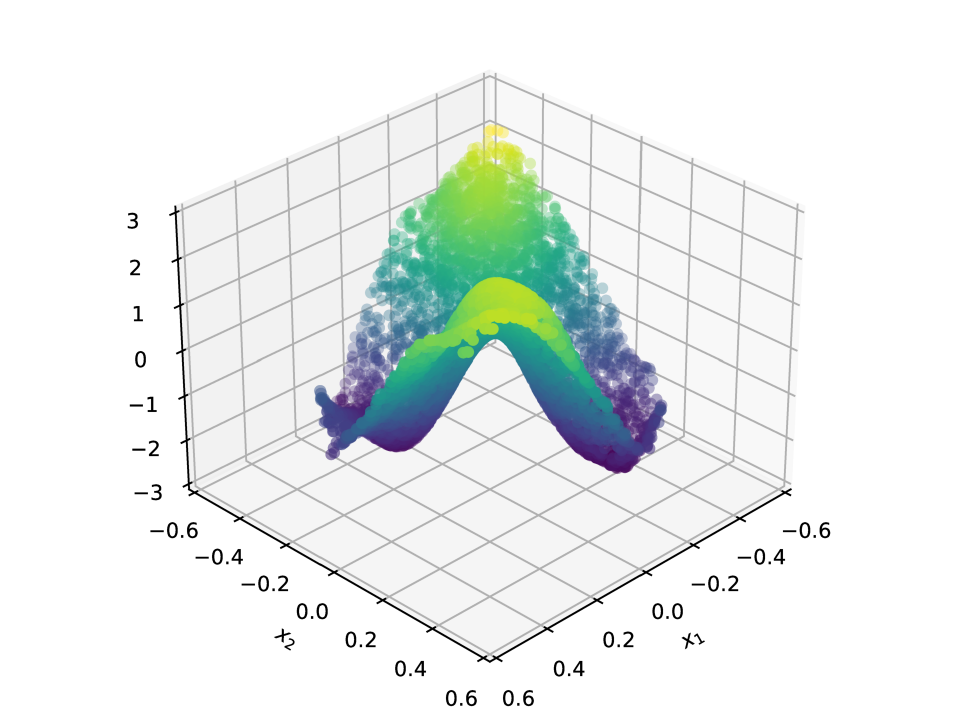}
        \caption{$R_{01}$ Patch 2}
    \end{subfigure}\\
    \begin{subfigure}{0.24\textwidth}
        \centering
        \includegraphics[width=0.98\textwidth]{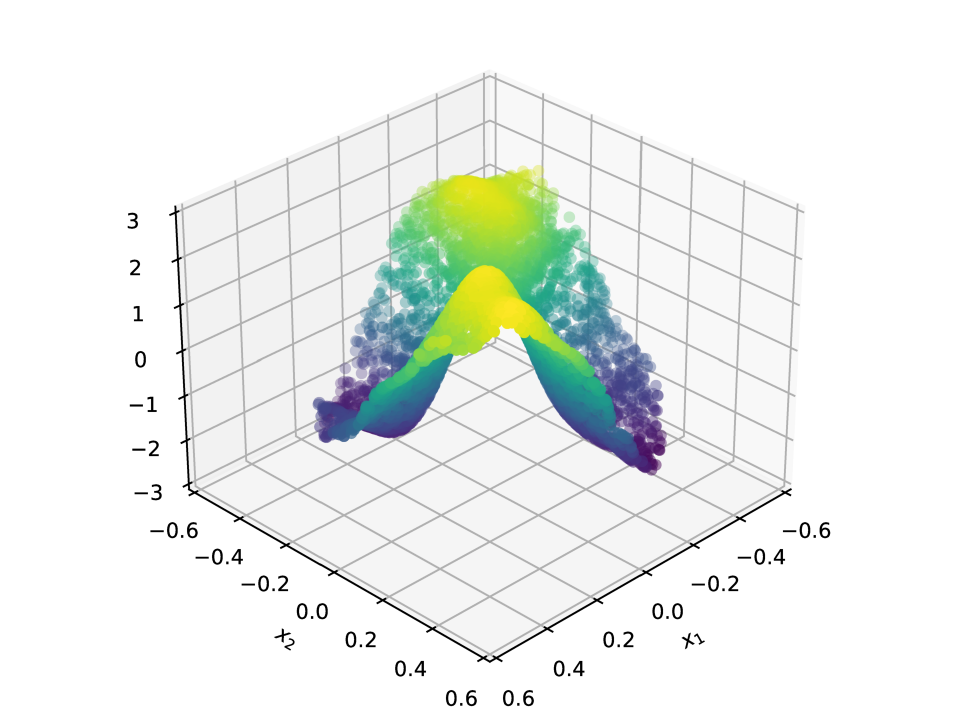}
        \caption{$R_{10}$ Patch 1}
    \end{subfigure} 
    \begin{subfigure}{0.24\textwidth}
        \centering
        \includegraphics[width=0.98\textwidth]{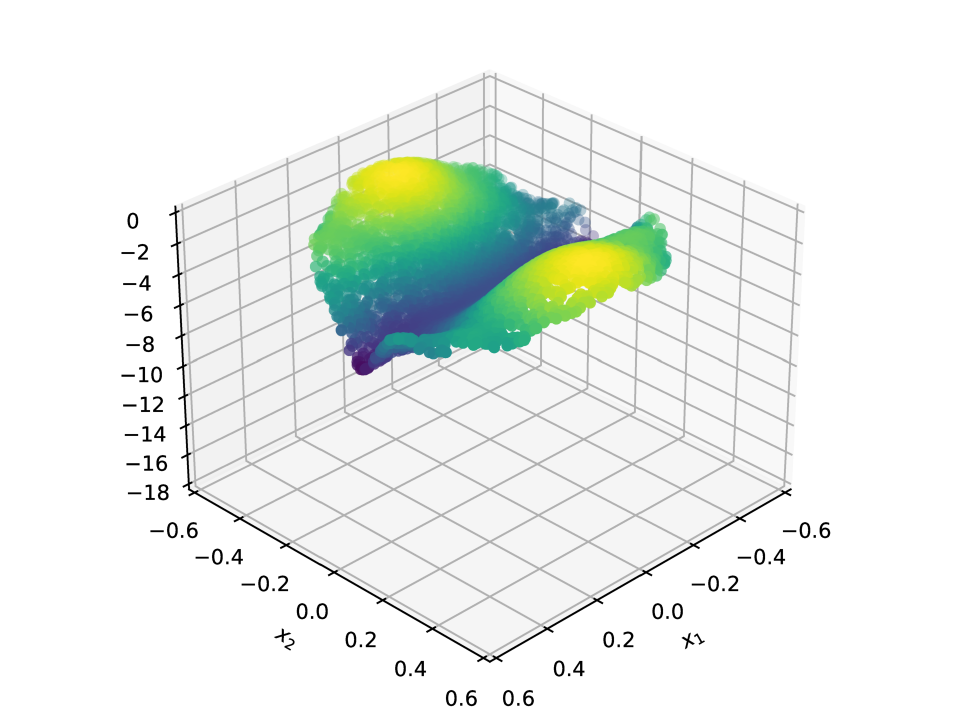}
        \caption{$R_{11}$ Patch 1}
    \end{subfigure} 
    \begin{subfigure}{0.24\textwidth}
        \centering
        \includegraphics[width=0.98\textwidth]{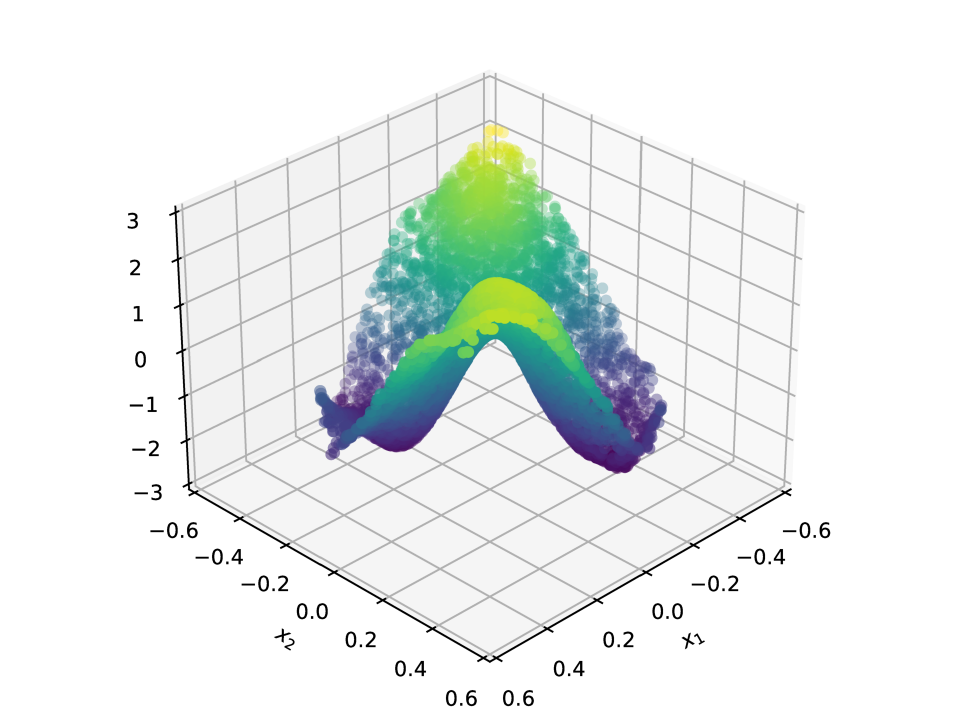}
        \caption{$R_{10}$ Patch 2}
    \end{subfigure} 
    \begin{subfigure}{0.24\textwidth}
        \centering
        \includegraphics[width=0.98\textwidth]{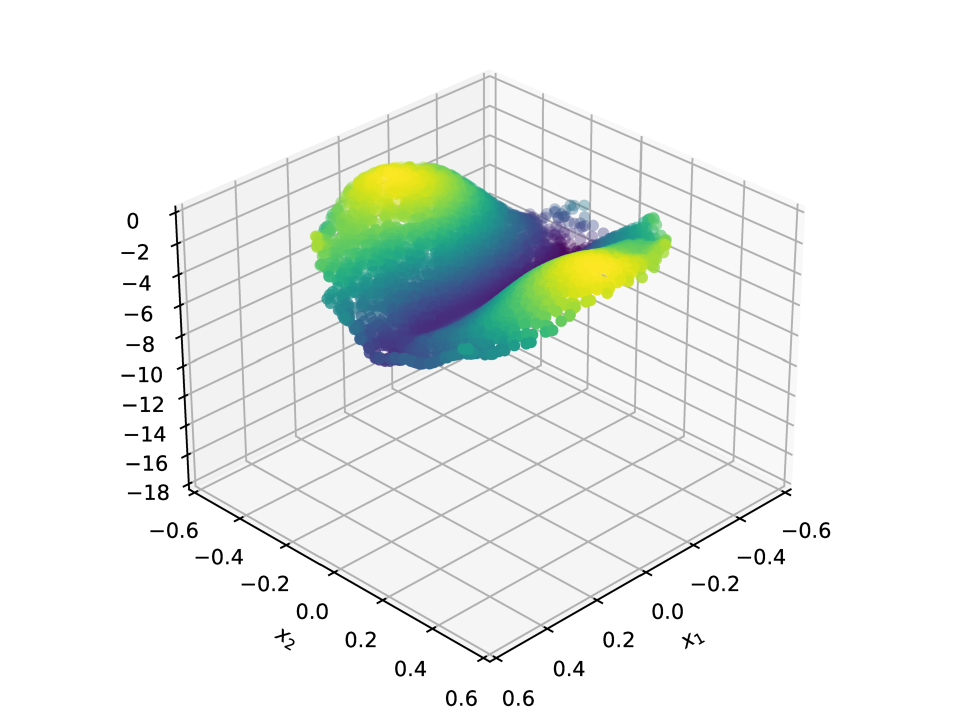}
        \caption{$R_{11}$ Patch 2}
    \end{subfigure}
    \caption{Visualisations of the Ricci tensors, $R_{ij}$, of the learnt metrics in 2d, on the 2 patches, trained for negative Einstein constant (such that $R_{ij} = -g_{ij}$).}
    \label{fig:vis_2dneg_R}
\end{figure*}

To extend the visual interpretation of the metric learning, as shown in Figures \ref{fig:vis_2dpos_g} \& \ref{fig:vis_2dpos_R}, here equivalent plots are shown for example runs from the 2d $\lambda \in \{0,-1\}$ investigations.
For $\lambda=0$, the metric components in both patches are shown in Figure \ref{fig:vis_2d0_g}, whilst the equivalent Ricci tensor components are shown in Figure \ref{fig:vis_2d0_R}.
Then for $\lambda = -1$, the metric components in both patches are shown in Figure \ref{fig:vis_2dneg_g}, whilst the equivalent Ricci tensor components are shown in Figure \ref{fig:vis_2dneg_R}. 

For $\lambda=0$ the model is clearly trying to set all the Ricci tensor components to 0, however it fails with clear instabilities it cannot avoid due to the geometric obstruction to existence of Ricci-flat metrics.
Whereas for $\lambda=-1$ the respective Ricci components look somewhat like inversions of the metric components as the model tries to match these Ricci components to the negative values of the metric.
However, again there are clear instabilities around the patch centre, and at the edges of the plotting restriction where the overlap region is defined, where the model expectedly cannot overcome these geometric obstructions.

A final comment, is that the shape of the components looks somewhat similar between the $\lambda$ values, for example with conical-like shapes for the $(0,0)$ components.
Upon further inspection of these components one can start to see the differing curvatures.
In Figure \ref{fig:vis_2dpos_g001} (for $\lambda=+1$) the cone outline from the centre along the $x_1$ axes the outline starts to curve up, whereas in Figure \ref{fig:vis_2d0_g001} (for $\lambda=0$) the outline is quite flat, and finally in Figure \ref{fig:vis_2dneg_g001} (for $\lambda=-1$) the outline curves downwards.
These opposing visual curvatures match the expected behaviour, and demonstrate the subtlety in the learning of Einstein metrics via this highly non-linear and extremely sensitive Einstein equation.

\newpage
%%%%%%%%%%%%%%%%%%%%%%%%%%%%%%%%%%%%%%%%%%%%%%%%%
\section{Details of Manifold Sampling}\label{app:sampling}
The boundary of the open ball patches represent the infinite limits of the stereographic real plane and where the sphere projections break down, unsurprisingly it is here that the greatest numerical instabilities are seen.
Conversely, points near the ball centre in patch 1 map to near the boundary in patch 2, and thus optimal sampling to avoid instabilities skews generation to the parts of the patch away from these extremities.
Additionally, since the patch gluing conditions require each patch only up the $r_m + \varepsilon$, to ensure consistent gluing at the overlap points should be dense near this midpoint.

From these motivations, the ball sampling procedure used polar coordinates for the patch, implementing a modified beta function for the radii, and sampled the angles uniformly; then transforming into the Euclidean coordinate inputs.
The beta function skews sampling to prioritise radii near to $r_m$; and to ensure the patches are sampled symmetrically, half the requested number of samples are generated using the same beta function for patch 2 and are transformed back to patch 1.
The general beta function is defined by the distribution
\begin{equation}\label{eq:beta_fn}
    f(r;\alpha,\beta) := \frac{r^{\alpha - 1}(1-r)^{\beta - 1}}{\int_0^1 t^{\alpha - 1}(1 - t)^{\beta - 1}dt}\;,
\end{equation}
for $r$ the sampled variable in the domain $[0,1]$, for us the radius of the sampled point in polar coordinates of the ball patch, and the parameters $\alpha,\beta>0$ control the distribution shape.

The mean of this distribution is $\frac{\alpha}{\alpha + \beta}$, therefore to encourage sampling to be symmetric between the patches we set this mean to equal the radial midpoint $r_m = \sqrt{2}-1$; such that rearranging sets $\beta = \alpha(\frac{1}{r_m}-1) \sim 1.41 \alpha$.
However, despite the sample mean now being symmetric under the patch change, the rate of sampling density change is still not symmetric.
Therefore to rectify this, half the sampled radii are transformed using \eqref{eq:radii_patchchange}, such that the full list of sampled radii are symmetric under the patch change and both patches are then sampled equivalently.
The value of $\alpha$ then determines how skewed the distribution is, when $\alpha=\beta=1$ the numerator of \eqref{eq:beta_fn} becomes 1 and the distribution is uniform; for testing samples we take the near uniform limit with $\alpha=1$ and $\beta$ defined as above.
In the limit $\alpha << 1$ the distribution skews to prioritise the bounds of the $[0,1]$ interval, whilst the $\alpha >> 1$ limit prioritises the middle of the interval.
The latter is desired to optimise overlap and avoid numerical instability, hence after some heuristic experimentation a value of $\alpha=4$ was selected for the training samples.

\begin{figure}[!t]
    \centering
    \begin{subfigure}{0.32\textwidth}
        \centering
        \includegraphics[width=0.98\textwidth]{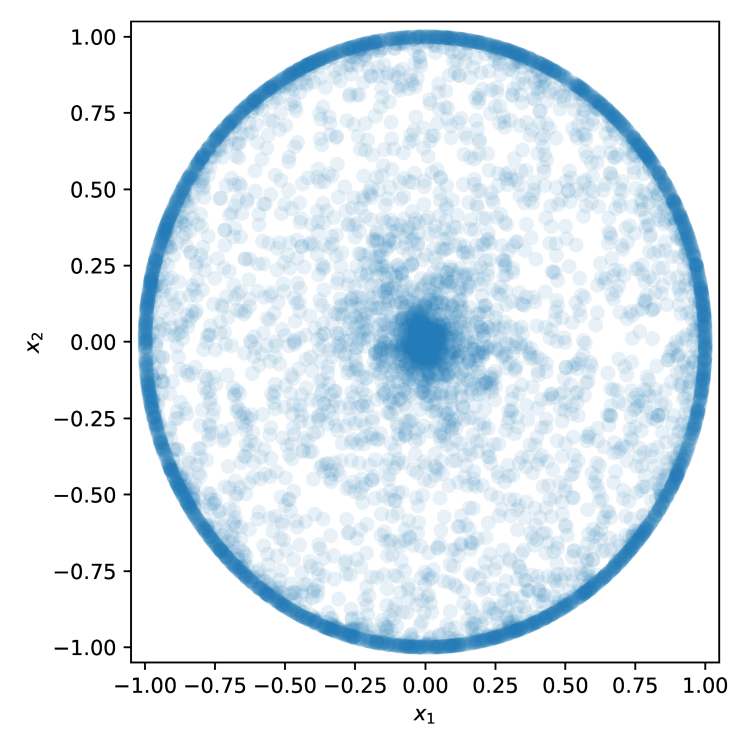}
        \caption{$\alpha=0.1$}
    \end{subfigure} 
    \begin{subfigure}{0.32\textwidth}
        \centering
        \includegraphics[width=0.98\textwidth]{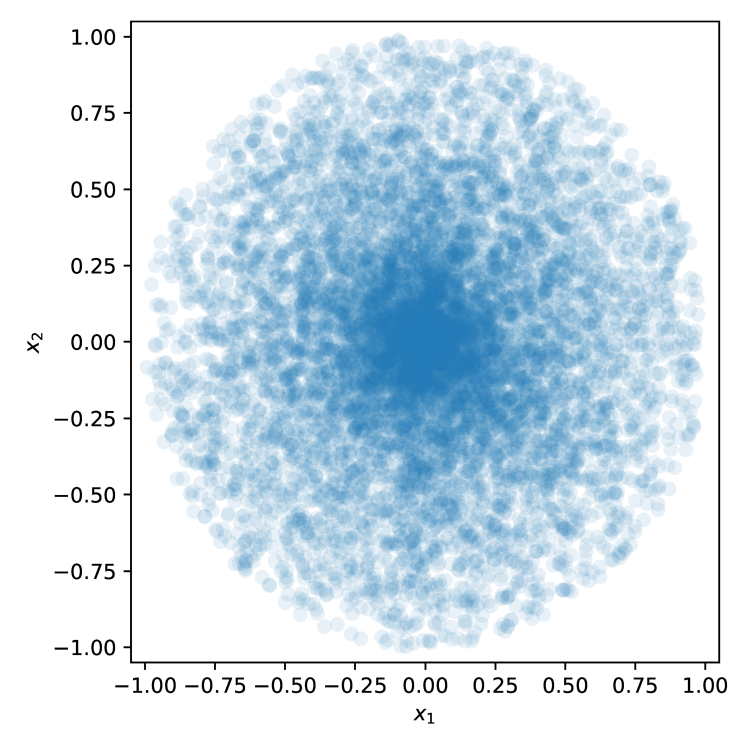}
        \caption{$\alpha=1$}
    \end{subfigure}
    \begin{subfigure}{0.32\textwidth}
        \centering
        \includegraphics[width=0.98\textwidth]{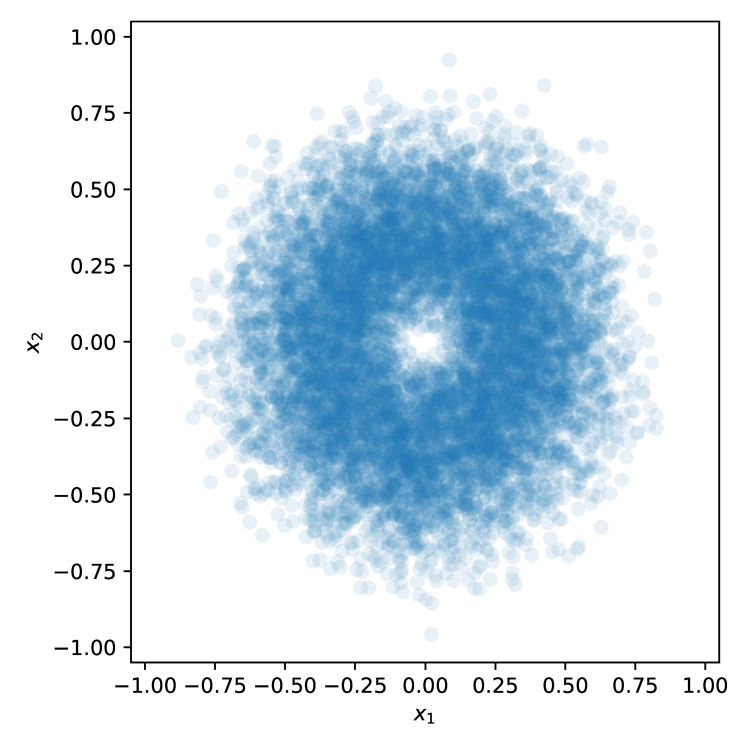}
        \caption{$\alpha=4$}
    \end{subfigure} 
    \caption{Point samples in a 2d ball patch using the modified Beta function sampling scheme. The scheme sets the $\beta$ value to centre sampling at $r_m$, and explicitly symmetrises such that these points in the other patch have the same distribution. Plots show the behaviour for varying $\alpha$.}
    \label{fig:sampling}
\end{figure}

To illustrate how the sampling in a patch varies with $\alpha$, Figure \ref{fig:sampling} shows a single patch sampled with $\alpha \in \{0.1, 1, 4\}$, due to the symmetric nature of the scheme the other patch sampling distribution looks identical.
The sampling code is highly vectorised to ensure hyper-efficient sample generation, and is released with the AInstein codebase. We include a Jupyter \cite{jupyter} notebook with interactive visualisations for varying $\alpha$.
We emphasise that $\alpha=4$ was used for training data, and $\alpha=1$ was used for testing data.

%%%%%%%%%%%%%%%%%%%%%%%%%%%%%%%%%%%%%%%%%%%%%%%%%
\section{Details of Data Filters}\label{app:filters}
\setcounter{equation}{1}

In order to vary priority of sample points in various loss components filters were designed to apply appropriate weightings based on the sample point radii. % or the absolute values of the metric outputs. %and to evade the ``zero-metric'' ($g_{ij} \sim 0$) as a learning attractor point,
Two filters were designed and used in the final model, as mentioned in §\ref{sec:bkg_ml}, and are detailed here.

\subsubsection*{Radial filter in the Einstein loss}
The radial filter in the Einstein loss is of the form
\begin{align}
    \exp\left[-\left(\frac{|x|- c_e}{w_e}\right)^{t_e}\right] \, ,
\end{align}
with parameters $(t_e, c_e, w_e)$, where $t_e$ is even. This is a Gaussian-shaped object, where $t_e$ controls how steep the edges are. Very large $t_e$ yields a very good approximation of the rectangular function. $c_e$ is the centre of the Gaussian, and is set it to be zero for simplicity in this case, since we are not concerned with negative values of the radius. $w_e$ controls the width, which therefore determines what portion of the ball is taken into account for this loss. A plot of this filter, with an illustration of what feature each parameter controls, is shown in Figure \ref{fig:filters_e}. The plot refers exactly to the parameters which were used to collect our results.

\subsubsection*{Radial filter in the overlap loss}
\begin{figure}[!t]
    \centering
    \begin{subfigure}{0.32\textwidth}  
        \centering
        \includegraphics[width=0.98\textwidth]{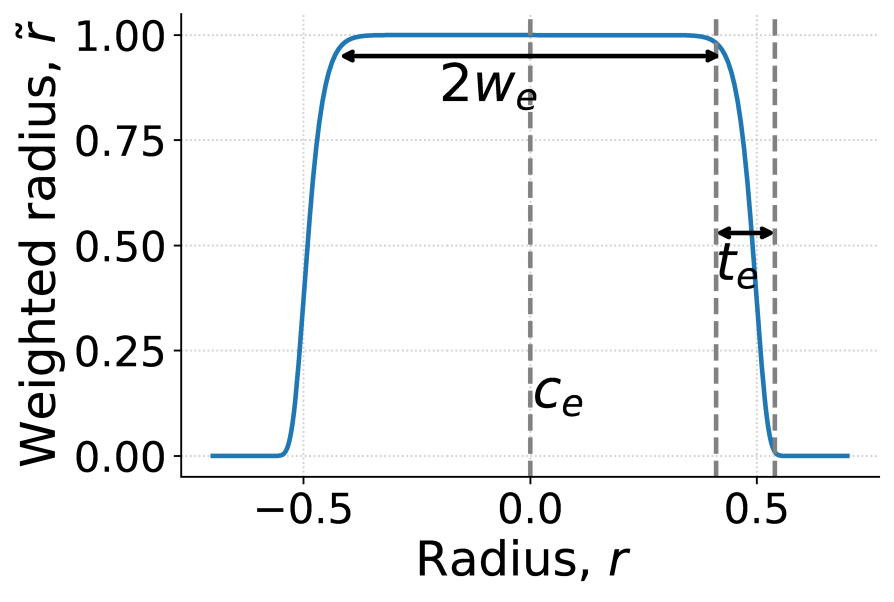}
        \caption{Einstein radial filter}
        \label{fig:filters_e}
    \end{subfigure} 
    \hfill
    \begin{subfigure}{0.32\textwidth}  
        \centering
        \includegraphics[width=0.98\textwidth]{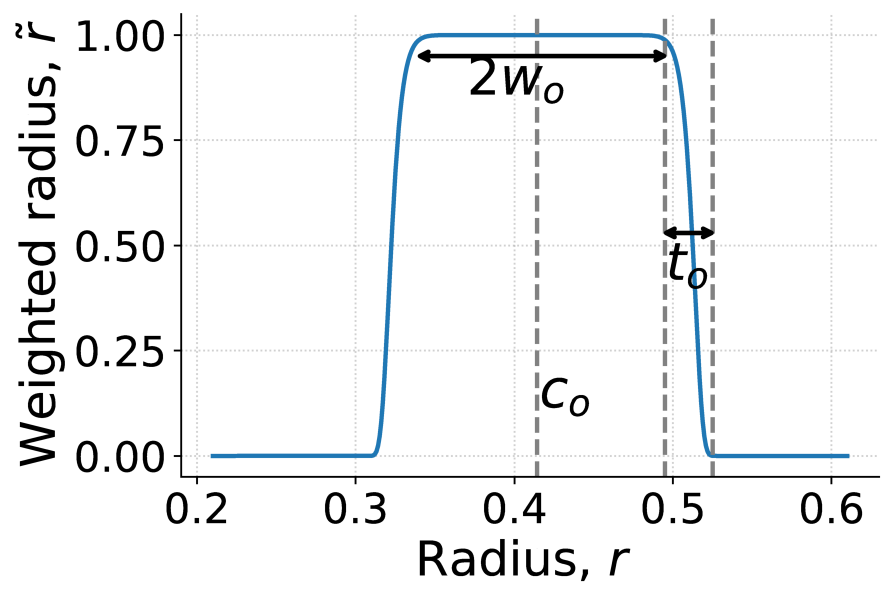}
        \caption{Overlap radial filter} 
        \label{fig:filters_o}
    \end{subfigure} 
    \hfill
    \begin{subfigure}{0.32\textwidth} 
        \centering
        \includegraphics[width=0.98\textwidth]{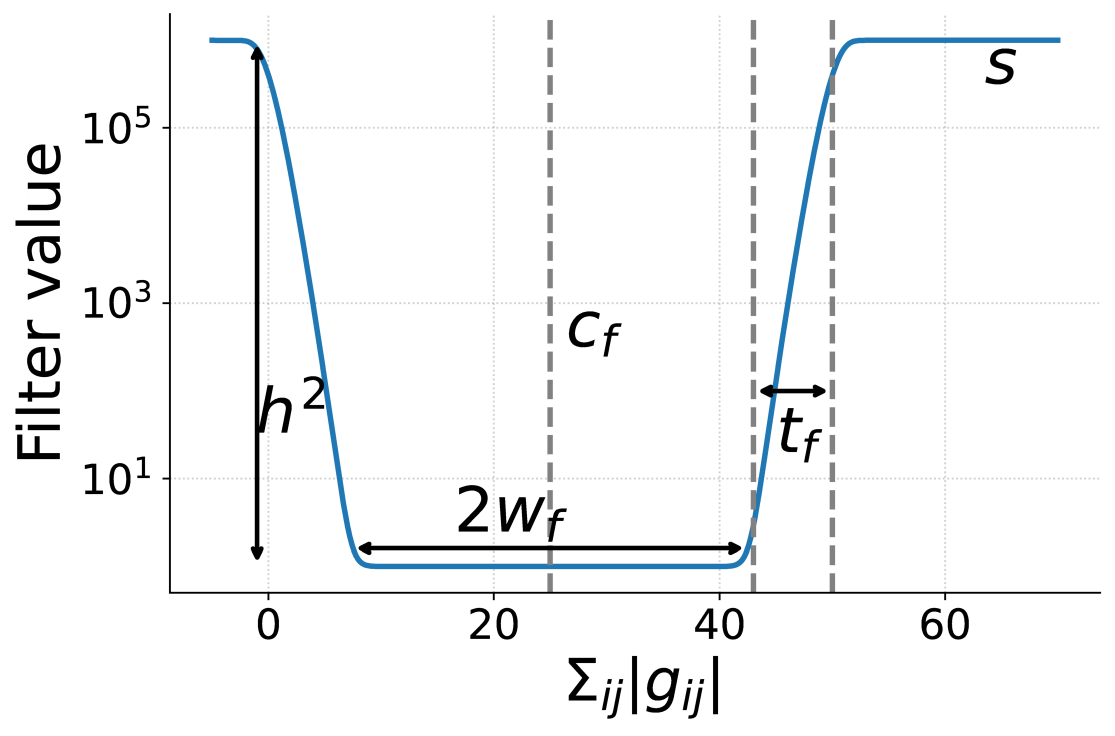}
        \caption{Finiteness filter} 
        \label{fig:filters_f}
    \end{subfigure} 
    \caption{Plots of the filter functions used in the loss components including the Einstein loss $\mathcal{L}^{\text{Einstein}}$ (Figure \ref{fig:filters_e}), the overlap loss $\mathcal{L}^{\text{Overlap}}$ (Figure \ref{fig:filters_o}), and the finiteness loss $\mathcal{L}^{\text{Finiteness}}$ (Figure \ref{fig:filters_f}).}
    \label{fig:filters}
\end{figure}

The radial filter used in the overlap loss has the same form, but it involves different choices of parameters:
\begin{align}
    \exp\left[-\left(\frac{|x|-c_o}{w_o}\right)^{t_o}\right] \, ,
\end{align}
now labelled $(t_o, c_o, w_o)$. As before, $c_o$ controls the centre of the Gaussian-like curve, $w_o$ its width and $t_o$ how vertical the walls are. In this case, however, the filter should isolate the overlap region (i.e. the annulus between $\frac{1 - (r_m +  \varepsilon)}{1 + (r_m +  \varepsilon)}$ and $r_m +  \varepsilon$), while setting to zero the other regions of the ball. A plot of the specific filter used in our runs is shown in Figure \ref{fig:filters_o}.

%%%%%%%%%%%%%%%%%%%%%%%%%%%%%%%%%%%%%%%%%%%%%%%%%
\section{Neural Network Hyperparameters}
\label{app:hyperparams}

In order to arrive at the set of hyperparameters stated above, we performed an extensive sweep for the 2d model with the experiment management tool Weights and Biases \cite{wandb}. The final training hyperparametes are presented in Table \ref{tab:hyperparams}. We release the code for this feature with the package, such that it may be readily utilised by those possessing an API key.

In this work, the activation function $\sigma(x)$ is chosen to be the Gaussian Error Linear Unit (GELU)\footnote{Indeed one may choose $\sigma(x) = \text{ReLU}(x)$ here, however the constant behaviour for $x \leq 0$ leads to numerical instability for this use-case; derivatives of the network must be taken to calculate the Ricci tensor.}. Likewise, for the finiteness loss filter we use the hyperparameter choices specified in Table \ref{tab:filters}.
\begin{table}[h!]
    \centering
    \begin{tabular}{|c|c|}
        \hline
        \textbf{Hyperparameter} & \textbf{Value} \\ \hline
        Training epochs & 500 \\ \hline
        Training samples & 10k (2D, 3D), 100k (4D, 5D) \\ \hline
        Batch size & 100 \\ \hline
        Learning rate (max,min) & (0.005, 0.001) \\ \hline
        Learning rate schedule & Cosine \\ \hline
        Optimizer & Adam (\cite{kingma2017adammethodstochasticoptimization}) \\ \hline
        Patch submodel layers & 3 Dense layers \\ \hline
        Neurons per layer & 64 \\ \hline
        Activation function & GELU \\ \hline
        Biases & On \\ \hline
    \end{tabular}
    \caption{Hyperparameters for the Einstein metric machine learning model training.}
    \label{tab:hyperparams}
\end{table}
\begin{table}[ht]
    \centering
    \begin{tabular}{|c|c|}
        \hline
        \textbf{Filter parameter} & \textbf{Value} \\ \hline
        $h$ & 1000 \\ \hline
        $c_f$ & 25 \\ \hline
        $w_f$ & 25 \\ \hline
        $t_f$ & 20 \\ \hline
        $s$ & 0.2 \\ \hline
    \end{tabular}
    \caption{Parameters for the finiteness loss filter.}
    \label{tab:filters}
\end{table}

%%%%%%%%%%%%%%%%%%%%%%%%%%%%%%%%%%%%%%%%%%%%%%%%%
%\section{Brief Introduction to Neural Networks}\label{app:ml_intro}

%%%%%%%%%%%%%%%%%%%%%%%%%%%%%%%%%%%%%%%%%%%%%%%%%
\end{document}